\documentclass[12pt]{iopart}

\newcommand{\bra}[1]{\left\langle #1\right|}
\newcommand{\ket}[1]{\left| #1\right\rangle}
\usepackage{amssymb}
\usepackage{bbold}
\usepackage{graphicx}
\usepackage{dcolumn}
\usepackage{amsfonts}
\usepackage{bm}
\usepackage{color}
\usepackage{hyperref}
\hypersetup{colorlinks=true,breaklinks,linkcolor=red,citecolor=blue}
\begin{document}

\title[Novel applications of the dispersive optical model]{Novel applications of the dispersive optical model}

\author{W. H. Dickhoff$^1$, R. J. Charity$^2$, and M. H. Mahzoon$^{1,3}$}

\address{${}^1$Department of Physics, Washington University, St. Louis, MO 63130, USA}
\ead{wimd@wuphys.wustl.edu}

\address{${}^2$Department of Chemistry, Washington University, St. Louis, MO 63130, USA}
\ead{charity@wustl.edu}

\address{${}^3$Department of Physics and Astronomy, Michigan State University, East Lansing, MI 48824, USA}
\ead{hossein@pa.msu.edu}

\vspace{10pt}
\begin{indented}
\item[]June 2016
\end{indented}

\begin{abstract}
A review of recent developments of the dispersive optical model (DOM) is presented.
Starting from the original work of Mahaux and Sartor, several necessary steps are developed and illustrated which increase the scope of the DOM allowing its interpretation as generating an experimentally constrained functional form of the nucleon self-energy. 
The method could therefore be renamed as the dispersive self-energy method.

The aforementioned steps include the introduction of simultaneous fits of data for chains of isotopes or isotones allowing a data-driven extrapolation for the prediction of scattering cross sections and level properties in the direction of the respective drip lines. 
In addition, the energy domain for data was enlarged to include results up to 200 MeV where available. 

An important application of this work was implemented by employing these DOM potentials to the analysis of the (\textit{d,p}) transfer reaction  using the adiabatic distorted wave approximation (ADWA).
We review these calculations which suggest that physically meaningful results are easier to obtain by employing DOM ingredients as compared to the traditional approach which relies on a phenomenologically-adjusted bound-state wave function combined with a global (nondispersive) optical-model potential.
Application to the exotic  ${}^{132}$Sn nucleus also shows great promise for the extrapolation of DOM potentials towards the drip line with attendant relevance for the physics of FRIB.
We note that the DOM method combines structure and reaction information on the same footing providing a unique approach to the analysis of exotic nuclei.

We illustrate the importance of abandoning the custom of representing the non-local  HF potential in the DOM by an energy-dependent local potential as it impedes the proper normalization of the solution of the Dyson equation.
This important step allows for the interpretation of the DOM potential as representing the nucleon self-energy permitting the calculations of spectral amplitudes and spectral functions above and below the Fermi energy.
The latter feature provides access to quantities like the momentum distribution, charge density, and particle number which were not available in the original work of Mahaux and Sartor.

When employing a non-local HF potential, but local dispersive contributions (as originally proposed by Mahaux and Sartor), we illustrate that it is impossible to reproduce the particle number and the measured charge density. 
Indeed, the use of local absorptive potentials leads to a substantial overestimate of particle number. However from detailed comparisons with self-energies calculated with \textit{ab initio} many-body methods that include both short- and long-range correlations, we demonstrate that it is essential to introduce non-local absorptive potentials in order to remediate these deficiencies.

We review the fully non-local DOM potential fitted to ${}^{40}$Ca where elastic-scattering data, level information, particle number, charge density and high-momentum-removal $(e,e'p)$ cross sections obtained at Jefferson Lab were included in the analysis. 
All these quantities are accurately described by assuming more or less traditional functional forms for the potentials but allowing for non-locality and the abandonment of complete symmetry around the Fermi energy for surface absorption which is suggested by \textit{ab initio} theory.
An important consequence of this new analysis is the finding that the spectroscopic factor for the removal of valence protons in this nucleus comes out larger by about 0.15 than the results obtained from the NIKHEF analysis of their $(e,e'p)$ data. 
This issue is discussed in detail and its implications  clarified.
Another important consequence of this analysis is that it can shed light on the relative importance of two-body and three-body interactions as far as their contribution to the energy of the ground state is concerned through application of the energy sum rule.


\end{abstract}

%
%
%
%
%

\section{Introduction}
How do the properties of protons and neutrons in the nucleus change from the 
valley of stability to the respective drip lines?
By simultaneously studying the propagation of  nucleons through the nucleus at positive energies for instance in elastic scattering, as well as the movement of nucleons in the ground state at negative energies, it is possible to shed light on this fundamental question relevant for the study of rare isotopes.
The latter information constrains the density distribution of both protons \textit{and} neutrons relevant for clarifying properties of neutron stars. 
In addition, a detailed knowledge of this propagation at positive energies allows for an improved description of other hadronic reactions, including those that purport to 
extract structure information, like transfer or knockout reactions.
Structure information associated with the removal of nucleons
from the target nucleus, is therefore subject of these 
studies and must be supplemented by the appropriate description of the 
hadronic reaction.
Consequently, a much tighter link between reaction and structure studies than is commonly practiced is an 
important goal of the research reported in this review. 

The properties of a nucleon that is strongly influenced by the presence of other nucleons have traditionally been studied in separate energy domains.
Positive-energy nucleons are described by fitted optical-model potentials mostly in local form~\cite{Varner91,Koning03}.
Bound nucleons have been analyzed in static potentials that lead to an independent-particle model (IPM) modified by the interaction between valence nucleons as in traditional shell-model calculations~\cite{Brown01,Caurier05}.
The link between nuclear reactions and nuclear structure is provided by considering these potentials as representing  different energy domains of one underlying nucleon self-energy.
To implement such a framework we have revived the seminal work of Mahaux and Sartor who emphasized the link between these traditionally separate fields in nuclear physics~\cite{Mahaux86,Mahaux91}.
The main idea behind the resulting dispersive optical model (DOM) approach is to employ the concepts of the 
Green's function formulation of the many-body problem~\cite{Dickhoff08} to allow experimental
data to constrain the static and dynamic content of the nucleon self-energy through the use of dispersion relations.
By employing dispersion relations, the method provides a critical link between the physics above and below the Fermi energy with each side being influenced by the absorptive potentials on \textit{both} sides.
Since the self-energy determines both the properties of the system when 
a nucleon is removed as well as when one is added to the ground state (of a 
target), a unique link between structure and elastic-scattering 
information can be forged.
Most applications of the DOM method have been limited to a single nucleus at a time~\cite{Mahaux87,Mahaux88,Johnson88,Mahaux89,Delaroche89,Tornow90,Chiba92,Wang93,Molina02,Nagadi03,Chen04,Morillon04,Capote05,Morillon07,Li08,Sin2016}.
Our initial foray into this project extended the application to a simultaneous analysis of different nuclei belonging to an isotope chain like the calcium isotopes~\cite{Bespalova2005,Charity06,Charity07,Mueller11,Bespalova2015}.
Such an approach is therefore ideally suited to study rare isotopes by providing data-constrained extrapolations into unknown territory which can subsequently be probed by new experiments in inverse kinematics.

Further insights into this approach are provided by \textit{ab initio} Green's function calculations or other many-body techniques that clarify the appropriate functionals that are needed to describe the essential features of the nucleon self-energy~\cite{Waldecker2011,Dussan11}.
The essential new step that has been motivated by these \textit{ab initio} calculations is the introduction of non-local absorptive potentials~\cite{Mahzoon14}.
When these are included it was possible to make substantial progress in describing  ground-state properties as will be discussed in detail in the present review.
Predictions of nucleon elastic cross sections and ground-state properties at larger nucleon asymmetries can then be made after data at smaller asymmetries constrain the potentials that represent the nucleon self-energy.
While microscopic calculations of the nucleon self-energy have made substantial progress~\cite{Barb:1,Barb:2,Dickhoff04,Barb09}, accurate representations of elastic-scattering data are not yet within reach, particular for heavier nuclei.
Even for light nuclei, \textit{ab initio} methods face severe limitations when higher-energy scattering processes are considered~\cite{Navratil16}.
The present approach can therefore provide an intermediate step by putting optical-model potentials on a more solid theoretical footing by insisting on a proper treatment of non-locality and causality through the use of dispersion relations.

We present in Sec.~\ref{sec:general} a review of the relevant elements of Green's function theory that provide the basis of the DOM.
Equations for applications of the DOM to the analysis of data related to the single-particle (sp) properties of nucleons in nuclei are collected in Sec.~\ref{sec:equations}.
In Sec.~\ref{sec:MS}, the approach introduced by Mahaux and Sartor is extended to higher energies and to sequences of isotopes and isotones, including in Sec.~\ref{sec:param} a presentation of the functionals that have been employed.
A comparison of the results using this original approach with elastic-scattering data is presented in Sec.~\ref{sec:localfits}.
Wider application of these DOM results is presented in Sec.~\ref{sec:transfer} where DOM ingredients are employed to describe transfer reactions.
Insight into the functional forms relevant for representing the nucleon self-energy in finite nuclei are provided by \textit{ab initio} calculations.
We review the implications of the Faddeev random-phase approximation (FRPA) results which emphasize the coupling of nucleons to excitations near the Fermi energy in Sec.~\ref{sec:FRPA}.
These correlations describe mostly the coupling to surface excitations of the nucleus and can therefore be associated with long-range correlations (LRC).
A method that is tailored to treat short-range correlations (SRC) in calculating the nucleon self-energy and its implications are discussed  in Sec.~\ref{sec:SRC}.

The recent development of the DOM to include non-local potentials in the analysis of nuclear structure and reactions is presented in Sec.~\ref{sec:nlocal}.
Initially, the non-locality of just the real HF-like term was reinstated as discussed in Sec.~\ref{sec:nlHF}.
Finally nonlocality was also restored to the imaginary parts of the self-energy and the corresponding results are discussed in Sec.~\ref{sec:nlIM} with special attention to nucleon spectral functions in Sec.~\ref{sec:specf}.
A discussion of the energy of the ground state based on knowledge of the experimentally constrained spectral functions is presented in Sec.~\ref{sec:energy} which provides insight into the relative contributions of two- and three-body interactions.
Finally, conclusions and an outlook are formulated in Sec.~\ref{sec:outlook}.


\section{Single-particle Green's function and the self-energy}
\subsection{Summary of general results}
\label{sec:general}
We start with a brief summary of relevant results from the Green's function formulation of the many-body problem~\cite{Dickhoff08}.
The single-particle (sp) propagator in a many-particle system is defined as
\begin{equation}
G (\alpha ,\beta ; t- t' ) = -\frac{i}{\hbar}
\bra{\Psi^A_0} {\mathcal{T}} 
[ a_{\alpha_H}(t) a^\dagger_{\beta_H}(t') ] \ket{\Psi^A_0} .
\label{eq:5.1}
\end{equation}
The expectation value with respect to the exact ground state of the system of
$A$ particles, samples an operator that represents both particle as
well as hole propagation. 
The state $\ket{\Psi^A_0}$ is the normalized, nondegenerate
Heisenberg ground state for the $A$-particle
system and $E^A_0$ the corresponding eigenvalue
\begin{equation}
\hat H \ket{\Psi^A_0} = E^A_0 \ket{\Psi^A_0} .
\label{eq:5.2}
\end{equation}
The particle addition and removal operators in the definition of the sp
propagator are given in the Heisenberg picture by
\begin{equation}
a_{\alpha_H}(t) = e^{\frac{i}{\hbar} \hat H t}
a_\alpha e^{-\frac{i}{\hbar}\hat H t} 
\label{eq:5.3a} 
\end{equation}
and
\begin{equation}
a^\dagger_{\alpha_H}(t) = e^{\frac{i}{\hbar}\hat H t}
a^\dagger_\alpha e^{-\frac{i}{\hbar}\hat H t} ,
\label{eq:5.3b}
\end{equation}
respectively and the labels $\alpha$ or $\beta$ refer to an appropriate complete set of quantum numbers associated with a sp basis.
The time-ordering operation $\mathcal{T}$
is defined here to include a sign change
when two fermion operators are interchanged and can be written, using step
functions, as
\begin{equation}
{\mathcal{T}}[a_{\alpha_H}(t) a^\dagger_{\beta_H}(t')]
= \theta(t-t') a_{\alpha_H}(t) a^\dagger_{\beta_H}(t')
- \theta(t'-t) a^\dagger_{\beta_H}(t') a_{\alpha_H}(t) .
\label{eq:5.4}
\end{equation}
The propagator depends only on the time difference $t-t'$.
In the following we employ the completeness of the exact eigenstates of $\hat H$ for both
the $A+1$ as well as the $A-1$ system, together with
\begin{equation}
\hat H \ket{\Psi^{A+1}_m}
= E^{A+1}_m \ket{\Psi^{A+1}_m} 
\label{eq:5.6a}
\end{equation}
and
\begin{equation}
\hat H \ket{\Psi^{A-1}_n}
= E^{A-1}_n \ket{\Psi^{A-1}_n} . 
\label{eq:5.6}
\end{equation}
It is convenient to introduce the Fourier transform of the
sp propagator which is more appropriate for practical calculations, 
but also brings out more clearly the information that is contained in the propagator 
\begin{equation}
G(\alpha,\beta;E) = \int_{-\infty}^\infty
d(t-t')\ e^{\frac{i}{\hbar}E(t-t')}\
G(\alpha,\beta; t-t') .
\label{eq:5.7}
\end{equation}
Using the integral representation of the
step function, this Fourier transform can be expressed as:
\begin{eqnarray}
G (\alpha ,\beta ; E )
& = & \sum_m \frac{\bra{\Psi^A_0} a_{\alpha}
\ket{\Psi^{A+1}_m} \bra{\Psi^{A+1}_m} a^\dagger_{\beta} \ket{\Psi^A_0}
}{ E - (E^{A+1}_m - E^A_0 ) +i\eta }  \nonumber \\
& +  & \sum_n \frac{\bra{\Psi^A_0} a^\dagger_{\beta} \ket{\Psi^{A-1}_n}
\bra{\Psi^{A-1}_n} a_{\alpha} \ket{\Psi^A_0} }{
E - (E^A_0 - E^{A-1}_n) -i\eta} 
. 
\label{eq:5.8}
\end{eqnarray}
This expression is known as the Lehmann representation~\cite{lehm} of the sp
propagator.
Note that \textbf{any} sp basis can be used in this
formulation of the propagator.
Continuum solutions in the $A\pm1$ systems are also implied in the completeness relations but are not explicitly included to simplify the notation.
At this point one assumes that a meaningful nuclear Hamiltonian $\hat{H}$ exists which contains a two-body component that describes nucleon-nucleon scattering and bound-state data up to a chosen energy, usually the pion-production threshold.
There is sufficient experimental evidence and theoretical insight suggesting that at least a three-body component should also be present in the Hamiltonian.
We will address this issue when the energy of the ground state is considered for which useful sum rules exist.

Maintaining the present general notation with regard to the sp quantum numbers, we introduce the spectral functions associated with particle and hole propagation.
At an energy $E$, the hole spectral function represents the combined 
probability density for
removing a particle with quantum numbers $\alpha$ from the ground state,
while leaving the remaining $A-1$ system at an energy
$E^{A-1}_n=E^A_0- E$. This quantity, 
which is related to the imaginary part of the diagonal
element of the sp propagator, is given by
\begin{eqnarray}
S_h (\alpha , E) &=& \frac{1}{\pi}\
\textrm{Im}\ G(\alpha ,\alpha ; E ) \hskip 4.cm \qquad 
E \le \varepsilon^-_F \nonumber \\
&=& \sum_{n}
\Bigl| \bra{\Psi^{A-1}_n} a_{\alpha} \ket{\Psi^A_0} \Bigr|^2
\delta(E-(E^A_0 - E^{A-1}_n)) . 
\label{eq:5.9}
\end{eqnarray}
Similarly, the probability density 
for the addition of a particle with quantum numbers
$\alpha$, leaving the $A+1$ system at energy $E^{A+1}_m=E^A_0+E$
, \textit{i.e.}, the particle spectral function, has the form
\begin{eqnarray}
S_p (\alpha , E )
&=& -\frac{1}{\pi}\ \textrm{Im}\
G(\alpha ,\alpha ; E ) \hskip 3.8cm  \qquad
E \ge \varepsilon^+_F \nonumber \\
&=& \sum_{m} \Bigl| \bra{\Psi^{A+1}_m} a^\dagger_{\alpha} \ket{\Psi^A_0}
\Bigr|^2 \delta(E-(E^{A+1}_m - E^A_0)) .
\label{eq:5.10}
\end{eqnarray}
The Fermi energies introduced in Eqs.~(\ref{eq:5.9}) and
(\ref{eq:5.10}) are given by
\begin{equation}
\label{eq:5.11a}
\varepsilon^-_F = E^A_0 - E^{A-1}_0
\end{equation}
and
\begin{equation}
\varepsilon^+_F = E^{A+1}_0 - E^A_0 ,
\label{eq:5.11b}
\end{equation}
respectively.

The occupation number of a sp state $\alpha$ can be generated
from the hole part of the spectral function by evaluating
\begin{eqnarray}
n (\alpha) &=& \bra{\Psi^A_0} a^\dagger_\alpha a_\alpha
\ket{\Psi^A_0} 
= \sum_{n} \Bigl| \bra{\Psi^{A-1}_n} a_\alpha \ket{\Psi^A_0} \Bigr|^2
\nonumber \\
& = & \int_{-\infty}^{\varepsilon^-_F} \!\!\! dE\ 
\sum_{n} \Bigl| \bra{\Psi^{A-1}_n}
a_{\alpha} \ket{\Psi^A_0} \Bigr|^2 
\delta(E-(E^A_0 - E^{A-1}_n)) \nonumber \\
&=& \int_{-\infty}^{\varepsilon^-_F} \!\!\! dE\ S_h(\alpha, E) .
\label{eq:5.13}
\end{eqnarray}
The depletion (or emptiness) number is determined by the particle part
of the spectral function
\begin{eqnarray}
d (\alpha) &=& \bra{\Psi^A_0} a_\alpha a^\dagger_\alpha 
\ket{\Psi^A_0} =  
\sum_{m} \Bigl| \bra{\Psi^{A+1}_m} a^\dagger_\alpha \ket{\Psi^A_0}
\Bigr|^2 \nonumber \\
& = & \int_{\varepsilon^+_F }^{\infty} \!\!\!
dE\ \sum_{m} \Bigl| \bra{\Psi^{A+1}_m}
a^\dagger_{\alpha} \ket{\Psi^A_0} \Bigr|^2
\delta(E-(E^{A+1}_m - E^A_0)) \nonumber \\
&=& \int_{\varepsilon^+_F}^{\infty} \!\!\! dE\ S_p(\alpha, E) .  
\label{eq:5.14}
\end{eqnarray}
An important sum rule exists for $n(\alpha)$ and $d(\alpha)$ which can be
deduced by employing the anticommutation relation for $a_\alpha$ and 
$a^\dagger_\alpha$
\begin{equation}
n(\alpha) + d (\alpha) =
\bra{\Psi^A_0} a^\dagger_\alpha a_\alpha \ket{\Psi^A_0} 
+ \bra{\Psi^A_0} a_\alpha a^\dagger_\alpha \ket{\Psi^A_0} 
= \langle \Psi^A_0 | \Psi^A_0 \rangle = 1 .
\label{eq:5.15}
\end{equation}
The partition between the occupation and emptiness of a sp
orbital in the correlated ground state is a sensitive measure of the
strength of correlations, provided a suitable sp basis is chosen.

The sp propagator generates the expectation value of any
one-body operator in the ground state
\begin{equation}
\bra{\Psi^A_0} \hat O \ket{\Psi^A_0} =
\sum_{\alpha,\beta} \bra{\alpha} O \ket {\beta} \bra{\Psi^A_0}
a^\dagger_\alpha a_\beta \ket{\Psi^A_0} 
= \sum_{\alpha,\beta} \bra{\alpha} O \ket {\beta} n_{\alpha\beta} .
\label{eq:5.16}
\end{equation}
Here, $n_{\alpha\beta}$ is the one-body density matrix element which can be
be obtained from the sp propagator
using the Lehmann representation
\begin{eqnarray}
n_{\beta\alpha} &=&
\int \! \frac{dE}{2\pi i}\ e^{iE\eta}\ G(\alpha,\beta;E) \label{eq:5.17} \\ 
&=& \sum_n \bra{\Psi^A_0} a^\dagger_\beta \ket{\Psi^{A-1}_n} 
\bra{\Psi^{A-1}_n} a_\alpha \ket{\Psi^A_0} 
= \bra{\Psi^A_0} a^\dagger_\beta a_\alpha \ket{\Psi^A_0} .
\nonumber
\end{eqnarray}
Note the convergence factor in the integral with an infinitesimal (positive)
$\eta$ which requires closing the
contour in the upper half of the complex $E$-plane.
Consequently, only the (nondiagonal) hole part of the spectral amplitude 
contributes.
Knowledge of $G$ in terms of $n_{\beta \alpha}$, therefore
yields the expectation value of any one-body operator in the correlated ground
state according to Eq.~(\ref{eq:5.16}).
An important recent application of this result concerns the nuclear charge density which is measured in detail for stable closed-shell nuclei, providing important constraints on the properties of the sp propagator. 
If neutron properties related to scattering are constrained and isospin symmetry is invoked, it is also be possible to make predictions for the neutron distribution and as a consequence the neutron skin.

Surprisingly,
the energy of the ground state can also be determined from the sp 
propagator provided that there are
only two-body interactions between the particles. 
We will assume this for now and discuss the influence of three-body interactions when results for the ground-state energy are discussed.
Two-body forces usually dominate in most systems, 
but  the
consideration of at least three-body forces is required to account for
all experimental details.
The energy sum rule for two-body interactions was first 
clarified in Ref.~\cite{gami58}
and later applied to finite systems in Refs.~\cite{kolta,koltb}.
It also requires
only the hole part of the propagator.
Employing the one-body density matrix and the hole spectral function, it is straightforward to demonstrate~\cite{Dickhoff08} that
the desired result can be expressed as
\begin{eqnarray}
E^A_0 &=& \bra{\Psi^A_0} \hat H \ket{\Psi^A_0} \nonumber \\
&=& \frac{1}{2} \sum_{\alpha,\beta} \bra{\alpha} T \ket {\beta} n_{\alpha\beta}
+ \frac{1}{2} \sum_\alpha
\int_{-\infty}^{\varepsilon^-_F} \!\! dE\ E\ S_h(\alpha,E) .
\label{eq:erule}
\end{eqnarray}

The perturbation expansion of the sp propagator is discussed in various textbooks (see \textit{e.g.} Refs.~\cite{Abrikosov1965,Dickhoff08}).
It is necessary to order the expansion into the so-called Dyson equation to obtain a meaningful nonperturbative link between the in-medium potential experienced by a nucleon, the so-called irreducible self-energy, and the propagator.
The resulting equation is given by
\begin{equation}
G(\alpha,\beta;E) = G^{(0)}(\alpha,\beta;E)
+ \sum_{\gamma,\delta} G^{(0)}(\alpha,\gamma;E)
\Sigma(\gamma,\delta;E) G(\delta,\beta;E) .
\label{eq:10.2}
\end{equation}
The noninteracting propagator $G^{(0)}$ can be chosen arbitrarily~\cite{Dickhoff04,Dickhoff08} and most often incorporates global conservation laws especially those associated with rotational symmetry and parity in nuclear applications.
When performing approximate calculations of the nucleon self-energy,  a noninteracting propagator is often chosen that corresponds to localized nucleons~\cite{Dickhoff04}. This  is accomplished by a corresponding term in the irreducible self-energy so that its effect is ultimately eliminated, at least in principle.
For applications to the DOM, it is more convenient to work in a coordinate or momentum-space sp basis suitably accompanied by conserved quantum numbers.
The corresponding quantum numbers are the orbital and total angular momentum of the nucleon which can therefore also be employed to label the propagator and the irreducible self-energy.

\subsection{Relevant equations for applications of the dispersive optical model in nuclei}
\label{sec:equations}
The nucleon propagator for the $A$-body ground state expressed in the sp basis with good radial position, orbital angular momentum (parity) and total angular momentum while suppressing the projection of the total angular momentum and the isospin quantum numbers can be obtained from Eq.~(\ref{eq:5.8}) as
\begin{eqnarray}
G _{\ell j}(r ,r' ; E) 
& = &  \sum_m \frac{\bra{\Psi^A_0} a_{r\ell j}
\ket{\Psi^{A+1}_m} \bra{\Psi^{A+1}_m} a^\dagger_{r' \ell j} \ket{\Psi^A_0}
}{ E - (E^{A+1}_m - E^A_0 ) +i\eta }  
\nonumber \\
& + &  \sum_n \frac{\bra{\Psi^A_0} a^\dagger_{r' \ell j} \ket{\Psi^{A-1}_n}
\bra{\Psi^{A-1}_n} a_{r \ell j} \ket{\Psi^A_0} }{
E - (E^A_0 - E^{A-1}_n) -i\eta} , 
\label{eq:prop}
\end{eqnarray}
where the continuum solutions in the $A\pm1$ systems are also implied in the completeness relations.
The numerators of the particle and hole components of the propagator represent the products of overlap functions associated with adding or removing a nucleon from the $A$-body ground state.
The resulting Dyson equation [see Eq.~(\ref{eq:10.2})] then has the following form
\begin{eqnarray}
\label{eq:dyson}
G_{\ell j}(r,r';E) &=& G^{(0)}_{\ell j}(r,r';E)  \\
&+ &\int \!\! d\tilde{r}\ \tilde{r}^2 \!\! \int \!\! d\tilde{r}'\ \tilde{r}'^2 G^{(0)}_{\ell j}(r,\tilde{r};E)
\Sigma_{\ell j}(\tilde{r},\tilde{r}';E) G_{\ell j}(\tilde{r}',r';E) . 
\nonumber
\end{eqnarray}
For the present discussion, the noninteracting propagator involves just the
 kinetic energy contributions.
The nucleon self-energy contains all linked diagrammatic contributions that are irreducible with respect to propagation represented by $G^{(0)}$.
All contributions to the propagator are then generated by the Dyson equation itself.
The solution of the Dyson equation generates all discrete poles corresponding to bound $A\pm1$ states explicitly given by Eq.~(\ref{eq:prop}) that can be reached by adding or removing a particle with quantum numbers $r \ell j$.
The hole spectral function is obtained from
\begin{equation}
S_{\ell j}(r;E) = \frac{1}{\pi}  \textrm{Im}\ G_{\ell j}(r,r;E)  
\label{eq:holes}
\end{equation}
for energies in the $A-1$-continuum.
For a given $\ell j$ combination, the total spectral strength per unit of energy at $E$  is 
\begin{equation}
S_{\ell j}(E) = \int_{0}^\infty dr\ r^2\ S_{\ell j}(r;E) ,
\label{eq:specs}
\end{equation}

For discrete energies, overlap functions for the addition or removal of a particle are generated as well. For discrete states in the $A-1$ system, one can show that the overlap function obeys a Schr{\"o}dinger-like equation~\cite{Dickhoff08}.
Introducing the notation
\begin{equation}
\psi^n_{\ell j}(r) = \bra{\Psi^{A-1}_n}a_{r \ell j} \ket{\Psi^A_0} ,
\label{eq:overlap}
\end{equation}
for the overlap function for the removal of a nucleon at $r$ with discrete quantum numbers $\ell$ and $j$, one finds
\begin{eqnarray}
\left[ \frac{ p_r^2}{2m} +
 \frac{\hbar^2 \ell (\ell +1)}{2mr^2}\right]  & \psi^{n}_{\ell j}(r) 
+   \int \!\! dr'\ r'^2 
\Sigma_{\ell j}(r,r';\varepsilon^-_n)  &\psi^{n}_{\ell j}(r') =
\varepsilon^-_n \psi^{n}_{\ell j}(r) ,
\label{eq:DSeq}
\end{eqnarray}
where
\begin{equation}
\varepsilon^-_n=E^A_0 -E^{A-1}_n 
\label{eq:eig}
\end{equation}
and, in coordinate space, the radial momentum operator is given by $p_r = -i\hbar(\frac{\partial}{\partial r} + \frac{1}{r})$.
Discrete solutions to Eq.~(\ref{eq:DSeq}) exist in the domain where the self-energy has no imaginary part.  These solutions are normalized by utilizing the inhomogeneous term in the Dyson equation.
For these so-called  quasihole solutions the  normalization or spectroscopic factor is given  by~\cite{Dickhoff08}
\begin{equation}
S^n_{\ell j} = \bigg( {1 - 
\frac{\partial \Sigma_{\ell j}(\alpha_{qh},
\alpha_{qh}; E)}{\partial E} \bigg|_{\varepsilon^-_n}} 
\bigg)^{-1} ,
\label{eq:sfac}
\end{equation}
which is the discrete equivalent of Eq.~(\ref{eq:specs}).
Discrete solutions in the domain where the self-energy has no imaginary part can therefore be obtained by expressing Eq.~(\ref{eq:DSeq}) on a grid in coordinate space and performing the corresponding matrix diagonalization. 
Likewise, the solution of the Dyson equation [Eq.~(\ref{eq:dyson})] for continuum energies in the domain below the Fermi energy, can be formulated as a complex matrix inversion in coordinate space.
This is advantageous in the case of a non-local self-energy representative of all microscopic approximations that include at least the Hartree-Fock (HF) approximation.
For particle removal below the Fermi energy $\varepsilon_F^-$,
the corresponding discretization is limited by the size of the nucleus as can be inferred from the removal amplitude given in Eq.~(\ref{eq:overlap}), which demonstrates that only coordinates inside the nucleus need to be considered.
Such a finite interval therefore presents no numerical difficulty.

The particle spectral function for a finite system can be generated by the calculation of the reducible self-energy $\mathcal{T}$.
In some applications relevant for elucidating correlation effects, a momentum-space scattering code~\cite{Dussan11} to calculate $\mathcal{T}$ was employed.
In an angular-momentum basis, iterating the irreducible self-energy $\Sigma$ to all orders, yields
\begin{eqnarray}\label{eq:redSigma1}
\mathcal{T}_{\ell j}(k,k^\prime ;E) &=& \Sigma_{\ell j}(k,k^\prime ;E) \\ \nonumber
  &+&  \!\!       \int \!\! dq\ q^2\ \Sigma_{\ell j}(k,q;E)\ G^{(0)}(q;E )\ \mathcal{T}_{\ell j}(q,k^\prime ;E) ,
\end{eqnarray}
where $G^{(0)}(q; E ) = (E - \hbar^2q^2/2m + i\eta)^{-1}$ is the free propagator.
The propagator is then obtained from an alternative form of the Dyson equation~\cite{Dickhoff08}
\begin{eqnarray}
\!\!\!\!\!\!\! G_{\ell j}(k, k^{\prime}; E) = \frac{\delta( k - k^{\prime})}{k^2}G^{(0)}(k; E)  
 \label{eq:gdys1}  
		+		      G^{(0)}(k; E)\mathcal{T}_{\ell j}(k, k^{\prime}; E)G^{(0)}(k'; E) .
\end{eqnarray}
The on-shell matrix elements of the reducible self-energy in Eq.~(\ref{eq:redSigma1}) are sufficient to describe all aspects of elastic scattering including differential, reaction, and total cross sections as well as polarization data~\cite{Dussan11}.
The connection between the nucleon propagator and elastic-scattering data can therefore be made explicit by identifying the nucleon elastic-scattering $\mathcal{T}$-matrix with the reducible self-energy obtained by iterating the irreducible one to all orders with $G^{(0)}$~\cite{Bell59,Villars67, BlaizotR86,Dickhoff08}.

The spectral representation of
the particle part of the propagator, referring to the $A+1$ system, appropriate for the treatment of the continuum and possible open channels can be 
generalized from the discrete formulation of Eq.~(\ref{eq:prop})~\cite{Mahaux91}, \textit{i.e.}, 
\begin{eqnarray}
\!\!\!\!\!\!\!\!\!\!\!\!\!\! G_{\ell j}^{p}(k ,k' ; E)  =  
\sum_n  \frac{ \phi^{n+}_{\ell j}(k) \left[\phi^{n+}_{\ell j}(k')\right]^*
}{ E - E^{*A+1}_n +i\eta }   \label{eq:propp} 
 +  
\sum_c \int_{T_c}^{\infty} dE'\  \frac{\chi^{cE'}_{\ell j}(k) \left[\chi^{cE'}_{\ell j}(k')\right]^* }{
E - E' +i\eta} ,
\end{eqnarray}
where now the formalism is in momentum rather than coordinate space.
Overlap functions for bound $A+1$ states are given by $ \phi^{n+}_{\ell j}(k)=\bra{\Psi^A_0} a_{k\ell j}
\ket{\Psi^{A+1}_n}$, whereas those in the continuum are given by $ \chi^{cE}_{\ell j}(k)=\bra{\Psi^A_0} a_{k\ell j} \ket{\Psi^{A+1}_{cE}}$ indicating the relevant channel by $c$ and the energy by $E$.
Excitation energies in the $A+1$ system are with respect to the $A$-body ground state $E^{*A+1}_n = E^{A+1}_n -E^A_0$.
Each channel $c$ has an appropriate threshold indicated by $T_c$ which is the experimental threshold with respect to the ground-state energy of the $A$-body system.
The overlap function for the elastic channel can be explicitly calculated by solving the Dyson equation and it is also possible to obtain the complete spectral density for $E>0$ 
\begin{eqnarray}
S_{\ell j}^{p}(k ,k' ; E) 
=
\sum_c \chi^{cE}_{\ell j}(k) \left[ \chi^{cE}_{\ell j}(k') \right]^* .
\label{eq:specp}
\end{eqnarray}
In practice, this requires solving the scattering problem twice at each energy so that one may employ
\begin{eqnarray}
\!\! S_{\ell j}^{p}(k ,k' ; E) 
= \frac{i}{2\pi} \left[ G_{\ell j}^{p}(k ,k' ; E^+) - G_{\ell j}^{p}(k ,k' ; E^-) \right]
\label{eq:specpp}
\end{eqnarray}
with $E^\pm =E\pm i\eta$, and only the elastic-channel contribution to Eq.~(\ref{eq:specp}) is explicitly known.
Equivalent expressions pertain to the hole part of the propagator $G_{\ell j}^{h}$~\cite{Mahaux91}.

The calculations are performed in momentum space according to Eq.~(\ref{eq:redSigma1}) to generate the off-shell reducible self-energy and thus the spectral density by employing Eqs.~(\ref{eq:gdys1}) and (\ref{eq:specpp}).
Because the momentum-space spectral density contains a delta-function associated with the free propagator, it is convenient for visualization purposes to also consider the Fourier transform back to coordinate space 
\begin{eqnarray}
S_{\ell j}^{p}(r ,r' ; E) = \frac{2}{\pi} \label{eq:specpr} 
  \int \!\! dk k^2 \! \int \!\! dk' k'^2 j_\ell(kr) S_{\ell j}^{p}(k ,k' ; E) j_\ell(k'r') ,
\end{eqnarray}
which has the physical interpretation for $r=r'$ as the probability density $S_{\ell j}(r;E)$ for adding a nucleon with energy $E$ at a distance $r$ from the origin for a given $\ell j$ combination.
By employing the asymptotic analysis to the propagator in coordinate space following \textit{e.g.} Ref.~\cite{Dickhoff08}, one may express the elastic-scattering wave function that contributes to Eq.~(\ref{eq:specpr}) in terms of the half on-shell reducible self-energy obtained according to
\begin{eqnarray}
\!\!\!\!\!\!\!\!\!\!\!\!\!\!  \chi^{el E}_{\ell j}(r) = \left[ \frac{2mk_0}{\pi \hbar^2} \right]^{1/2} \bigg\{ j_\ell(k_0r)  \label{eq:elwf} 
 +  \left. \int \!\! dk k^2 j_\ell(kr) G^{(0)}(k;E) \mathcal{T}_{\ell j}(k,k_0;E) \right\} ,
\end{eqnarray}
where $k_0=\sqrt{2 m E}/\hbar$ is related to the scattering energy in the usual way.

The depleted strength in the continuum of  mostly-occupied orbits (or at negative energies for mostly-empty orbits) is obtained by double folding the spectral density in Eq.~(\ref{eq:specpr}) in the following way
\begin{eqnarray}
\!\!\! S_{\ell j}^{n+}(E) 
=  \int \!\! dr r^2 \!\! \int \!\! dr' r'^2 \phi^{n-}_{\ell j}(r) S_{\ell j}^{p}(r ,r' ; E) \phi^{n-}_{\ell j}(r') ,
\label{eq:specfunc}
\end{eqnarray}
using an overlap function 
\begin{equation}
\sqrt{S^n_{\ell j}} \phi^{n-}_{\ell j}(r)=\bra{\Psi^{A-1}_n} a_{r\ell j} \ket{\Psi^{A}_0} , 
\label{eq:overm}
\end{equation}
corresponding to a bound orbit with $S^n_{\ell j}$,  the relevant spectroscopic factor, and $\phi^{n-}_{\ell j}(r)$  normalized to 1~\cite{Dussan11}.

The occupation number of an orbit is given by an integral over a corresponding folding of the hole spectral density
\begin{eqnarray}
\!\!\!\!\!\!\! S_{\ell j}^{n-}(E) 
= \!\! \int \!\! dr r^2 \!\! \int \!\! dr' r'^2 \phi^{n-}_{\ell j}(r) S_{\ell j}^{h}(r ,r' ; E) \phi^{n-}_{\ell j}(r') ,
\label{eq:spechr}
\end{eqnarray}
where $S_{\ell j}^{h}(r,r';E)$ provides equivalent information below the Fermi energy as $S_{\ell j}^{p}(r,r';E)$ above.
An important sum rule is valid for the sum of the occupation number $n_{n \ell j}$ for the orbit characterized by 
$n \ell j$
\begin{equation}
n_{n \ell j} = \int_{-\infty}^{\varepsilon_F} \!\!\!\! dE\ S_{\ell j}^{n-}(E)
\label{eq:nocc}
\end{equation}
and its corresponding depletion number $d_{n \ell j}$
\begin{equation}
d_{n \ell j} = \int_{\varepsilon_F}^{\infty} \!\!\!\! dE\ S_{\ell j}^{n+}(E) ,
\label{eq:depl} 
\end{equation}
as discussed in general terms in Sec.~\ref{sec:general}.
It is simply given by~\cite{Dickhoff08}
\begin{eqnarray}
\!\! 1 =  n_{n \ell j} + d_{n \ell j} \!\! =\bra{\Psi^A_0} a^\dagger_{n \ell j} a_{n \ell j} +a_{n \ell j}a^\dagger_{n \ell j}  \ket{\Psi^A_0} ,
\label{eq:sumr} 
\end{eqnarray}
reflecting the properties of the corresponding anticommutator of the operators $a^\dagger_{n \ell j}$ and $a_{n \ell j}$.
It is convenient to employ the average Fermi energy
\begin{equation}
\varepsilon_F \equiv \frac{1}{2} \left[
\varepsilon_F^+  - \varepsilon_F^- \right] =  \frac{1}{2} \left[ (E^{A+1}_0-E^A_0) + (E^A_0 - E^{A-1}_0) \right]
\label{eq:FE}
\end{equation}
in Eqs.~(\ref{eq:nocc}) and (\ref{eq:depl})~\cite{Mahaux91}.

Strength above $\varepsilon_F$, as expressed by Eq.~(\ref{eq:specfunc}), reflects the presence of the imaginary self-energy at positive energies.
Without it, the only contribution to the spectral function comes from the elastic channel.
The folding in Eq.~(\ref{eq:specfunc}) then involves integrals of orthogonal wave functions and yields zero.
Because it is essential to describe elastic scattering with an imaginary potential, 
it automatically ensures that the elastic channel does not exhaust the spectral density and therefore some spectral strength associated with IPM bound orbits also occurs in the continuum.

For completeness we include here the relation between the reducible on-shell self-energy for neutrons and relevant data pertaining to elastic scattering.
In the language of many-body theory, the elastic nucleon-nucleus scattering is determined by the on-shell matrix element of the reducible self-energy $\mathcal{T}_{\ell j}(k,k^\prime ;E)$, since it is directly related to the $\mathcal{S}$-matrix by~\cite{Dickhoff08}
\begin{eqnarray}\label{Smatrix}
\bra{k_0}\mathcal{S}_{\ell j}(E)\ket{k_0}&\equiv&e^{2i\delta_{\ell j}} \\
&=&1-2\pi i\left(\frac{m k_0}{\hbar^2}\right)\bra{k_0}\mathcal{T}_{\ell j}(E)\ket{k_0}, \nonumber
\end{eqnarray}
where $E$ is 
the energy relative to the center-of-mass. The phase shift, $\delta_{\ell j}$, defined by Eq.~(\ref{Smatrix}) is in general a complex number.
Its real part yields the usual phase shift and its imaginary part is associated with the inelasticity of the scattering process and denoted by
\begin{equation}
\eta_{\ell j}=e^{-2\textrm{Im}(\delta_{\ell j})} .
\label{eq:inel}
\end{equation}
In general, the coupling to more complicated excitations in the self-energy implies a complex potential responsible for the loss of flux in the elastic channel, characterized by the inelasticities $\eta_{\ell j}$.

Because self-energy calculations at positive energy are rare, it is perhaps useful to include some relevant results in terms of the phase shifts $\delta_{\ell j}$ for the quantities that will be discussed later .
The scattering amplitude for initial and final spin projections $m_s'$ and $m_s$ is 
\begin{equation}
f_{m_s',m_s}(\theta , \phi ) = -\frac{4m\pi^2}{\hbar^2} \langle 
{\bm k}'m_s' | \mathcal{T}(E) | {\bm k} m_s \rangle ,
\label{eq:23.16}
\end{equation} 
with wave vectors of magnitude $k_0$.
The matrix structure is usually represented by
\begin{equation}
[f(\theta,\phi)] = \mathcal{F}(\theta) I + \bm{\sigma} \cdot \hat{\bm{n}} 
\mathcal{G}(\theta) ,
\label{eq:23.17}
\end{equation}
based on rotational invariance and parity conservation.
Here $I$ is the unity matrix, $\hat{\bm{n}}= \bm{k} \times \bm{k}'/|\bm{k} \times \bm{k}'|$,
and $\bm{\sigma}$ is  the Pauli spin matrix.
From Eq.~(\ref{Smatrix}), one can show
\begin{eqnarray}
\mathcal{F}(\theta) =
\frac{1}{2ik} \sum_{\ell = 0}^{\infty}
\left[ (\ell +1)\left\{e^{2i\delta_{\ell +}}-1\right\} 
+
\ell \left\{e^{2i\delta_{\ell -}}-1\right\} \right] P_\ell (\cos \theta )
\label{eq:23.17b}
\end{eqnarray}
and
\begin{equation}
\mathcal{G}(\theta)=
\frac{\sin \theta}{2k} \sum_{\ell = 1}^{\infty}
\left[ e^{2i\delta_{\ell +}}
-      e^{2i\delta_{\ell -}} \right] P_\ell' (\cos \theta ) .
\label{eq:23.17c}
\end{equation}
We employ the notation $\delta_{\ell \pm} \equiv \delta_{\ell j=\ell
\pm \frac{1}{2}}$ and $P_\ell'$ denotes the derivative of the Legendre 
polynomial with respect to $\cos \theta$.
The unpolarized cross section is
\begin{equation}
\left(\frac{d\sigma}{d\Omega}\right)_{unpol}=|\mathcal{F}|^2 +
|\mathcal{G}|^2 .
\label{eq:23.17d}
\end{equation}
For polarization measurements with an initially unpolarized beam one
obtains a polarization\index{polarization} along $\hat{\bm{n}}$, 
{\textit{i.e.}} perpendicular to the scattering plane characterized by
\begin{equation}
P(\theta) = \frac{2 \textrm{Re}\{\mathcal{F}(\theta)\mathcal{G}^*(\theta)\}}
{|\mathcal{F}|^2 +|\mathcal{G}|^2} .
\label{eq:23.17e}
\end{equation}
It is also common to denote this quantity as the analyzing power $A_y$.
A third independent observable called the spin-rotation 
parameter\index{spin-rotation parameter} was 
introduced in Ref.~\cite{Glauber79}
\begin{equation}
Q(\theta) = \frac{2 \textrm{Im}\{\mathcal{F}(\theta)\mathcal{G}^*(\theta)\}}
{|\mathcal{F}|^2 +|\mathcal{G}|^2} .
\label{eq:23.17f}
\end{equation}

Employing the partial wave expansions~(\ref{eq:23.17b}) and (\ref{eq:23.17c})
and the orthogonality of the Legendre polynomials,  the total elastic cross section is
\begin{eqnarray}
\sigma^{el}_{tot} & = &
\frac{\pi}{k^2} \sum_{\ell = 0}^{\infty} \frac{
\left| (\ell +1)\left\{e^{2i\delta_{\ell +}}-1\right\}
+ \ell \left\{e^{2i\delta_{\ell -}}-1\right\} \right|^2}{2\ell +1} 
\nonumber \\
& + & \frac{\pi}{k^2} \sum_{\ell = 0}^{\infty} \frac{\ell (\ell +1)\left|
e^{2i\delta_{\ell +}}
-      e^{2i\delta_{\ell -}} \right|^2 }{2\ell +1} .
\label{eq:23.17h}
\end{eqnarray}
We can define partial elastic cross sections $\sigma^{el}_{\ell}$ such that 
\begin{equation}
\sigma^{el}_{tot} = \sum_{\ell =0}^\infty \sigma^{el}_{\ell} ,
\label{eq:23.17z}
\end{equation}
where
\begin{equation}
\sigma^{el}_{\ell}= \frac{\pi}{k^2}\left[ (\ell +1)
\left| e^{2i\delta_{\ell +}}-1 \right|^2 + \ell \left|
e^{2i\delta_{\ell -}}-1 \right|^2 \right] .
\label{eq:23.17i}
\end{equation}
With complex potentials, and therefore complex phase shifts, it is possible
to calculate the total reaction cross 
section
\begin{equation}
\sigma^r_{tot} = \sum_{\ell =0}^\infty \sigma^{r}_{\ell} ,
\label{eq:23.17y}
\end{equation}
with
\begin{equation}
\sigma^{r}_{\ell}= \frac{\pi}{k^2}\left[ (2\ell +1) - (\ell +1)
\left| e^{2i\delta_{\ell +}}\right|^2 - \ell \left| 
e^{2i\delta_{\ell -}} \right|^2 \right] .
\label{eq:23.17k}
\end{equation}
These results are derived by using the optical theorem that yields the total 
cross section from the imaginary part
of the forward scattering amplitude~\cite{Gottfried04}
\begin{equation}
\sigma_{T} = \sigma^{el}_{tot} + \sigma^{r}_{tot} .
\label{eq:23.17j}
\end{equation}
Both $\sigma_{T}$ and $\sigma_{el}$ are infinite for proton scattering as the Coulomb contribution to the self-energy has an infinite range.

It is clear that at positive energies the problem is completely reduced to solving the integral equation for the reducible self-energy given in Eq.~(\ref{eq:redSigma1}).
It should be noted that the solution in momentum space automatically treats the non-locality of the reducible self-energy in coordinate space.
In practice, the integral equation is solved in two steps. 
First the integral equation is solved by only including the principal value part of the noninteracting propagator.
 Subsequently, it is straightforward to employ the resulting reaction matrix to take into account
the contribution of the $\delta$-function associated with the imaginary part of the noninteracting propagator. 
The analysis for protons includes the Coulomb potential which must be treated in coordinate space and requires the solution of the corresponding differential equation with appropriate matching, outside the range of the nuclear potential, to regular and irregular Coulomb functions~\cite{ThNu09}.

The (irreducible) nucleon self-energy in general obeys a dispersion relation between its real and imaginary parts given by~\cite{Dickhoff08}
\begin{eqnarray} 
\!\!\!\!\!\!\!\! \mbox{Re}\ \Sigma_{\ell j}(r,r';E)\!& =& \! \Sigma^s_{\ell j} (r,r')\! \label{eq:disprel} \\
\!\!\!\!\!\!\!\! &-& \! {\cal P} \!\!
\int_{\varepsilon_T^+}^{\infty} \!\! \frac{dE'}{\pi} \frac{\mbox{Im}\ \Sigma_{\ell j}(r,r';E')}{E-E'}  
+{\cal P} \!\!
\int_{-\infty}^{\varepsilon_T^-} \!\! \frac{dE'}{\pi} \frac{\mbox{Im}\ \Sigma_{\ell j}(r,r';E')}{E-E'} , \nonumber
\end{eqnarray}
where $\mathcal{P}$ represents the principal value.
The static contribution $\Sigma^s_{\ell j}$ arises from the correlated HF term involving the exact one-body density matrix and the dynamic parts start and end at corresponding thresholds in the $A\pm1$ systems that have a larger separation than the corresponding difference between the Fermi energies for addition $\varepsilon_F^+$ and removal $\varepsilon_F^-$ of a particle.
The latter feature is particular to a finite system and generates possibly several discrete quasiparticle and hole-like solutions of the Dyson equation in Eq.~(\ref{eq:DSeq}) in the domain where the imaginary part of the self-energy vanishes.

The standard definition of the self-energy requires that its imaginary part is negative, at least on the diagonal, in the domain that represents the coupling to excitations in the $A+1$ system, while it is positive for the coupling to $A-1$ excitations.
This translates into an absorptive potential for elastic scattering at positive energy, where the imaginary part is responsible for the loss of flux in the elastic channel.
The energy-independent part of the self energy $\Sigma^s_{\ell j} (r,r')$ can be eliminated by calculating  Eq.~(\ref{eq:disprel}) at $E=\varepsilon_F$ and subtracting the result from Eq.~(\ref{eq:disprel}).
After some rearranging we thus obtain the so-called subtracted dispersion relation, 
\begin{eqnarray} 
\!\!\!\!\! \mbox{Re}\ \Sigma_{\ell j}(r,r';E)\! &=& \!  \mbox{Re}\ \Sigma_{\ell j} (r,r';\varepsilon_F) \hspace{2.0cm}  \label{eq:sdisprel} \\
&-& \! {\cal P} \!\!
\int_{\varepsilon_T^+}^{\infty} \!\! \frac{dE'}{\pi} \mbox{Im}\ \Sigma_{\ell j}(r,r';E') \left[ \frac{1}{E-E'}  - \frac{1}{\varepsilon_F -E'} \right]  \nonumber  \\
&+& {\cal P} \!\!
\int_{-\infty}^{\varepsilon_T^-} \!\! \frac{dE'}{\pi} \mbox{Im}\ \Sigma_{\ell j}(r,r';E') \left[ \frac{1}{E-E'}
-\frac{1}{\varepsilon_F -E'} \right]  .
\nonumber
\end{eqnarray}
The beauty of this representation was recognized by Mahaux and Sartor~\cite{Mahaux86,Mahaux91} since it allows for a link with empirical information both at the level of the real part of the non-local self-energy at the Fermi energy (probed by a multitude of HF calculations) and also through empirical knowledge of the imaginary part of the optical-model potential (constrained by experimental data) that consequently yields a dynamic contribution to the real part by means of Eq.~(\ref{eq:sdisprel}).
In addition, the subtracted form of the dispersion relation emphasizes contributions to the integrals from the energy domain nearest to the Fermi energy on account of the $E'$-dependence of the integrands of Eq.~(\ref{eq:sdisprel}). 
Recent DOM applications reviewed in this paper include experimental data up to 200 MeV of scattering energy and are therefore capable of determining the nucleon propagator in a wide energy domain as well as all negative energies.

\section{DOM implementation based on the Mahaux and Sartor approach}
\label{sec:MS}
The presentation of the self-energy in the previous section requires further   assumptions before experimental data can be employed to constrain it.
While a considerable collection of experimental data related to elastic nucleon-nucleus scattering exists, relevant knowledge of the imaginary part of the self-energy below the Fermi energy apart from the level structure of valence nucleons, has mostly been probed with electron-induced proton knockout reactions~\cite{FM84,Lapikas93} and $(p,2p)$ reactions~\cite{Jacob66,Jacob73}.

\begin{figure}[tpb]
\begin{center}
\includegraphics[scale=.6]{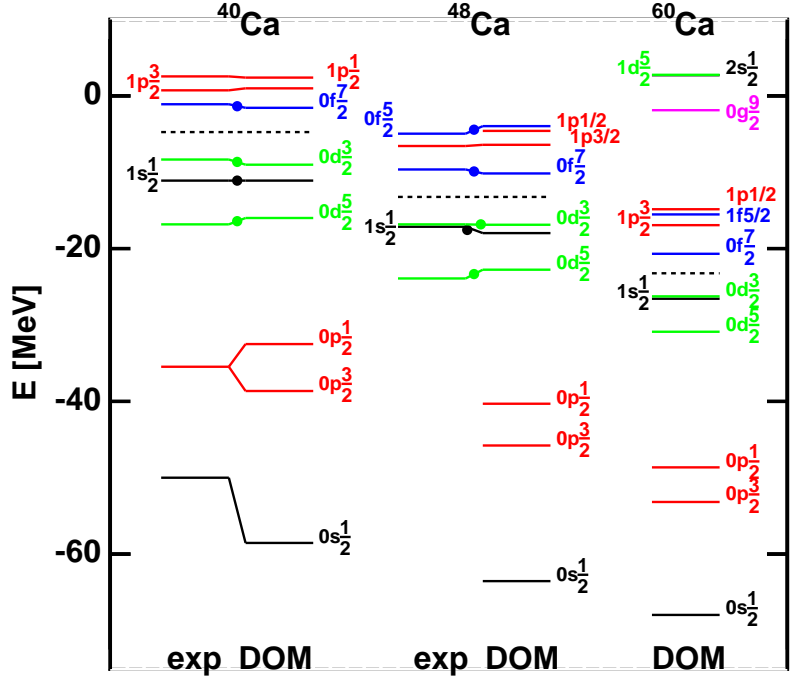}
\caption{Comparison of experimental (exp) and
fitted (DOM) sp-level energies for protons in
$^{40}$Ca and $^{48}$Ca taken from Ref.~{\protect{\cite{Charity06}}}. The levels with solid dots were included in the fits. The dashed lines indicate the Fermi energy.  }
\label{fig:Calev}
\end{center}
\end{figure}
Since the empirical knowledge has relied on local representations of the imaginary part of the optical-model potential, it is natural to make a similar assumption for the DOM version as was proposed by Mahaux and Sartor~\cite{Mahaux91}. 
In addition, a separation in terms of surface (low-energy) and volume (higher-energy) absorption has been incorporated in accordance with standard practice of global optical-model fits~\cite{Becchetti69,Varner91,Koning03}.
The seminal work of Mahaux and Sartor emphasized the double-closed-shell nuclei ${}^{40}$Ca and ${}^{208}$Pb~\cite{Mahaux86,Mahaux88, Mahaux89,Mahaux91}. 
A limited energy window of scattering energies was included in the fit.
The more recent work of our St. Louis group~\cite{Charity06,Charity07} extended this domain to 200 MeV.
By implementing a simultaneous fit to ${}^{40}$Ca and ${}^{48}$Ca it became possible to extrapolate the potentials towards $N$ and $Z$ closer to the drip lines thereby transforming the DOM framework to a tool of relevance for FRIB-related physics. 
We illustrate this point in Fig.~\ref{fig:Calev} from Ref.~\cite{Charity06} where the fit to the levels of the closed-shell Ca nuclei implies a prediction of a further compression of the proton levels in ${}^{60}$Ca near the Fermi energy based on the extrapolated potential. 
Such a compression suggests that proton pairing in exotic Ca nuclei near the neutron drip line may become relevant.

Following Perey and Buck \cite{Perey62}, the central part of the non-local energy-independent term of the subtracted dispersion relation of Eq.~(\ref{eq:sdisprel}) given by
$\textrm{Re}~\Sigma\left( \bm{r},\bm{r}^{\prime };\varepsilon_{F}\right)$ can be 
approximated by a local energy-dependent term which Mahaux and Sartor designate
as the HF potential $\mathcal{V}_{HF}(r,E)$. 
Strictly this is not 
a HF potential, but it does describe the effects of the mean field.
The energy derivative of $\mathcal{V}_{HF}$ is a measure of non-locality,
which is related to the momentum-dependent effective mass 
\begin{equation}
\frac{\widetilde{m}\left( r,E\right) }{m}=1-\frac{d\mathcal{V}_{HF}(r,E)}{dE}%
,
\end{equation}%
where $m$ is the nucleon mass. 

A consequence of the local approximation is that one needs to use a  
scaled imaginary potential
\begin{equation}
\mathcal{W} = \frac{\widetilde{m}\left( r,E\right) }{m} \textrm{Im}\, \Sigma
\label{eq:nlcorr}
\end{equation}
and a similarly scaled dispersive correction~\cite{Mahaux91,Dickhoff08}.
The imaginary part of the self-energy is also approximated by a local potential 
and thus the dispersive correction denoted by $\Delta \mathcal{V}$, is correspondingly local. Mahaux and Sartor 
argue that this modifies $\Delta \mathcal{V}$ by a smooth function of 
energy  can easily be compensated by correspondingly smooth modification 
of $\mathcal{V}_{HF}$. 
In addition to the momentum-dependent effective mass, two other effective
masses can be defined. The total effective mass is given by 
\begin{equation}
\frac{m^{\ast }(r,E)}{m}=1-\frac{d}{dE}  \left[ 
\mathcal{V}_{HF}(r,E) + \Delta\mathcal{V}(r,E) \right],
\end{equation}%
while the energy-dependent effective mass is
\begin{equation}
\frac{\overline{m}\left( r,E\right) }{m}=1-\frac{m}{\widetilde{m}\left(
r,E\right) }\frac{d\Delta \mathcal{V}(r,E)}{dE}.
\end{equation}

At the highest energies considered, relativistic effects become
relevant. We have included a corresponding lowest-order correction in
solving the radial wave equation~\cite{Nadasen81} 
\begin{equation}
\left[ \frac{d^{2}}{d\rho ^{2}}+\left( 1-\frac{\widetilde\Sigma\left(
\rho ,E\right) }{E_{tot}-M-m}-\frac{\ell \left( \ell +1\right) }{\rho ^{2}}%
\right) \right] u\left( \rho \right) =0  \label{Eq:radial}
\end{equation}%
with $\rho =k\,r$, where $k=\frac{M}{E_{tot}}\sqrt{T\left( T+2m\right) }$, $%
T $ is the laboratory kinetic energy, $E_{tot}$ is the total energy in the
center-of-mass frame, and $M$ is the target mass. The scaled potential is 
\begin{equation}
\widetilde{\Sigma}=\gamma \,\Sigma,\gamma =\frac{2\left(
E_{tot}-M\right) }{E_{tot}-M-m}.  \label{Eq:scaledPot}
\end{equation}%
If $u_{n\ell j}\left( r\right) $ are bound-state solutions to the radial
wave equation with energy $\varepsilon_{n\ell j}$, then the normalized wave functions corrected for non-locality
are given by 
\begin{equation}
\overline{u}_{n\ell j}\left( r\right) =\sqrt{\frac{\widetilde{m}\left(
r,\varepsilon_{n\ell j}\right) }{m}}u_{n\ell j}\left( r\right) .
\end{equation}

Mahaux and Sartor~\cite{Mahaux91} developed the following approximations
 to determine bound-state properties. 
For valence states, the spectroscopic factor, relative to the 
IPM value, is given by
\begin{equation}
S_{n\ell j}=\int_{0}^{\infty }\overline{u}_{n\ell j}^{2}\left( r\right) 
\frac{m}{\overline{m}(r,\varepsilon_{n\ell j})}dr .
\label{eq:SF}
\end{equation}%
For hole states, the occupation probability is approximated by 
\begin{eqnarray}
n_{n\ell j}=\int_{0}^{\infty }\overline{u}_{n\ell j}^{2}\left( r\right)  \\
\times \left[ 1+\frac{m}{\widetilde{m}\left( r,\varepsilon_{n\ell j}\right) }\frac{1}{%
\pi }\int_{\varepsilon_{F}}^{\infty }\frac{\mathcal{W}(r,E^{\prime })}{\left(
E^{\prime }-\varepsilon_{n\ell j}\right) ^{2}}dE^{\prime }\right] dr,
\label{eq:occHole}
\end{eqnarray}%
while for bound particle states, the same approximation gives 
\begin{eqnarray}
n_{n\ell j}=-\int_{0}^{\infty }\overline{u}_{n\ell j}^{2}\left( r\right)  \\
\times \left[ \frac{m}{\widetilde{m}\left( r,\varepsilon_{n\ell j}\right) }\frac{1}{%
\pi }\int_{-\infty }^{\varepsilon_{F}}\frac{\mathcal{W}(r,E^{\prime })}{\left(
E^{\prime }-\varepsilon_{n\ell j}\right) ^{2}}dE^{\prime }\right] dr.
\label{eq:occPart}
\end{eqnarray}
A comparison of these approximations with results obtained from Eqs.~(\ref{eq:sfac}) and (\ref{eq:nocc}) will be made when results from a non-local implementation of the DOM are discussed.
We note however that a full treatment of the self-energy in the Dyson equation as discussed in Sec.~\ref{sec:equations} allows one to extract the full spectral distribution of bound orbits both above and below the Fermi energy without any approximation for a given choice of the dispersive self-energy.

\subsection{Parametrization of the potentials}
\label{sec:param}

The parametrization of the real and imaginary optical-model potentials 
is the central aspect of a DOM analysis. 
We will now consider the essential ingredients of a local DOM representation of the relevant potentials as employed recently in Ref.~\cite{Mueller11}.
We note again that this comprises a fit which contains only local potentials as is common in the usage of optical-model potentials.
The number of free parameters in
the fits must be sufficient to allow one to describe the important physics 
but not too large or fitting becomes impractical in term of CPU usage.
The functional forms have their origin in empirical knowledge used by Mahaux and Sartor and implemented in previous studies~\cite{Charity06,Charity07}.
These functional forms are partly based on theoretical expectations but also on 
confrontation with data.

The imaginary potential is composed of the sum of volume, surface, and
imaginary spin-orbit components, 
\begin{eqnarray}
\mathcal{W}\left( r,E\right) & = & -W^{vol}\left( E,r\right) \nonumber \\
& + & 4a^{sur}W^{sur}\left( E\right) \frac{d}{dr}f(r;r^{sur},a^{sur})+\mathcal{W}%
_{so}\left( r,E\right) .  \label{Eq:imag}
\end{eqnarray}%
with Woods-Saxon form factors: 
\begin{equation}
f(r;R^{i},a^{i})=
\left[1+\exp \left(\frac{r-R^{i}}{a^{i}}\right) \right]^{-1} .
\end{equation}%

Standard optical-model fits to elastic-scattering data at a single energy 
require a surface-type absorption at low bombarding energies $E\ll$50~MeV 
and a volume-type absorption at high bombarding energies $E\gg$ 50~MeV.
However, fits encompassing a large range of energies often have a 
significant, but small, surface absorption component extending 
to energies much larger than 50 MeV \cite{Tornow90,Koning03,Charity07}. 
One can reconcile these 
statements by noting that the addition of a small surface component to a 
volume-type component acts to increase the radius of the volume component. 
If the radius of the volume potential is increased by $\delta R$, 
we find, after a Taylor expansion, that
\begin{eqnarray}
f(r;R+\delta R,a)& \sim & f(r,R,a) + \frac{df}{dR} \delta R \\
                & \sim & f(r;R,a) - \frac{df}{dr} \delta R  \label{Eq:expand}
\end{eqnarray}
and thus the first-order correction is a surface-type component.

Thus a gradually decaying surface term above $E$=50~MeV can be understood 
as being associated with a volume-type component whose radius 
decreases with energy.
Such a feature is found in the JLM potential~\cite{Jeukenne74,Bauge98} which 
is derived from infinite-matter calculations coupled with the 
local-density approximation to get the potential in the surface region. 
We have assumed the radius of the imaginary volume potential to decay with 
energy as
\begin{equation}
R^{vol}(E) = R^{vol}_0 + \delta R \exp \left( -\frac{|E-\varepsilon_F|}{E_R} \right).
\end{equation}
However, an energy dependence of the radius was not used in the fits as it 
would require the dispersive correction to be calculated for each $r$ value 
which would be very CPU intensive. Instead we make use of the expansion of
Eq.~(\ref{Eq:expand}) to obtain 
\begin{eqnarray}
\!\!\!\!\!\!\!\!\! W^{vol}(E,r) = W^{vol}_{0}(E) f(r;R^{vol}_0,a^{vol}) 
- 4 a^{vol} W^{vol}_{sc}(E)
 \frac{d}{dr} f(r;R^{vol}_{sc},a^{vol})
\label{Eq:Wvol}
\end{eqnarray} 
where $W^{vol}_{0}(E)$ is the energy dependence of the depth of the volume
 component and the surface-correction, which accounts for the energy dependence
of the radius, is
\begin{equation}
W^{vol}_{sc}(E) = W^{vol}_{0}(E) \frac{\delta R}{4 a^{vol}} \exp \left( 
- \frac{|E-\varepsilon_F|}{E_R} \right).
\end{equation}

In the work reported in Ref.~\cite{Mueller11} we also have a surface component that extents well beyond
$E$=50 MeV, however unlike other studies it is not tied to the ``real'' 
surface component at lower energies which is important if we are going to 
separate the asymmetry dependencies of the surface and volume  components. 
It is also useful 
to maintain a distinction between the ``real'' surface potential at 
low energies which is associated with LRC and the 
surface-correction at high energies associated with SRC.  

The phase space of particle states for $E\gg \varepsilon_{F}$ is significantly larger
than that of hole levels for $E\ll \varepsilon_{F}$. Therefore the contributions from
two-particle-one-hole states (and more complicated states) for $E\gg \varepsilon_{F}$ to the self-energy will be larger than that for
two-hole-one-particle states (and more complicated states) at $E\ll \varepsilon_{F}$. Thus at energies well removed
from $\varepsilon_{F}$, the form of the imaginary volume potential should no longer be
symmetric about $\varepsilon_{F}$. Hence the following form was assumed for 
the depth of volume potential 
\begin{eqnarray}
\!\!\!\!\!\!\!\!\!\! W^{vol}_0(E) &=&  \Delta W_{NM}(E)  \label{eq:volume} \\
\!\!\!\!\!\!\!\!\!\! &+& \left\{  
\begin{array}{ll}
0  &  |E-\varepsilon_F| < E^{vol}_{p} \\
A^{vol} \left( 1 \pm C^{vol}\frac{N-Z}{A} \right)\frac{\left(|E-\varepsilon_F|-E^{vol}_p\right)^4}
{\left(|E-\varepsilon_F|-E^{vol}_p\right)^4 + (B^{vol})^4} & 
 |E-\varepsilon_F| > E^{vol}_{p}
\end{array}
\right. 
\nonumber
\end{eqnarray}
where $\Delta W_{NM}(E)$ is the energy-asymmetric correction modeled after
nuclear-matter calculations. Apart from this correction, the parametrization
is similar to the Jeukenne and Mahaux form~\cite{Jeukenne83} used in many DOM\
analyses. For the asymmetry term, the + and - values refer to protons 
and neutrons respectively. This form of the asymmetry potential is 
consistent with the Lane potential~\cite{Lane62} and for SRC can be justified based on the difference between the 
\textit{n}-\textit{p} and the \textit{n}-\textit{n} or \textit{p}-\textit{p}
in-medium nucleon-nucleon cross sections~\cite{Charity07}. 
Nuclear-matter calculations of 
occupation probabilities which should be closely associated with the volume 
component, also suggest that this form is valid except for  
extreme asymmetry values~\cite{Frick05,Rios09,Rios14}.

We set the parameter $E^{vol}_p$=11~MeV to force the imaginary 
potential to be zero just in the vicinity of the Fermi energy (see later).
The radii of the volume and surface correction components $W^{vol}_0$ and $W^{vol}_{sc}$ in Eq.~(\ref{Eq:Wvol}) are taken to be identical:
\begin{equation}
R^{vol}_{0} = r^{vol}_0 A^{1/3} .
\end{equation}

The energy-asymmetric correction was taken as :%
\begin{eqnarray}
\!\!\!\!\!\!\!\!\!\!\!\!\! \Delta W_{NM}(E)=
\left\{
\begin{array}{ll}
\alpha A^{vol} \left[ \sqrt{E}+\frac{\left( \varepsilon_{F}+E_{a}\right) ^{3/2}}{2E}-\frac{3}{2}
\sqrt{\varepsilon_{F}+E_{a}}\right] & E-\varepsilon_{F}>E_{a} \\ 
- A^{vol} \frac{(\varepsilon_F-E-E_a)^2}{(\varepsilon_F-E-E_a)^2+(E_a)^2} & E-\varepsilon_{F}<-E_{a} \\ 
0 & \textrm{otherwise}
\end{array}
\right.
\label{eq:Wnm}
\end{eqnarray}
which is similar to the form suggested by Mahaux and Sartor~\cite{Mahaux91}.
Following an earlier study~\cite{Charity07}, we have taken 
$\alpha$=0.08~MeV$^{-1/2}$ and $E_a$=60~MeV. 

The ``true''  imaginary surface potential is taken to have the form 
\begin{eqnarray}
\!\!\!\! W^{sur}(E)=
\left\{
\begin{array}{ll}
0 & |E-\varepsilon_F| < E^{sur}_{p} \\
\frac{A^{sur}}{1+\exp\left( \frac{|E-\varepsilon_F|-C^{sur}}{D^{sur}} \right)} 
\frac{\exp\left( \frac{|E-\varepsilon_F|-E^{sur}_p}{B^{sur}}\right)-1}
{\exp\left( \frac{|E-\varepsilon_F|-E^{sur}_p}{B^{sur}}\right)+1}
&  |E-\varepsilon_F| > E^{sur}_{p} \\
\end{array}
\right.
\label{eq:surface}
\end{eqnarray}
where for protons and neutrons ($i=n,p$) the $E^{sur}_{p}$ parameter 
is related to 
the experimental particle-hole energy gaps $\Delta_i$ via  
\begin{equation}
E^{sur(i)}_{p} = f_{\Delta} \left[ \frac{\Delta_i}{2} +  \min \left(\Delta_p,\Delta_n\right) \right] \\
\Delta_i = \varepsilon_F^{(i)+} - \varepsilon_F^{(i)-} 
\label{eq:delta_np}
\end{equation}

In the IPM, $f_{\Delta}$=1 and $E^{sur}_p$ 
represents the minimum particle energy above the Fermi value, for which a 
particle can couple to a two-particle-one-hole excitation. Similarly it 
also the maximum energy, relative to the Fermi value, for which a hole can 
couple to  a two-hole-one-particle excitation.  Thus between these two 
limits, damping of sp states is not possible and   
the imaginary potential should exhibit a region of width 2$E^{sur}_p$ 
where it is exactly zero. Many-body correlations reduce the width 
of this gap and thus we include  the fitting parameter $f_{\Delta}$.   
Mahaux and Sartor had also explored imaginary potentials which were 
zero in the immediate vicinity of the Fermi energy~\cite {Mahaux91}, however, they assumed
a somewhat different energy dependence.

The mass dependence was taken as
\begin{equation}
R^{sur} = r^{sur}_0 A^{1/3}  \\
\end{equation}
and the parameter $A^{sur}$ of Eq.~(\ref{eq:surface}) individually fit for each nucleus 
and nucleon type.

The HF potential is parametrized in the following way 
\begin{eqnarray}
\mathcal{V}_{HF}\left( r,E\right) &=&-V_{HF}^{Vol}\left( E\right)
\,f(r;r^{HF},a^{HF})   \nonumber \\
&+& 4V_{HF}^{sur}\frac{d}{dr}f(r;r^{HF},a^{HF})+V_{c}\left( r\right) +\mathcal{V%
}_{so}(r,E),
\label{eq:HFl}
\end{eqnarray}%
where the Coulomb $V_{C}$ and real spin-orbit $\mathcal{V}_{so}$ terms have
been separated from the volume and surface components. The volume component
contains the energy-dependence representing non-locality, which is 
approximated by the cubic equation 
\begin{equation}
V^{HF}_{vol}(E) =   V^{HF}_0 
- \alpha^{vol} \left( E-\varepsilon_{F} \right) - \beta^{vol} \left( E-\varepsilon_{F} \right)^2 - \gamma^{vol} \left( E-\varepsilon_{F} \right) ^3 .
\label{Eq:HFvol}
\end{equation}
The value of $V^{HF}_0$ is constrained for each nucleus and nucleon type 
by obtaining the correct Fermi energy. 
This is essentially independent of the imaginary potential and their 
dispersive corrections, \textit{i.e.}, the dispersive corrections have equal but 
opposite effects on $\varepsilon_F^+$ and $\varepsilon_F^-$ and so cancel in the calculation of the
Fermi energy in Eq.~(\ref{eq:FE}).  

The HF surface component was found necessary to fit high-energy
elastic-scattering data~\cite{Charity07} and was parametrized as  
\begin{eqnarray}
V_{HF}^{sur}(E) =
\left\{
\begin{array}{ll}
0 & \textrm{if } x < 0 \\
\alpha^{sur} \frac{x^2}{x^2+(\gamma^{sur})^2} & \textrm{if } x > 0
\end{array}
\right.
\end{eqnarray}
where
\begin{equation}
x = E - \varepsilon_{F} - \beta^{sur}.
\end{equation}
The mass and asymmetry dependence were taken as 
\begin{equation}
\alpha^{vol} = \alpha^{vol}_{0}   \pm \alpha^{vol}_{NZ}  
\frac{N-Z}{A} \\
\label{eq:alphaNZ}
R^{vol} = r^{vol} A^{1/3} .  \\
\end{equation}
The Coulomb potential was assumed to have a sharp-surface sphere with radius
\begin{equation}
R_{C}=r_{C} A^{1/3}.
\end{equation}

At high energies, optical-model potentials generally include an imaginary spin-orbit
potential~\cite{VanOers71}. Given that this term is usually assumed to be
zero for lower energies, this implies that the imaginary spin-orbit term is
energy dependent. As such, it should give rise to a dispersive correction to
the real component. Given these considerations, the total spin-orbit
potential was taken as 
\begin{eqnarray}
\!\!\!\!\!\!\!\!\!\!\!\!\!\!\! \mathcal{U}^{so}(r,E) & = & \mathcal{V}^{so}(r,E)+i\mathcal{W}^{so}(r,E)  \label{eq:SOrbit}
\\
\!\!\!\!\!\!\!\!\!\!\!\!\!\!\! & = & \Delta \mathcal{V}^{so}\left( r,E\right) +\left( \frac{h}{m_{\pi }c}\right)
^{2}[V^{so}+iW^{so}(E)]  
\frac{1}{r}\frac{d}{dr}f(r;R^{so},a^{so})\;\frac{\bm{\ell}\cdot \bm{s}%
}{2}, \nonumber
\end{eqnarray}%
where $\left( \hbar /m_{\pi }c\right) ^{2}$=2.0~fm$^{2}$ and $\Delta 
\mathcal{V}_{so}$ is the dispersive correction determined from the imaginary
component $\mathcal{W}_{so}$. As the imaginary spin-orbit component is
generally needed only at high energies, we chose the form 
\begin{equation}
W^{so}(E)= A^{so}  \frac{(E-\varepsilon_F)^4}{(E-\varepsilon_F)^4+(B^{so})^4} .
\end{equation}%
The dispersive correction $\Delta V_{so}(E)$ associated with this component
gives an approximately linear decrease in
magnitude of the total real spin-orbit strength over the energy region of
interest. 
The mass and asymmetry dependence of the spin orbit was taken as
\begin{equation}
V^{so} = V^{so}_0 \pm V^{so}_{NZ}\frac{N-Z}{A} \\
R^{so} = r^{so} A^{1/3}  .
\end{equation}
We refer to Ref.~\cite{Mueller11} for a detailed list of the parameters that were obtained in different regions of the chart of nuclides.
Some representative results are summarized in the next section.

\subsection{Results for local DOM potentials applied to chains of isotopes and isotones}
\label{sec:localfits}
\begin{figure*}[tbp]
\begin{center}
\includegraphics[scale=.6]{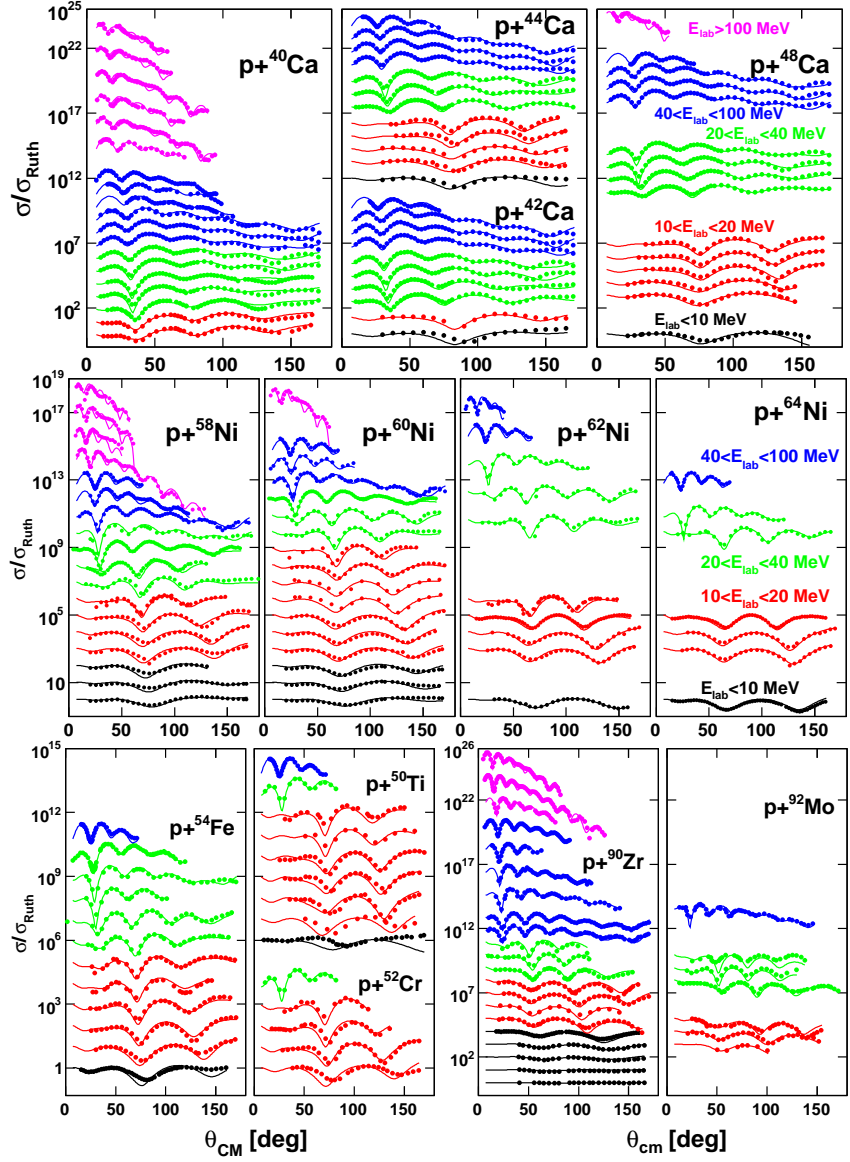}
\caption{Fitted proton elastic-scattering differential 
cross sections expressed as a ratio to the Rutherford scattering value.
Data from each energy is offset along the vertical axis for clarity. 
Lowest energy on the bottom and highest energy on the top for each frame. 
Individual energies can be obtained from tables given in Ref.~\cite{Mueller11}. Results for different ranges of 
energies are plotted with different colors as indicated in figure.}
\label{fig:sigCaNiP}
\end{center}
\end{figure*}

\begin{figure*}[tbp]
\begin{center}
\includegraphics[scale=.7]{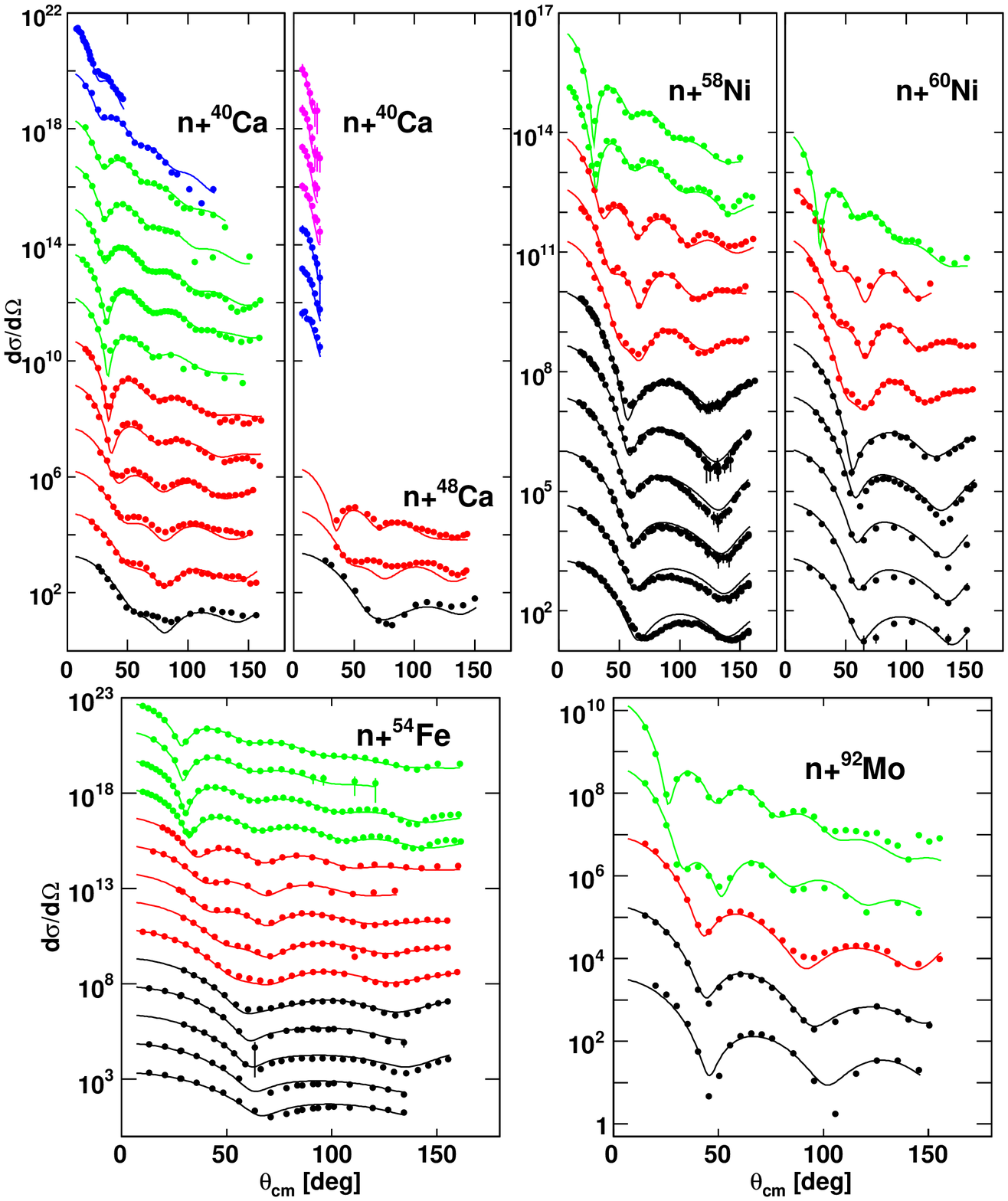}
\caption{Fitted neutron elastic-scattering differential 
cross sections for reactions of Ca, Ni isotopes and $^{54}$Fe and $^{92}$Mo.
Data from each energy is offset along the vertical axis for clarity. 
Lowest energy on the bottom and highest energy on the top for each frame. 
Individual energies can be obtained from tables in Ref.~\cite{Mueller11}.
The color convention is the same as employed in Fig.~\ref{fig:sigCaNiP}.
}
\label{fig:sigCaNiN}
\end{center}
\end{figure*}

\begin{figure*}[tbp]
\begin{center}
\includegraphics[scale=.7]{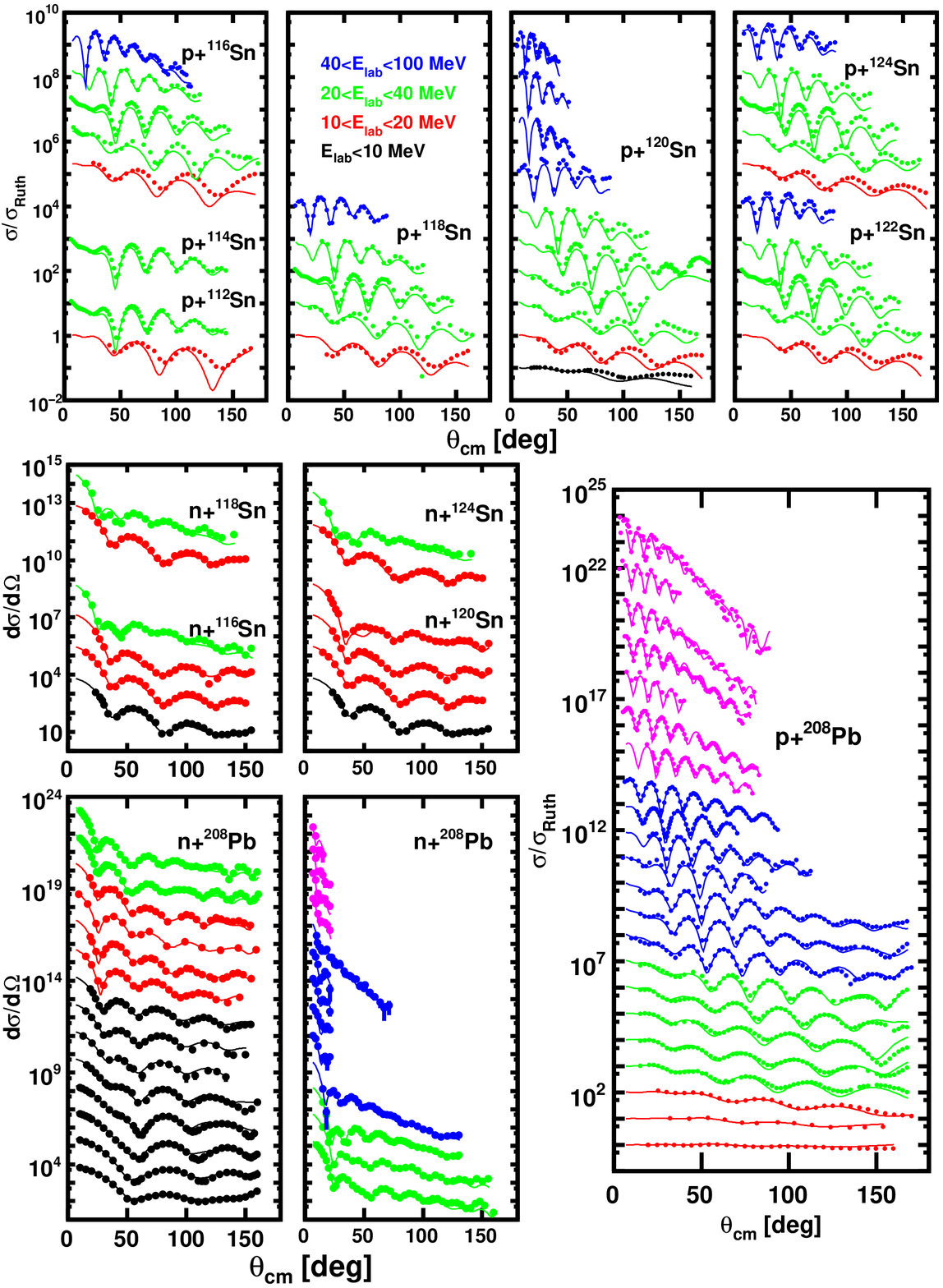}
\caption{Fitted elastic-scattering differential 
cross sections 
for proton and neutron reactions on Sn isotopes and $^{208}$Pb.
The color convention is the same as employed in Fig.~\ref{fig:sigCaNiP}.}
\label{fig:sigSnPb}
\end{center}
\end{figure*}

Global fits to the data were performed for four regions:
1) Ca, Ni isotopes and $N=28$ isotones, 2) $N=50$ isotones, 3) Sn isotopes, 
and 4) $^{208}$Pb.
References to experimental data can be found in Ref.~\cite{Mueller11}.
The fitted elastic-scattering 
differential cross sections
are shown in Figs.~\ref{fig:sigCaNiP} to \ref{fig:sigSnPb}.

In many cases parameters like diffuseness and those related to spin-orbit potentials are kept the same in all regions.
Parameters related to volume absorption are also kept fixed for all regions.
Other parameters reveal a consistency between fits to the different regions.
Fitting data from semi-magic nuclei as far as elastic scattering cross sections is concerned, appears just as successful as for double-closed shell systems.

\begin{figure*}[tbp]
\begin{center}
\includegraphics[scale=.7]{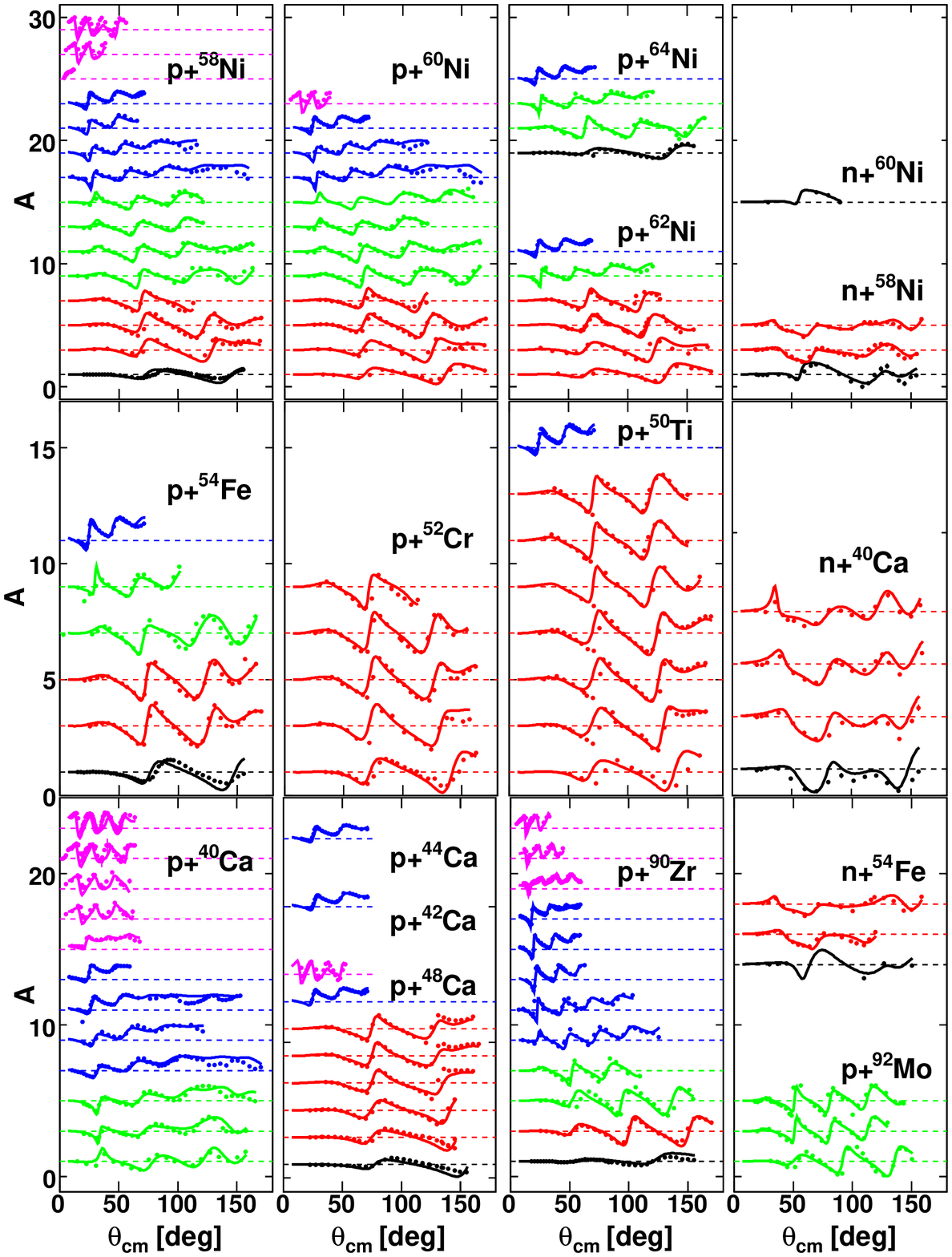}
\caption{Fitted analyzing powers
for proton and neutron reactions on $Z$=20, 28 and $N$=28, 50 target nuclei.
For clarity, successively larger energies have been shifted further up along the vertical axis. The dashed lines indicate zero analyzing power for each 
energy. The color convention is the same as employed in Fig.~\ref{fig:sigCaNiP}.}
\label{fig:analCaNi}
\end{center}
\end{figure*}

\begin{figure*}[tbp]
\begin{center}
\includegraphics[scale=.7]{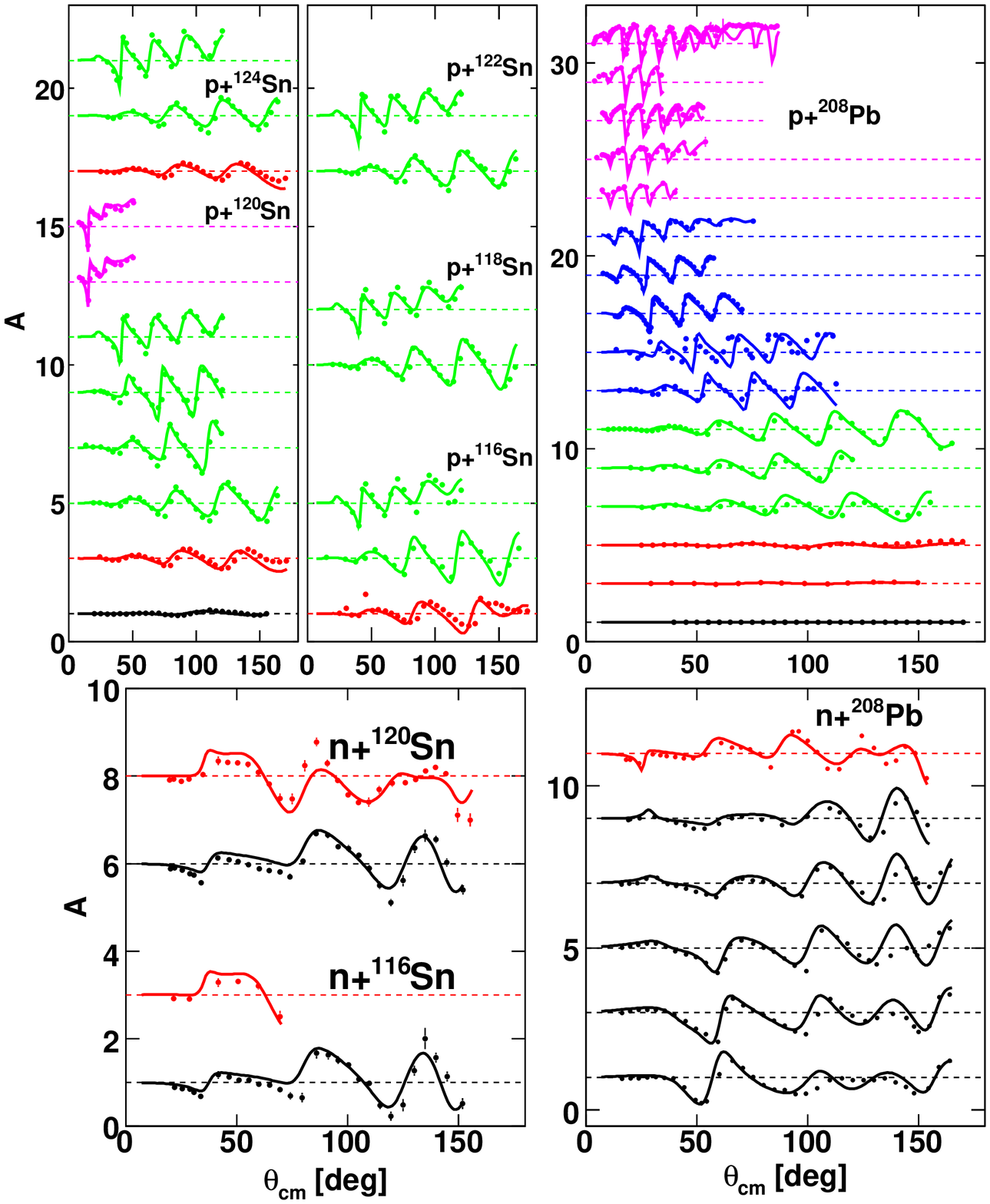}
\caption{As for Fig.~\ref{fig:analCaNi}, but reactions on 
Sn isotopes and $^{208}$Pb.}
\label{fig:analSnPb}
\end{center}
\end{figure*}

Results for the fitted
analyzing powers are displayed in Figs.~\ref{fig:analCaNi} and 
\ref{fig:analSnPb}. 
The quality of the description is comparable to that obtained with global optical-model potentials that are not dispersive. 
This holds also for spin-rotation parameters that are not shown here but given in Ref.~\cite{Charity07}.

\begin{figure*}[tbp]
\begin{center}
\includegraphics[scale=.6]{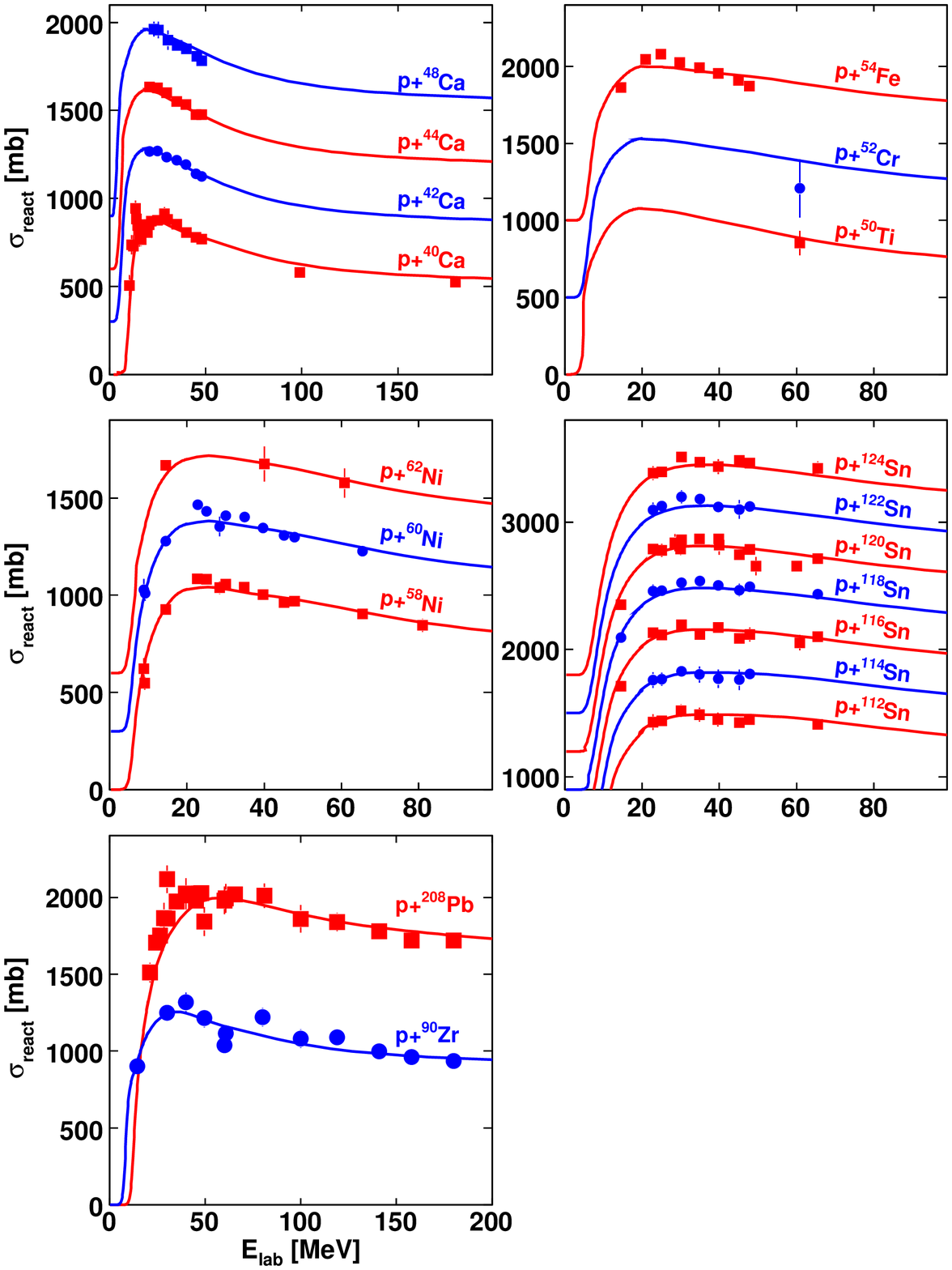}
\caption{Fits to proton total reaction cross sections.
For clarity, data and curves have been progressively shifted up along 
the vertical axis. } 
\label{fig:react}
\end{center}
\end{figure*}

\begin{figure}[tpb]
\begin{center}
\includegraphics[scale=.6]{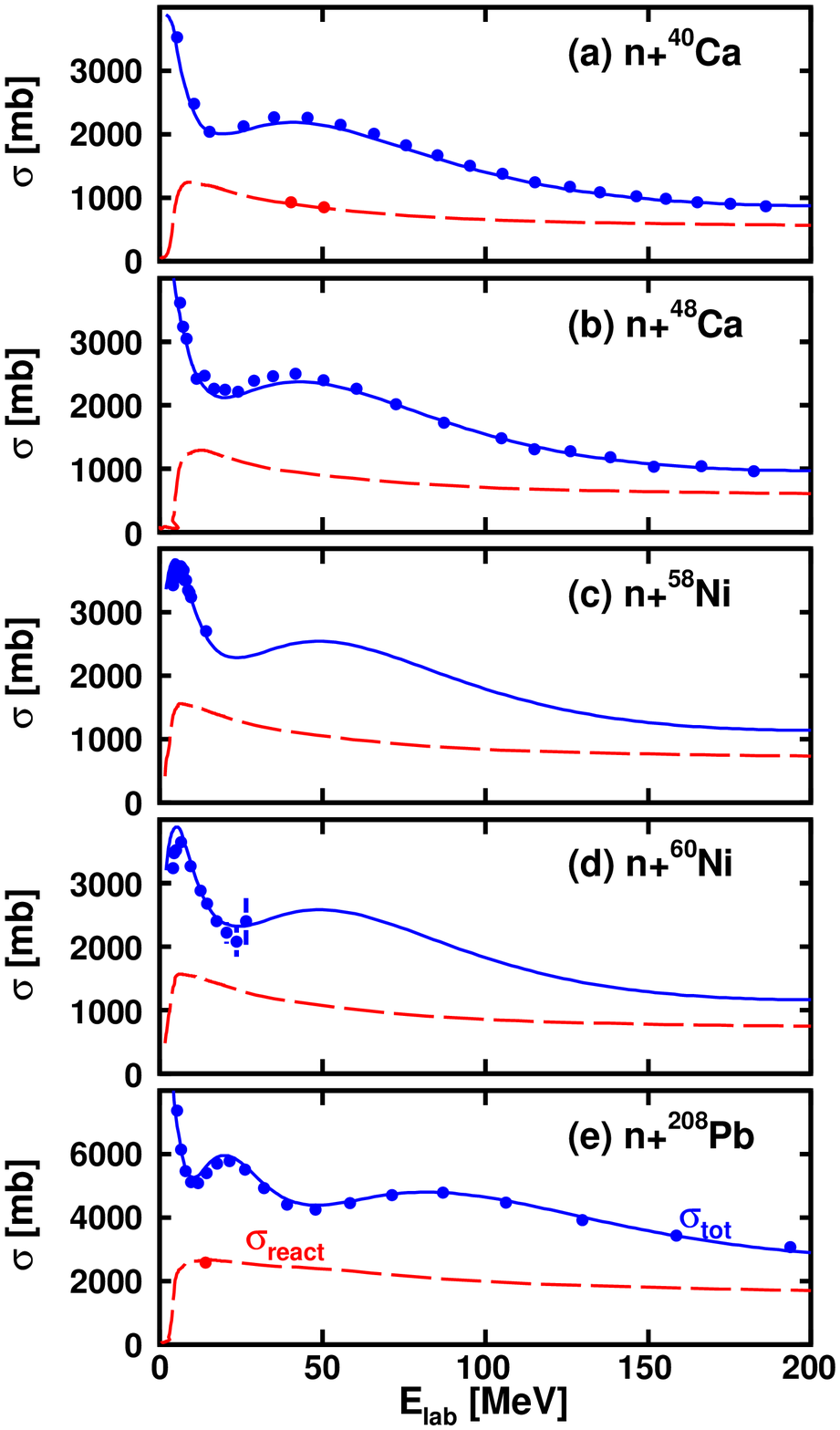}
\caption{Comparison of experimental (data points) and fitted (curves)
 cross sections for the indicated 
neutron-induced reactions. The solid and dashed curves are for the total and reaction cross section, respectively. 
 For clarity, some of the experimental points for the 
total reaction cross sections have been suppressed.
}
\label{fig:nReaction}
\end{center}
\end{figure}

Fitted  reaction cross sections for protons are shown in 
Fig.~\ref{fig:react} while fitted reaction and total cross sections for 
neutrons can be found in Fig.~\ref{fig:nReaction}.
The availability of total neutron cross sections out to high energy is particularly helpful as neutron elastic  differential cross sections at such energies are only widely available for $^{208}$Pb.
The quality of the fits are 
at least as good, if not better, than other global optical-model fits.

In most DOM analyses until recently, a comparison with level energies, spectroscopic factors [see Eq.~(\ref{eq:SF})], occupation numbers [see Eqs.~(\ref{eq:occHole}) and (\ref{eq:occPart})] is presented~\cite{Charity06,Charity07,Mueller11} based on the approximations introduced by Mahaux and Sartor~\cite{Mahaux91}.
Since the more recent applications of the DOM no longer need these approximations, we will discuss results for spectroscopic factors and occupation numbers when we present the introduction of non-local ingredients in subsequent sections.
In addition, results for observables like the nuclear charge density can then be presented as well.

\section{Application of DOM ingredients to the analysis of transfer reactions}
\label{sec:transfer}
Transfer reactions provide an important tool to study the properties of exotic nuclei.
Impinging a radioactive beam on a deuteron or proton target permits us to study neutron addition or removal from the beam nucleus.
Analyses of the corresponding cross sections has recently started to employ the ingredients of the DOM~\cite{Nguyen2011}. 
In these cases, the reaction theory is based on the adiabatic-wave approximation (ADWA) with inclusion of finite-range effects~\cite{Tandy74} which is capable of treating deuteron breakup in a practical way.
\begin{table}[b]
\caption{Spectroscopic factors obtained from the finite-range ADWA analysis. The deuteron kinetic energy $E_d$ (lab. frame) is in MeV. References to experimental data can be found in Ref.~\cite{Nguyen2011}\label{sfact}}
\begin{center}
\begin{tabular}{lcccc}
\hline
\hline
Nucleus &                   E$_d$ &  CH89+WS & DOM+WS & DOM \\ \hline
{$^{48}$Ca} & 2 &     0.94 &  0.72 &  0.66 \\
                            & 13 &  0.82 &    0.67 &  0.61  \\
                            & 19.3 & 0.77 &  0.68 &  0.62  \\
                            & 56 &  1.1 &     0.70 &  0.62  \\ \hline
$^{132}$Sn                  & 9.46 & 1.1 &   1.0 &   0.72 \\ 
\hline \hline
\end{tabular}
\end{center}
\end{table}

We discuss here the case of the $^{48}$Ca$(d,p)^{49}$Ca reaction at $E_d=2, 13, 19.3$ and $56$~MeV,  as well as the $^{132}$Sn$(d,p)^{133}$Sn reaction at $E_d=9.46$~MeV.
The traditional method of analyzing these reactions employs global optical-model potentials for both the protons and neutrons. For example the Chapel-Hill parametrization CH89~\cite{Varner91} is commonly used.
In addition, the overlap function is obtained from solving the Schr{\"{o}}dinger equation for a Woods-Saxon (WS) potential whose depth is suitably adjusted to obtained the correct binding energy for the added neutron.
A spectroscopic factor is then extracted by rescaling the cross section such that the maximum is adequately described.
Such an approach can yield useful information on relative spectroscopic factors~\cite{Lee06,Lee10}.
Nevertheless, for these reactions and energies the traditional approach often yields inconsistent results with regard to the probabilistic nature of a spectroscopic factor as defined in  Sec.~\ref{sec:equations}
 which should strictly generate a value between 0 and 1.
We illustrate this point in Table~\ref{sfact} which reports on finite-range ADWA calculations performed for three interaction models: optical-model potentials from CH89~\cite{Varner91} and the neutron
overlap function from an adjusted Woods-Saxon potential (CH89+WS); optical-model potentials from DOM but the neutron overlap function as before (DOM+WS);
and finally both the optical-model potentials and the neutron overlap function from this local approximation of the DOM corrected
for non-locality (DOM) (see Ref.~\cite{Nguyen2011}).
Note that the normalizations are calculated with an overlap function
normalized to unity. 

\begin{figure}
\centering
\includegraphics[width=0.6\textwidth]{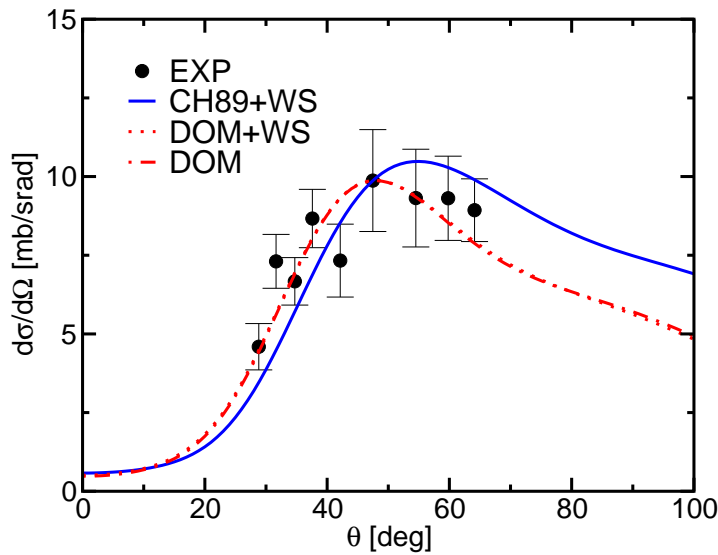}
\caption{Angular distributions for the $^{132}$Sn$(d,p)^{133}$Sn reaction at a deuteron energy of $E_d = 9.46$ MeV are shown normalized at the peak of the experimental cross section. The curves for the DOM+WS and DOM calculations largely overlap and are difficult to separate.}
\label{fig-sn132}
\end{figure}

The first thing to note is that fits with the DOM ingredients are able to describe the experimental cross sections  as illustrated in Ref.~\cite{Nguyen2011} and Fig.~\ref{fig-sn132}. In fact there is essentially no difference between the angular distributions predicted with the DOM, DOM+WS and CH89+WS ingredients for the $^{48}$Ca($d,p$) reactions whereas  for the ${}^{132}$Sn($d,p$) case as illustrated in Fig.~~\ref{fig-sn132}, the DOM and DOM+WS cases differ somewhat from the standard calculation, but still follow the experimental data quite well.
While the angular distributions predicted using the DOM do not differ considerably from those using CH89, the normalization of the cross sections do.
An experimental spectroscopic factor is then determined by taking the ratio of $d\sigma/d\Omega$(exp) over $d\sigma/d\Omega$(theory)  at the first peak of the angular distribution for all but sub-barrier energies where the normalization is determined at backward angles.
If one focuses first on the traditional CH89+WS approach, one sees that a wide range of spectroscopic factors can be obtained depending
on the beam energy.
For example with ${}^{48}$Ca, the spectroscopic factors range from $S=0.77$ to 1.1.
However, this range is significantly reduced when DOM optical potentials are used.
The consistency of the extracted spectroscopic factors employing DOM ingredients is therefore very encouraging.
The extrapolation of traditional non-dispersive potentials like CH89 towards the drip line therefore generates a different result than the corresponding extrapolation of the dispersive DOM potential generated with the ingredients of Sec.~\ref{sec:param}.
Further extensions of combining finite-range ADWA reaction theory with DOM potentials that include a full treatment of non-locality are in progress and a first analysis is reported in Ref.~\cite{Ross15}.

\section{Comparison with \textit{ab initio} calculations of the nucleon self-energy}
  
 Further progress in gaining insights into the character of DOM potentials has come from studies of \textit{ab initio} calculations of the nucleon self-energy.
We report here on two types of calculations with very different emphasis published in Refs.~\cite{Waldecker2011} and \cite{Dussan11}.
In Ref.~\cite{Waldecker2011} an analysis is made of self-energies calculated for ${}^{40}$Ca, ${}^{48}$Ca, and ${}^{60}$Ca using the  Faddeev random-phase approximation (FRPA) developed and applied in Refs.~\cite{Barb:1,Barb:2,Dickhoff04}.
Such calculations first proceed by constructing an effective interaction calculated in  a configuration space that is large enough to accommodate most of the effects of low-lying collective excitations on the motion of individual particles.
This interaction is a standard version of the $\mathcal{G}$-matrix which propagates noninteracting particles outside the chosen configuration space to all orders and therefore adequately treats SRC.
Subsequently, calculated phonons in the random-phase approximation in both the combined particle-particle and hole-hole space as well as particle-hole space are then employed in a Faddeev summation to all orders to generate a suitable treatment of the effect of collective motion in the nucleon self-energy.
Due to computational restrictions, the configuration space is limited to 8 or 10 major shells and therefore ceases to provide meaningful results beyond energies that are further removed from the Fermi energy than about 60-80 MeV.
The emphasis of these FRPA calculations for the Ca isotopes is therefore on the effect of low-energy or LRC.

Contrary to this approach, the results of Ref.~\cite{Dussan11} focus on volume effects that are usually studied in the treatment of SRC.
Also here a $\mathcal{G}$-matrix is first constructed but in nuclear matter at some suitably chosen density and starting energy.
This interaction is then corrected  for the propagation of two-particle-one-hole ($2p1h$) and one-particle-two-hole ($1p2h$) states in the finite system ($^{40}$Ca) by propagating the difference between the finite-nucleus and nuclear-matter propagators in second order.
Since these intermediate states do not contain additional interactions at low energy, the resulting self-energy only contains the dynamical effects associated with SRC and therefore provides complementary information to the results of Ref.~\cite{Waldecker2011} which emphasize LRC.

\subsection{Self-energy calculations for LRC}
\label{sec:FRPA}
For a $J = 0$ target nucleus, all partial waves $(\ell , j, \tau )$ are decoupled, were $\ell$,$j$ label the projectile nucleon's orbital and total angular momentum and $\tau$ represents its isospin. The irreducible self-energy in coordinate space (for either a proton or a neutron) can be written in terms of the harmonic-oscillator basis used in the FRPA calculation, as follows:
\begin{eqnarray}
\!\!\!\!\!\!\!\!\!\!\!\!\!\!\!\!\!\!\!\!\!
\Sigma( \bm{x}, \bm{x^\prime}; E ) = \sum_{\ell j m_j \tau} {\cal I }_{\ell j m_j}( \Omega, \sigma ) 
\left[ \sum_{n_a, n_b} R_{n_a \ell}(r) \Sigma^\star_{ab}(E)R_{n_b \ell}( r^\prime )\right] ( {\cal I }_{\ell j m_j}( \Omega^\prime, \sigma^\prime ) )^* ,
\label{eq:selfr}
\end{eqnarray}
where $\bm{x} \equiv \bm{r}, \sigma, \tau$.  
The spin variable is represented by $\sigma$, $n$ is the principal quantum number of the harmonic oscillator, and $a\equiv(n_a, \ell , j, \tau)$ (note that for a $J = 0$ nucleus the self-energy is independent of $m_j$).
The standard harmonic-oscillator function is denoted by $R_{n \ell}(r)$, while ${\cal I }_{\ell j m_j}( \Omega, \sigma )$ represent the $j$-coupled angular-spin function. 

In Ref.~\cite{Waldecker2011} the harmonic oscillator projection of the self-energy was calculated directly as
\begin{eqnarray}
\Sigma_{ab}( E ) = \Sigma^{\infty}_{ab}(E) + \tilde{\Sigma}_{ab}(E) 
= \Sigma^{\infty}_{ab}(E) + \sum_{r}{\frac{m_{a}^r( m_{b}^r )^*}{E -\varepsilon_r \pm i\eta}} .
\label{eq:selfHO}
\end{eqnarray}
The term with the tilde is the dynamic part of the self-energy due to LRC calculated in FRPA~\cite{Barb:1,Barb:2,Dickhoff04,Barb07,Barb09}, and $\Sigma^{\infty}_{ab}(E)$ is the correlated HF term which acquires an energy dependence through the energy dependence of the $\mathcal{G}$-matrix effective interaction. $\Sigma^{\infty}_{ab}(E)$ is the sum of the strict-correlated HF diagram (which  is energy independent) and the dynamical contributions due to short-range interactions outside the chosen model space.
The self-energy was further decomposed in a central ($0$) part and a spin-orbit ($ls$) part according to
\begin{eqnarray}
\Sigma^{\ell j_>} &=& \Sigma^{\ell}_{0}+\frac{\ell}{2}\Sigma^{\ell}_{\ell s} \label{eq:ls1} \, ,\\
\Sigma^{\ell j_<} &=& \Sigma^{\ell}_{0} - \frac{ \ell + 1}{2}\Sigma^{\ell}_{\ell s}  \, ,
\label{eq:ls2}
\end{eqnarray}
with $j_{>,<}\equiv \ell \pm \frac{1}{2}$.
The corresponding static terms are denoted by $\Sigma^{\infty, \ell}_0$ and $\Sigma^{\infty, \ell}_{\ell s}$, and the corresponding dynamic terms are denoted by $\tilde{\Sigma}^{\ell}_0$ and $\tilde{\Sigma}^{\ell}_{\ell s}$.

The FRPA calculation employs a discrete sp basis in a large model space which results in a large number of poles in the self-energy given in Eq.~(\ref{eq:selfHO}).
For a comparison with optical-model potentials at positive energies, it is appropriate to smooth out these contributions by employing a finite width for these poles as the optical-model potential was always intended to represent an average smooth behavior of the nucleon self-energy~\cite{Mahaux91}.
In addition, it makes physical sense to at least partly represent the escape width of the continuum states by this procedure.
Finally, further spreading of the intermediate states to more complicated states ($3p2h$ and higher excitations that are not included in the calculation) can also be accounted for by this procedure.
Thus, before comparing to the DOM potentials, the dynamic part of the microscopic self-energy was smoothed out using a finite, energy-dependent width for the poles 
\begin{equation}
\tilde{\Sigma}_{n_a, n_b}^{\ell j}(E) = \sum_{r}\frac{m_{n_a}^{r} m_{n_b}^{r}}{ E - \varepsilon_r \pm i\eta } \longrightarrow \sum_{r}\frac{m_{n_a}^{r} m_{n_b}^{r}}{ E - \varepsilon_r \pm i\Gamma(E) } \, .
\label{eq:width}
\end{equation}
Solving for the real and imaginary parts we obtain
\begin{eqnarray}
\lefteqn{
\tilde{\Sigma}_{n_a, n_b}^{\ell j}(E) = \sum_{r}\frac{(E - \varepsilon_r )}{ ( E - \varepsilon_r )^2 + [\Gamma(E)]^2 }m_{n_a}^{r} m_{n_b}^{r} 
 } & &	
 \label{eq:smooth} \\
&\qquad +& i\left[\theta( \varepsilon_F - E )\sum_{h}\frac{\Gamma}{( E - \varepsilon_h )^2 + \Gamma(E)^2}m_{n_a}^{h}m_{n_b}^h  \right. \nonumber \\
&\qquad -& \left.  \theta( E - \varepsilon_F )\sum_{p}\frac{\Gamma}{( E - \varepsilon_p )^2 + [\Gamma(E)]^2}m_{n_a}^{p}m_{n_b}^p\right] ,
\nonumber 
\end{eqnarray}
where, $r$ implies a sum over both particle and hole states, $h$ denotes a sum over the hole states only, and $p$ a sum over the particle states only. 
For the width, the following form was used~\cite{Brown81}:
$$
\Gamma( E ) = \frac{1}{\pi}\frac{ \mathbb{A} \, (E- \varepsilon_F)^2}{(E - \varepsilon_F)^2 - \mathbb{B}^2}
$$
with $\mathbb{A}$=12~MeV and $\mathbb{B}$=22.36~MeV.
This generates a narrow width near $\varepsilon_F$ that increases as the energy moves away from the Fermi surface, in accordance with observations.

In the DOM representation of the optical-model potential, the self-energy is recast in a subtracted form as
\begin{equation}
\Sigma^\star_{ab}( E ) = \Sigma^{\infty}_{ab, \,S} + \tilde{\Sigma}_{ab}(E)_S, 
\label{eq:suba}
\end{equation}
where
\begin{eqnarray}
       \Sigma^{\infty}_{ab \, S} &= &\Sigma^\star_{ab}(\varepsilon_F) \, ,
\label{eq:SigSubHF} \\
\tilde{\Sigma}_{ab}(E)_S    &= &\Sigma^\star_{ab}(E) - \Sigma^\star_{ab}(\varepsilon_F) \, .
\label{eq:SigSubDyn}
\end{eqnarray}		
We reiterate that the (real) $\Sigma^\infty_{ab, \, S}$ and the imaginary part of $\tilde{\Sigma}_{ab}(E)_S$ are 
parametrized in the DOM potential. The real part of \hbox{$\tilde{\Sigma}_{ab}(E)_S$} is then fixed by the dispersion relation.    
The imaginary potentials of the DOM are defined in the same way as the self-energies in Eq.~(\ref{eq:selfHO}) and the potentials can therefore be compared directly apart from the non-locality correction of Eq.~(\ref{eq:nlcorr}) for the DOM.

In fitting optical potentials, it is usually found that volume integrals are better constrained by the experimental data~\cite{Mahaux91,Greenless68}. For this reason, they have been considered as a reliable measure of the total strength of a potential. For a non-local and $\ell$-dependent potential of the form~(\ref{eq:selfr}) it is convenient to consider separate integrals for each angular momentum component, $\Sigma^{\ell}_0(r, r^\prime)$ and $\Sigma^{\ell}_{\ell s}(r, r^\prime)$, which correspond to the square brackets in Eq.~(\ref{eq:selfr}) and decomposed according to Eqs.~(\ref{eq:ls1}) and (\ref{eq:ls2}).
 Denoting the central real part of the optical potential by $V$,  and the central imaginary part by $W$, we then calculate:
\begin{eqnarray}
J_W^{\ell}(E) = 4\pi\int{drr^2\int{dr^\prime r^{\prime 2} \textrm{Im } \Sigma^{\ell}_0(r, r^\prime ; E)}}
\label{eq:intgs_W} \\
J_V^{\ell}(E) = 4\pi\int{drr^2\int{dr^\prime r^{\prime 2} \textrm{Re } \Sigma^{\ell}_0(r, r^\prime ; E)}} .
\label{eq:intgs_V}
\end{eqnarray}

The correspondence between the above definitions and the volume integrals used for the (local) DOM potential in Refs.~\cite{Charity06,Charity07} can be seen by casting a spherical local potential $U(r)$ into a non-local form $U(\bm r, \bm r^\prime ) = U(r) \delta(\bm r - \bm r^\prime)$. Expanding this in spherical harmonics gives
\begin{equation}
 U(\bm r, \bm r^\prime ) = \sum_{lm} U^\ell(r, r^\prime)Y^*_{\ell m}(\Omega^\prime)Y_{\ell m}(\Omega) \, ,
\end{equation}
with the $\ell$ component
\begin{equation}
 U^\ell(r, r^\prime ) = \frac {U(r)}{r^2} \delta(r - r^\prime) \, ,
\end{equation}
which does not depend on $\ell$.
The definition in Eqs.~(\ref{eq:intgs_W}) and (\ref{eq:intgs_V}) for the volume integrals lead to
\begin{eqnarray}
 J^\ell_U &=& 4\pi  \int{ drr^2\int{dr^\prime r^{\prime 2} U^{\ell}(r, r^\prime)}} 
\\
      &=& 4\pi\int{ U(r) r^2dr} = \int{ U(r) d\bm{r}}  \, \hbox{, \hspace{.2in} for any $\ell$}
\nonumber
\end{eqnarray}
and reduces to the usual definition of volume integral for local potentials.
Thus, FRPA results can be compared directly to the corresponding quantities determined in previous studies of the DOM.

\begin{figure}
\begin{center}
\includegraphics[scale=0.6]{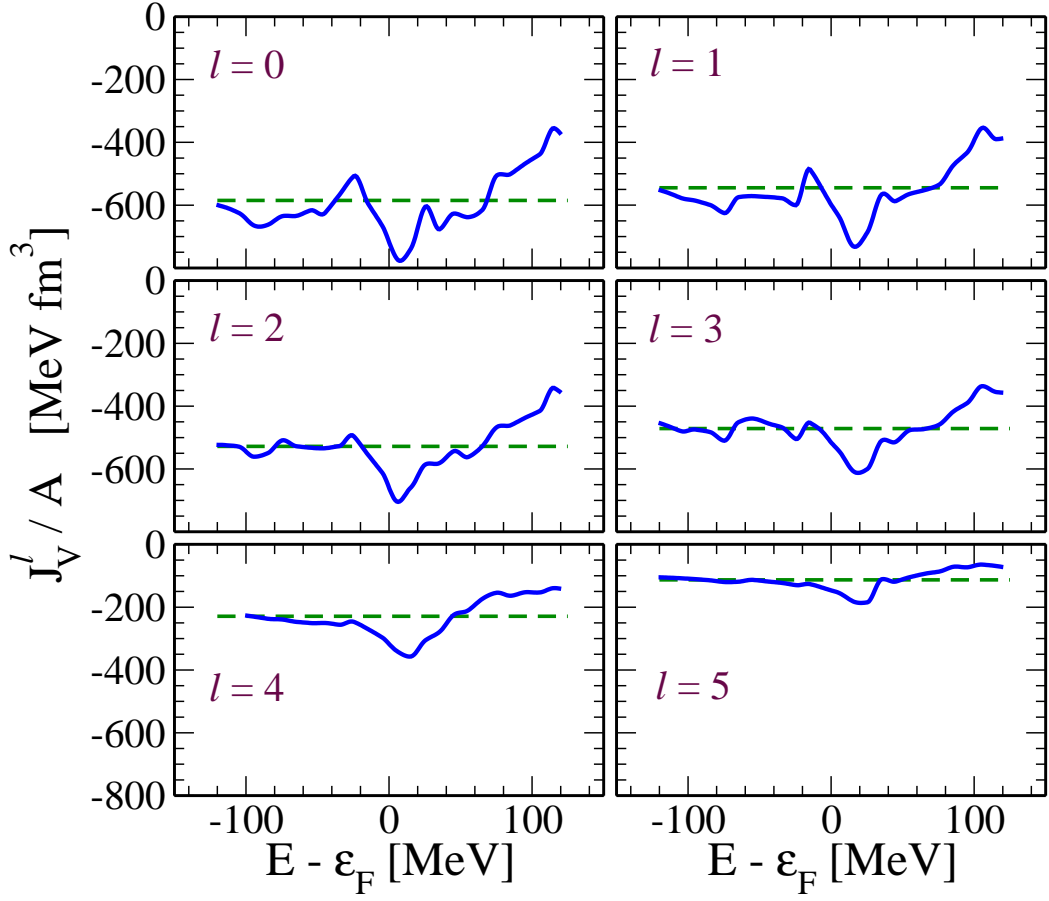}
\caption{Volume integrals of Re $\Sigma^{\ell}_0$ for neutrons in ${}^{40}$Ca. The horizontal, dashed lines are $J_F^{\ell}$, the volume integrals of $\Sigma^{\infty, \ell}_0(\varepsilon_F)$. }
\label{fig:Jv_nca40}
\end{center}
\end{figure}
Figure~\ref{fig:Jv_nca40} gives an overall example of the features of the real 
part of the self-energy~($J_V^\ell$). 
These results are shown for neutrons in ${}^{40}$Ca, employing the AV18 interaction~\cite{Wiringa1995}
and are separated in partial waves up to $\ell$=5.
In this figure, the difference between $J_V^\ell$ and $J_F^\ell$, the volume integral of  
$\Sigma^{\infty, \ell}_0(\varepsilon_F)$,
decreases with increasing $\ell$ which reflects a similar reduction of the imaginary parts,  $J_W^\ell$, to which $J_V^\ell$ are linked through
the dispersion relation.
The effect may be partly explained by the truncated model space, since the higher $\ell$-channels also have fewer orbits. 
However, most of this decrease must arise from the $\ell$-dependence implied by the non-locality of the potential. This $\ell$-dependence suggests the importance of including non-local features in DOM potentials.

\begin{figure}[bt]
\begin{center}
\includegraphics[scale=0.6]{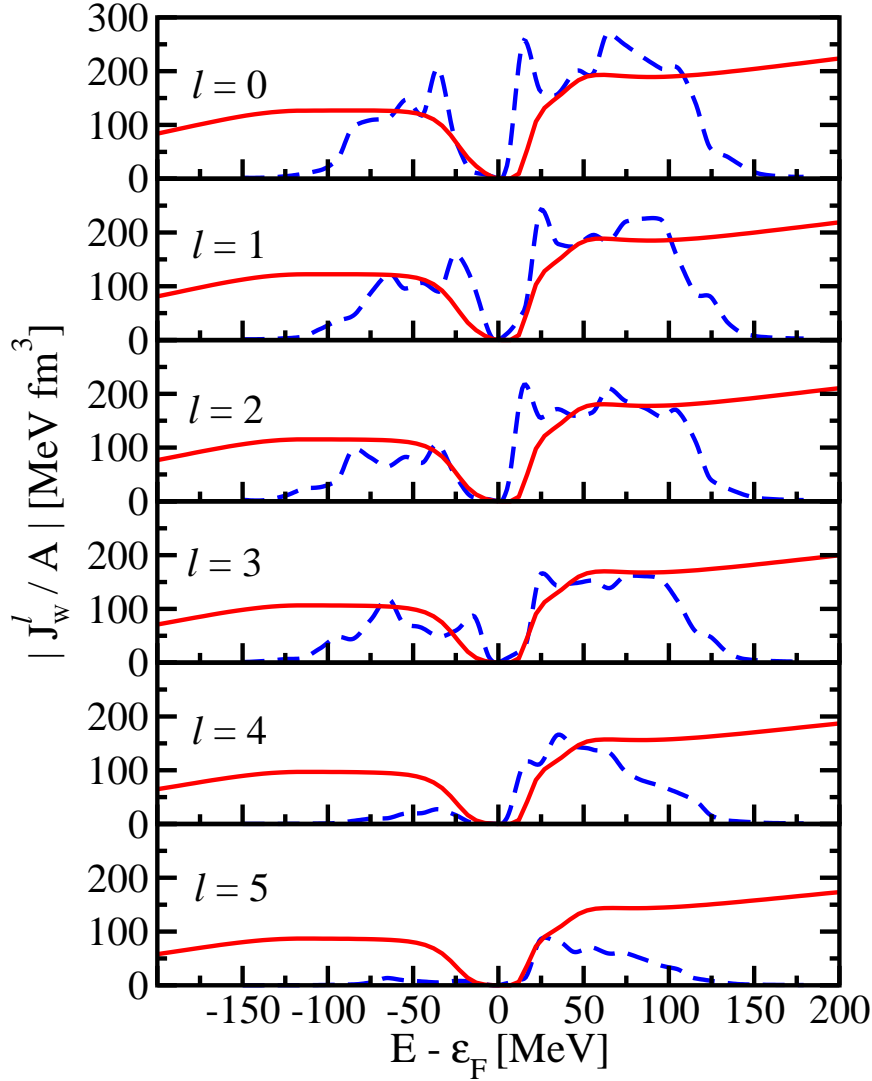}
\caption{ Imaginary volume integrals $J^{\ell}_W$ from Eq.~(\ref{eq:intgs_W}) of the ${}^{40}$Ca self-energy for neutrons. The dashed curves represent the FRPA results. The results of the non-local DOM fit discussed in Sec.~\ref{sec:nlIM} is shown by the solid lines.} 
\label{fig:Jw_review}
\end{center}
\end{figure}

The simplifying assumptions of a symmetric absorption around $\varepsilon_F$~\cite{Mahaux91,Charity06,Charity07,Mueller11} and locality (and therefore $\ell$-independence) in the DOM generates unrealistic occupation of higher $\ell$-values below the Fermi energy~\cite{Dickhoff10}.
Such features are not obtained in the FRPA
as illustrated in Fig.~\ref{fig:Jw_review}, where for $\ell$-values up to 5, the volume integrals of the FRPA calculation are displayed by the dashed lines.
A subtantial $\ell$-dependence below the Fermi energy as documented by the FRPA results suggests that a substantial non-local absorption is at work which should be incorporated in the DOM.
In  Fig.~\ref{fig:Jw_review}, we also include the result of the non-local fit discussed in Sec.~\ref{sec:nlIM} as the solid lines which confirms this assessment.
Higher $\ell$-values are less relevant below the Fermi energy and this is clearly illustrated by the FRPA results in this figure.
Since the absorption above the Fermi energy is strongly constrained by elastic-scattering data, it is encouraging that the $\ell$-dependent FRPA result is reasonably close to the DOM fit in the domain where the FRPA is expected to be relevant on account of the size of the chosen configuration space. 
Note that the calculated $J_W$ decreases quickly at energies $E-\varepsilon_F >$100~MeV due to the truncation the model space. Instead, the DOM result correctly shows that it remains sizable even at higher energies.
Also at negative energies, the FRPA results do not adequately describe the admixture of high-momentum components that occur at large missing energies [see Sec.~\ref{sec:nlIM}].

Improving the DOM analysis of elastic-scattering data above the Fermi energy and observables related to quantities below the Fermi energy  requires sensitively to the treatment of non-locality in the imaginary part of the self-energy~\cite{Dickhoff10}.
To gain some insight into the properties of the FRPA self-energy, a few simple fits were performed to represent the central part of the imaginary part of the  FRPA self-energy in coordinate space at a given energy assuming the following form of the potential
\begin{equation}
W_{NL}(\bm{r},\bm{r}') = W_0 \sqrt{f(r)}\sqrt{f(r^\prime)}H\left(\frac{\bm{r} - \bm{r}^\prime}{\beta}\right) .
\label{eq:nlocal}
\end{equation}
We deviate from the standard Perey prescription for non-locality by employing square-root factors of the function $f(r)$ which is still represented by the conventional Woods-Saxon form factor.
The function $H$ determines the degree of non-locality and is assumed to be a Gaussian following Ref.~\cite{Perey62}
\begin{equation}
H\left(\frac{\mathbf{r} - \mathbf{r}^\prime}{\beta}\right) = \frac{1}{\pi^{3/2}\beta^3}\textrm{exp}\left(\frac{|\mathbf{r} - \mathbf{r}^\prime|^2}{\beta^2}\right) .
\label{eq:gauss}
\end{equation}
When the angular dependence in $H$ is projected out, an analytic solution is obtained for each orbital angular momentum $\ell$ as
\begin{eqnarray}
\!\!\!\!\!\!\!\! W^{\ell}_{NL}(r,r') = W_0 \sqrt{f(r)}\sqrt{f(r^\prime)}\frac{4}{\pi^{1/2}\beta^3} 
\textrm{exp}\left(-\frac{r^2 + r^{\prime2}}{\beta^2}\right)i^\ell(-1)^\ell j_\ell(iz),
\label{ell-VanNeck}
\end{eqnarray} 
where $z = 2rr^\prime / \beta^2$ and $j_\ell$ is a spherical Bessel function with a purely imaginary argument. 
The fact that an analytic projection is possible provided the motivation of the choice of Eq.~(\ref{eq:nlocal}).

\begin{figure}{tb}
\begin{center}
\includegraphics[width=3.3in]{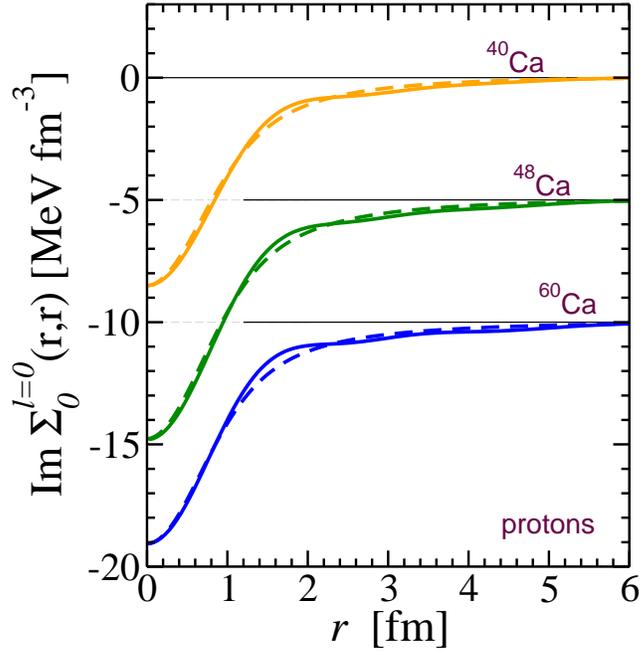}
\caption{Diagonal part of the FRPA imaginary self-energy for protons at $E-\varepsilon_F$=44~MeV (solid curve) and the corresponding parametrized self-energy (dashed curve). The results are shown for $\ell = 0$ and are offset by 5 MeV for each subsequent nucleus.\label{fig:diagx} }
\end{center}
\end{figure}
We have chosen to fit the imaginary part of the FRPA self-energy at an energy of $E-\varepsilon_F$=44 MeV, where surface physics dominates.
In practice, we consider only a fit to the $\ell = 0$ self-energy since it represents the partial wave with the best convergence properties associated with the limited model space considered.
In Fig.~\ref{fig:diagx},  the diagonals of the central imaginary part of the FRPA self-energy in coordinate space with $\ell=0$ for ${}^{40}$Ca, ${}^{48}$Ca, and ${}^{60}$Ca are displayed by solid lines. Results for the heavier isotopes 
are offset on the $y$ axis for clarity.
The corresponding $\ell=0$ projections of Eq.~(\ref{eq:nlocal}) given by Eq.~(\ref{ell-VanNeck}) are shown by the dashed curves.
The fit according to Eq.~(\ref{eq:nlocal}) appears to be quite satisfactory suggesting that a reasonable representation of the microscopic self-energy is possible and may provide a useful starting point for DOM functionals.

\begin{table}[tp]
\begin{center}
\caption{ Parameters from non-local fits to the imaginary part of the proton self-energy at $E-\varepsilon_F$=44 MeV for ${}^{40}$Ca, ${}^{48}$Ca, and ${}^{60}$Ca using Eq.~(\ref{eq:nlocal}). $r_0$ and $a_0$ are standard parameters of the Wood-Saxon form factor $f(r)$.}
\label{tbl:parms1}
\begin{tabular}{ccccccc}
\hline \hline
Isotope & $W_0$ & $r_0$ & $a_0$ & $\beta$ & $|J_W / A|$ & $|J_W / A|_{FRPA}$ \\
       & [MeV]  & [fm]  &  [fm] &  [fm]   & [Mev fm$^3$] & [MeV fm$^3$] \\ 
\hline 
${}^{40}$Ca & 14.1 & 1.23 & 1.23 & 1.54 & 187 & 188 \\ 
${}^{48}$Ca & 16.1 & 1.32 & 1.30 & 1.54 & 242 & 241 \\  
${}^{60}$Ca & 13.6 & 1.50 & 1.50 & 1.49 & 266 & 268\\ \hline \hline
\end{tabular}
\end{center}
\end{table}

The properties of the imaginary part of the FRPA self-energy in terms of its non-locality content are summarized in Table~\ref{tbl:parms1} for the three different nuclei.
The parameters are fitted to reproduce the essential properties of the self-energy including the volume integral for $\ell=0$.
We observe that the values for the diffuseness are larger than standard ones and increase with neutron number.
The radius parameter exhibits a similar nonstandard trend.
The value of the non-locality parameter $\beta$ is also substantially larger than typically assumed for real non-local potentials.

\subsection{Self-energy calculation for SRC in ${}^{40}\textrm{Ca}$}
\label{sec:SRC}
 Following Refs.~\cite{Muther94,Muther95,Polls95}, the calculation of the nucleon self-energy with emphasis on SRC proceeds in two steps.
A diagrammatic treatment of SRC always involves the summation of ladder diagrams.
When only particle-particle (pp) intermediate states are included, the resulting effective interaction is the so-called $\mathcal{G}$-matrix.
The corresponding calculation for a finite nucleus (FN) can be represented in operator form by
\begin{equation}
\mathcal{G}_{FN}(E)= V + V G^{pp}_{FN}(E) \mathcal{G}_{FN}(E) ,
\label{eq:gmatfn}
\end{equation}
where the noninteracting propagator $G^{pp}_{FN}(E)$ represents two particles above the Fermi sea of the finite nucleus taking into account the Pauli principle.
The simplest implementation of $G^{pp}_{FN}$ involves plane-wave intermediate states (possibly orthogonalized to the bound states).
Even such a simple assumption leads to a prohibitive calculation to solve Eq.~(\ref{eq:gmatfn}) and generate the relevant real and imaginary part of the self-energy over a wide range of energies both above and below the Fermi energy.
Such a direct solution has not yet been attempted, confining the use of the $\mathcal{G}$-matrix as an effective interaction at negative energy.
For our purpose the appropriate strategy was developed in Refs.~\cite{Bonatsos89,Borromeo92}  which first calculates a $\mathcal{G}$-matrix in nuclear matter at a fixed density and fixed energy according to
\begin{equation}
\mathcal{G}_{NM}(E_{NM})= V + V G^{pp}_{NM}(E_{NM}) \mathcal{G}_{NM}(E_{NM}) .
\label{eq:gmatnm}
\end{equation}
The energy $E_{NM}$ is chosen below twice the Fermi energy of nuclear matter for a kinetic energy sp spectrum and the resulting $\mathcal{G}_{NM}$ is therefore real.
Formally solving Eq.~(\ref{eq:gmatfn}) in terms of $\mathcal{G}_{NM}$ can be accomplished by 
\begin{eqnarray}
\mathcal{G}_{FN}(E) = \mathcal{G}_{NM} 
\label{eq:gmatfnp} 
+   \mathcal{G}_{NM} \left\{G^{pp}_{FN}(E) - G^{pp}_{NM} \right\} \mathcal{G}_{FN}(E) ,
\end{eqnarray}
where the explicit reference to $E_{NM}$ is dropped.
The main assumption that allows a managable self-energy calculation is to drop all terms higher than second order in $\mathcal{G}_{NM}$, leading to
\begin{eqnarray}
\mathcal{G}_{FN}(E) = \mathcal{G}_{NM} - \mathcal{G}_{NM}  G^{pp}_{NM}  \mathcal{G}_{NM}
+   \mathcal{G}_{NM} G^{pp}_{FN}(E) \mathcal{G}_{NM} ,
\label{eq:gmatfnq}
\end{eqnarray}
where the first two terms are energy-independent.
Since a nuclear-matter calculation already incorporates all the important effects associated with SRC, it is reasonable to assume that the lowest-order iteration of the difference propagator in Eq.~(\ref{eq:gmatfnq}) represents an accurate approximation to the full result.

Equation~(\ref{eq:gmatnm}) generates an appropriate solution
of two-body short-range dynamics, but the resulting matrix elements
require further manipulation before becoming useful for the finite nucleus~\cite{Dussan11}.
The self-energy contribution of the lowest-order term $\mathcal{G}_{NM}$ in Eq.~(\ref{eq:gmatfnq}) is similar to a Brueckner-Hartree-Fock (BHF) self-energy. 
While strictly speaking the genuine 
BHF approach involves self-consistent sp wave functions, as in the HF 
approximation, the
main features associated with using the $\mathcal{G}_{NM}$-matrix of Eq.~(\ref{eq:gmatnm})
are approximately the same when employing a summation over the occupied harmonic-oscillator states of ${}^{40}$Ca. Hence we will use the BHF abbreviation.
The correction term involving the second-order contribution in $\mathcal{G}_{NM}$ calculated in nuclear matter is also static and can be obtained from the second term in the nuclear-matter Bethe-Goldstone equation by replacing the bare interaction by $\mathcal{G}_{NM}$. The corresponding self-energy is also real and generated by summing over the occupied oscillator states in the same way as for the BHF term.

The second-order term containing the correct energy-dependence for $\mathcal{G}_{FN}$ in Eq.~(\ref{eq:gmatfnq}) can now be used to construct the self-energy contribution, representing the coupling to 2p1h states.
In the calculation, harmonic oscillator states for the occupied (hole) 
states and plane waves for the intermediate unbound particle 
states are assumed, incorporating the correct energy and density dependence 
characteristic of a finite nucleus $\mathcal{G}_{FN}$-matrix.
In a similar way, one can construct the second-order self-energy contribution which has an imaginary part below the Fermi energy and includes the coupling to $1p2h$ states.
Calculations of this kind require several basis transformations, including the one from relative and center-of-mass momenta with corresponding orbital angular momenta to two-particle states with individual momenta and orbital angular momentum. 
Complete details can be found in Refs.~\cite{Borromeo92,Muther95}.
In practice the imaginary parts associated with the dynamic self-energy contributions are employed to obtain the corresponding real parts by employing the appropriate dispersion relation.
The resulting (irreducible) self-energy is then 
\begin{eqnarray}
\Sigma & = & \Sigma_{BHF} + \Delta\Sigma \label{eq:defsel}  \\
& = & \Sigma_{BHF} + 
\left( \textrm{Re}~\Sigma_{2p1h} - \Sigma_{c} + \textrm{Re}~\Sigma_{1p2h} 
\right) 
+ i \left( \textrm{Im}~\Sigma_{2p1h} + \textrm{Im}~\Sigma_{1p2h} \right) \; 
 \nonumber
\end{eqnarray}
in obvious notation.
This self-energy is employed in the sp basis denoted by states $\ket{\left\{ k (\ell \frac{1}{2})  j m_j\right\}}$, characterized by wave vector, orbital, spin, total angular momentum and its projection (suppressing isospin).
Isospin conservation for ${}^{40}$Ca is employed but the Coulomb contribution is included for protons.

\begin{figure}[btp]
\begin{center}
\includegraphics[scale=0.5]{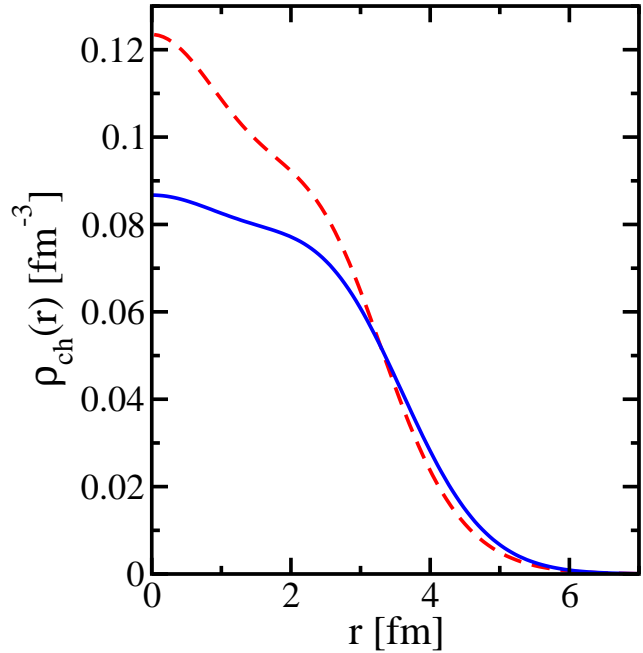}
\caption{ Charge density distribution for ${}^{40}$Ca from the CDBonn self-energy (dashed) compared to experiment (solid). \label{fig:cd}}
\end{center}
\end{figure}

The influence of treating SRC with this approach is illustrated in Ref.~\cite{Dussan11} by a presentation of the momentum content of the spectral functions and the associated momentum distribution demonstrating a presence of a modest 10\% of the nucleons with momenta not contained in the mean field.
The calculations were performed for the charge-dependent Bonn (CDBonn) interaction of 
Refs.~\cite{cdbonn,cdbonna}.
Here we focus on the results most relevant for the DOM issues of this review.
It is however useful to consider the quality of the charge density of ${}^{40}$Ca calculated in the present framework 
(corrected for the experimental charge distribution of a single proton and a single neutron as in Ref.~\cite{Brown79}). 
The final charge distribution is shown by the dashed line in Fig.~\ref{fig:cd} and compared to the experimental result obtained from the Fourier-Bessel analysis of Ref.~\cite{deVries1987}.

The mean square radius of the CDBonn self-energy is
\begin{equation}
\label{eq:msr}
\langle r^2 \rangle = \frac{1}{Ze} \int_0^\infty \!\!\!\! dr\ r^4 \rho_{ch}(r) ,
\end{equation}
yielding a value of 3.29 fm compared to the experimental result of 3.45 fm taken from Ref.~\cite{deVries1987}.
We note that microscopic calculations usually underestimate the experimental results (see \textit{e.g.} 
Ref.~\cite{Muther95} for ${}^{16}$O).
Better agreement with the experimental charge distribution likely requires
incorporating LRC correlations, especially since these are responsible for further depleting the $s_{1/2}$ orbit near the Fermi energy thereby reducing the charge near the origin.

The results discussed in Ref.~\cite{Dussan11} for elastic nucleon scattering demonstrate several important features.
First and not surprisingly, the results are not in agreement with data at energies dominated by surface physics.
As higher energies, there is some agreement with the DOM analysis when the contributions to the scattering cross section are limited to the partial waves considered in Ref.~\cite{Dussan11}.
The inclusion of higher partial waves going up to much larger values than are used in bound-state calculations are therefore needed for a proper treatment of the optical potential that is directly calculated for a finite nucleus.
Nevertheless, the resulting self-energy yields relevant insight into the properties of the nucleon self-energy that should be incorporated in a DOM analysis.
We therefore performed a few simple fits to represent the central part of the imaginary component of the  CDBonn self-energy in coordinate space at a given energy assuming the same form as used previously [Eqs.~(\ref{eq:nlocal}) to (\ref{ell-VanNeck})].

We have chosen to fit the imaginary part at 65 MeV partly because we expect that only at such energies does the imaginary part of these microscopic self-energy has real relevance since the role of LRC is expected to be diminished.
In practise, this means that only the $\ell = 0$ self-energy needs to be represented in terms of Eq.~(\ref{ell-VanNeck}).
If the choice of Eq.~(\ref{eq:nlocal}) is appropriate, the other $\ell$-values will be adequately represented as well.
A useful quantity to gauge the characteristic of an absorptive potential is the volume integral as discussed in Sec.~\ref{sec:FRPA}.
For local potentials this quantity is well-constrained by experimental cross sections~\cite{Charity07,Mueller11}.

\begin{figure}[bt]
\begin{center}
\includegraphics[scale=0.4]{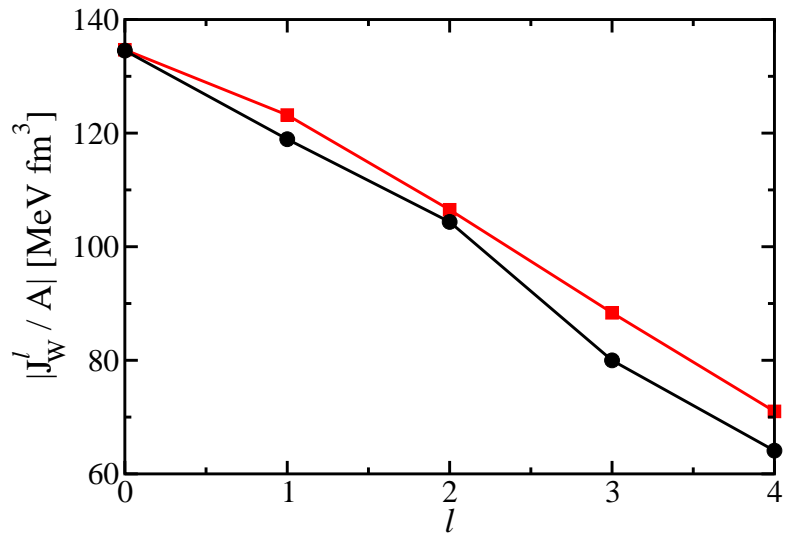}
\caption{ Imaginary volume integrals for the CDBonn self-energy at 65 MeV (black circular data points), and the corresponding result for the parametrized self-energy (red square data points).\label{fig:higherl}}
\end{center}
\end{figure}

Once a fit to the $\ell =0$ component of the self-energy has been made, 
the implied $\ell$-dependence of the chosen non-local potential leads to predictions for higher $\ell$-values.
The result of the corresponding volume integrals per nucleon are shown in Fig.~\ref{fig:higherl} as a function of the $\ell$-values considered for the CDBonn self-energy.
We employ dots for the CDBonn results and circles for the predictions based on Eq.~(\ref{eq:nlocal}).
The agreement appears very satisfactory and may permit the extraction of  the properties of the CDBonn self-energy for even higher $\ell$-values without recourse to an explicit calculation.

The properties of the imaginary part of the CDBonn self-energy in terms of its non-locality content are summarized in Table~\ref{tbl:parms} for four different energies, one below and three above the Fermi energy.
In all cases a substantial imaginary part of the CDBonn self-energy is present at the chosen energies.
The parameters are fitted at each energy to reproduce the essential properties of the self-energy including the volume integral for $\ell=0$, as discussed above for the case of 65 MeV.
\begin{table}[tb]
\begin{center}
\caption{ Parameters from non-local fits to the imaginary part of the proton self-energy at four different energies using Eq.~(\ref{eq:nlocal}).}
\label{tbl:parms}
\begin{tabular}{ccccccc}
\hline \hline
Energy & $W_0$ & $r_0$ & $a_0$ & $\beta$ & $|J_W / A|$ & $|J_W / A|_{CDBonn}$ \\
{ [MeV]} & [MeV] & [fm] & [fm] & [fm] & [MeV fm$^3$] & [MeV fm$^3$] \\
 \hline 
-76 & 36.30 & 0.90 & 0.90 & 1.33 & 193 & 193 \\ 
49 & 6.51 & 1.25 & 0.91 & 1.43 & 73 & 73 \\ 
65 & 13.21 & 1.27 & 0.70 & 1.29 & 135 & 135 \\ 
81 & 23.90 & 1.22 & 0.67 & 1.21 & 215 & 215\\ \hline \hline
\end{tabular}
\end{center}
\end{table}

We observe that the values for the diffuseness are consistent with standard values for the higher energies but are substantially larger at 49 and -76~MeV.
The radius parameter is quite small below the Fermi energy but yields rather standard values at positive energy.
The value of the non-locality parameter is quite a bit larger than typically assumed for real non-local potentials. For example,
wave-function corrections for non-locality in the analysis of $(e,e'p)$ reactions typically assume values of $\beta = 0.85$~fm~\cite{denherder}.
The DOM analysis of Ref.~\cite{Dickhoff10a} introduced a non-local HF potential to allow the calculation of additional properties below the Fermi energy from the spectral functions that are the solutions of the Dyson equation.
The adjusted non-locality parameter in that work corresponded to 0.91 fm [see Sec.~\ref{sec:nlHF}].

We note that with increasing energy, the non-locality parameter decreases suggesting a trend to a more localized potential.
Since for a local potential there is no $\ell$-dependence of the volume integral, we have investigated the behavior of $J_W^\ell$ for different $\ell$-values in a wide energy domain which confirms this conjecture~\cite{Dussan11}.
The degree of non-locality is largest below the Fermi energy with a substantial separation between the different $\ell$-values. 
At positive energies, the volume integrals for different $\ell$ at first exhibit a spread although not as large as below the Fermi energy.
However above 300 MeV, the curves apparently become similar suggesting a trend to a more local self-energy.

An analysis of the non-locality of the imaginary part to the CDBonn self-energy therefore reveals that its main properties can also be well represented by a Gaussian non-locality.
Typical non-locality parameters are somewhat larger than those found in the literature.
Volume integrals indicate that non-locality is very important below the Fermi energy.
Above the Fermi energy, it is initially substantial but appears to weaken at higher energies.

The lessons learned from these studies of LRC and SRC reported in Refs.~\cite{Waldecker2011} and \cite{Dussan11} have subsequently been incorporated in the DOM analysis requiring a full treatment of non-locality. Consequently, this shifts the traditional paradigm of local optical potentials into one in which dispersion relations and non-locality will become the mainstay.
A first such application is discussed in the next section.

\section{Introduction of non-local potentials}
\label{sec:nlocal}
In the present section we present the steps that have recently been implemented in the DOM approach to incorporate non-local potentials.
The main reason for also including non-local imaginary potentials apart from their theoretical omnipresence, is the necessity to restrict the admixture of higher $\ell$-values into the properties of the ground state.
In particular, particle number and the charge density of the nucleus provide strict constraints on the presence of these higher $\ell$-values as discussed in the following.

\subsection{Proper treatment of non-locality in the HF potential}
\label{sec:nlHF}
The first step to include non-locality in the DOM was made in Ref.~\cite{Dickhoff10}.
The spin-independent part of the HF potential in coordinate space can be denoted by $\Sigma_{HF}(\bm{r},\bm{r}')$ employing the HF label that was introduced by Mahaux and Sartor~\cite{Mahaux91} even though this term is not a strict HF contribution [see Sec.~\ref{sec:equations}]. 
The usual treatment of $\Sigma_{HF}(\bm{r},\bm{r}')$ is to assume that it can be replaced by a local but energy-dependent potential~\cite{Perey62,Fiedeldey66,Mahaux91,Dickhoff08}.
The corresponding form then can be written as
\begin{equation}
\Sigma_{HF}(\bm{r},\bm{r}') \Rightarrow \mathcal{V}_{HF}(r,E) \delta(\bm{r}-\bm{r}') ,
\label{eq:loceq}
\end{equation}
where 
\begin{equation}
\mathcal{V}_{HF}(r,E) = V_{HF} (E)f(r;r_{HF},a_{HF})
\label{eq:HFmstar}
\end{equation}
contains the usual Woods-Saxon form factor.
The factorized linear energy dependence can be parametrized by the corresponding effective mass below the Fermi energy and can be represented by
\begin{equation}
V_{HF}(E) =  V_{HF}(\varepsilon_F)+\left[1-\frac{m^*_{HF}}{m}\right] \left(E-\varepsilon_F\right) ,
\label{eq:HFEdep}
\end{equation}
which can be combined with the Woods-Saxon form factor to generate $m^*_{HF}(r)$.
This version is inspired by the Skyrme implementation of the HF potential~\cite{Mahaux91}.
More generally, one may identify this effective mass with an energy-dependent version of the effective mass $\widetilde{m}^*(r;E)$ that governs the non-locality of the self-energy and is sometimes referred to as the $k$-mass.
It was shown in Ref.~\cite{NY81} that this effective mass is critical to reconcile the phenomenological (local) imaginary part of the optical potential with the microscopic one~\cite{Dickhoff08} and to explain the observed nucleon mean free path.
For finite nuclei, this implies that the DOM version of its local imaginary part $\mathcal{W}$ is related to the self-energy by
\begin{equation}
\mathcal{W}(r;E) = \frac{\widetilde{m}^*(r;E)}{m} \mbox{Im} \Sigma(r;E) .
\label{eq:effmass}
\end{equation}
This clarifies that the use of a non-local HF self-energy in the DOM framework has to be accompanied by enhancing the imaginary part with a corresponding factor $m/\widetilde{m}^*(r;E)$.
Results discussed in Ref.~\cite{Dickhoff10} indeed corroborate the necessity of including this factor \textit{e.g.} to obtain identical spectroscopic factors as those in Ref.~\cite{Charity07}.
It is therefore possible to employ the same parameters as in the fit of~\cite{Charity07} and only replace the energy-dependent local equivalent HF potential by a suitable energy-independent non-local one.
We have chosen the standard form introduced in Ref.~\cite{Perey62} to represent
\begin{equation}
\Sigma_{HF}(\bm{r},\bm{r}') = V_{NL} f({\scriptstyle{\frac{1}{2}}}|\bm{r}+\bm{r}'|;r_{NL},a_{NL}) H\left(\frac{\bm{r}-\bm{r}'}{\beta}\right) ,
\label{eq:NL}
\end{equation}
where the degree of non-locality is expressed by a Gaussian governed by the parameter $\beta$ as in Eq.~(\ref{eq:gauss}).
This non-local form requires four parameters ($V_{NL},r_{NL},a_{NL}$, and $\beta$), which is the same number required to represent $\mathcal{V}_{HF}(r,E)$ in Eq.~(\ref{eq:HFmstar}).
We reiterate that this non-local representation is essential in obtaining properly normalized spectral functions and spectroscopic factors.
In the following we will discuss some results for ${}^{40}$Ca with this non-local version of the DOM with emphasis on energies below the Fermi energy.
We note that all fitted DOM parameters from Ref.~\cite{Charity07} have been kept fixed while only the local HF potential with its spurious energy dependence has been replaced by Eq.~(\ref{eq:NL}) and the application of the dispersion relation Eq.~(\ref{eq:sdisprel}) has been modified according to Eq.~(\ref{eq:effmass}).

As in the usual DOM fit, the location of the main fragments of the $0d_{3/2}$ and $1s_{1/2}$ valence hole levels was used to constrain the parameters of the non-local HF potential.
Since the complete one-body density matrix can be obtained with a non-local HF potential, it was also possible to constrain the parameters by the mean square radius of the charge distribution that is well known experimentally~\cite{deVries1987}.
\begin{table}[bpt]
\begin{center}
\caption{Parameters for the local energy-dependent Woods-Saxon potential and the non-local version with Gaussian non-locality for ${}^{40}$Ca.}
\label{Tbl:parmss}%
\begin{tabular}{ccc}
\hline \hline
 & local &non-local\\ 
\hline 
Depth [MeV] & -56.5 & -92.3 \\
Radius [fm] & 1.19 & 1.05 \\
Diffuseness [fm]& 0.70 & 0.70 \\
$\widetilde{m}^*_{HF}/m$ & 0.57 & -\\
Non-locality [fm] & -& 0.91 \\
\hline \hline
\end{tabular}
\end{center}
\end{table}
An additional problem that can be cured by the non-local version of the HF potential is associated with the linear energy dependence of the local version as shown in Eq.~(\ref{eq:HFEdep}).
Typical DOM fits generate rather deeply bound $0s_{1/2}$ states, often well below the peaks seen in (\textit{e,e}$^{\prime }$\textit{p}) and (\textit{p,2p}) experiments.
With a non-local potential it is possible to use the peak of the deeply bound $0s_{1/2}$ state as an additional constraint and avoid the problem.
The resulting parameters are collected in Table~\ref{Tbl:parmss} including those for the local potential.
All other parameters and the detailed shapes chosen for the imaginary parts of the DOM potential for ${}^{40}$Ca can be found in Ref.~\cite{Charity07}.
When adjusting the parameters of the non-local potential, it was found that it was possible to incorporate the constraint of the mean square radius of the charge distribution while generating quasihole fragments at energies that are at least as good as the original fit.

\begin{figure}[tbp]
\begin{center}
\includegraphics*[ scale=.5]{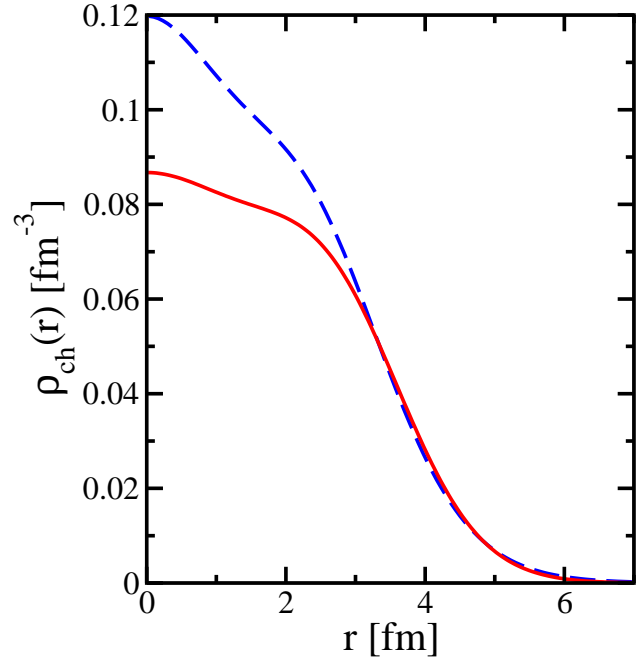}
\caption{Experimental charge density of ${}^{40}$Ca~\protect\cite{deVries1987} (solid curve) compared with the DOM result (dashed curve).}
\label{fig:chd}
\end{center}
\end{figure}

By integrating the imaginary part of the propagator given in Eq.~(\ref{eq:prop}) for each $\ell j$ combination up to the Fermi energy it is possible to obtain the one-body density matrix element
\begin{eqnarray}
n_{\ell j}(r',r)  =   \frac{1}{\pi} \int_{-\infty}^{\varepsilon_F} dE\ \mbox{Im}\ G_{\ell j}(r,r';E) = 
\bra{\Psi^A_0}a^\dagger_{r' \ell j} a_{r\ell j} \ket{\Psi^A_0} .
\label{eq:dmat}
\end{eqnarray}
For protons, the point charge distribution is thus obtained from the diagonal matrix elements of the one-body density matrix
\begin{equation}
\rho_p(r) = \frac{e}{4\pi} \sum_{\ell j} (2j+1) n_{\ell j}(r,r) .
\label{eq:chd}
\end{equation}
For a comparison with the experimental charge density of ${}^{40}$Ca it is necessary to fold this distribution with the proton charge density using the procedure outlined in Ref.~\cite{Brown79}.
The mean square radius of the resulting charge distribution is obtained from Eq.~(\ref{eq:msr}) 
and has been employed to constrain the non-local HF potential to generate good agreement with the experimental result for ${}^{40}$Ca.
The parameters in Table~\ref{Tbl:parmss} generate a value of 3.45 fm identical to the experimental result from the Fourier-Bessel analysis given in Ref.~\cite{deVries1987}.

We  compare the calculated charge density with the experimental one in Fig.~\ref{fig:chd}.
It is obvious that there is still a significant discrepancy from the experimental result near the origin which requires further analysis.
We note that the calculated charge density was normalized to $Z=20$ by dividing by the calculated proton number which exceeded 20 by more than 10 \% even when the orbital angular momentum was limited to $\ell = 5$.
The need to cut these higher $\ell$-contributions can only be accomplished by including a non-local absorption below the Fermi energy. When implementing 
 non-local imaginary potentials into the DOM as discussed in  Sec.~\ref{sec:nlIM}, it is essential that particle number is included in the fit.

\begin{figure}[tpb]
\begin{center}
\includegraphics[scale=.6]{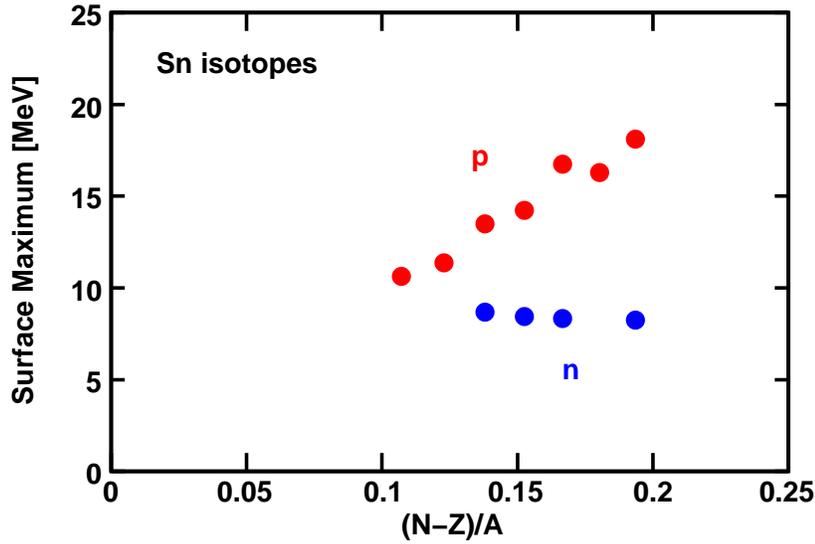}
\caption{The maximum value of the fitted 
imaginary surface potential obtained for Sn isotopes in Ref.~\cite{Mueller11} for protons and neutrons.} 
\label{fig:WasySn}
\end{center}
\end{figure}

The inclusion of non-locality in the HF potential was also applied to a series of Sn isotopes to allow a reasonable description of the radius of the charge distribution~\cite{Mueller11}.
The most important $(N-Z)/A$ dependence in the fit illustrated in Sec.~\ref{sec:localfits} is associated with the magnitude of the surface absorption.
The asymmetry dependence of the $W^{sur}_{max}$, the maximum value of 
imaginary surface potential, is plotted in Fig.~\ref{fig:WasySn} for all Sn 
isotopes studied. The protons show a substantial increase with
$(N-Z)/A$ whereas for neutrons there is almost no change. The proton data could be 
 fit by a linear relationship, however a linear extrapolation
 to $^{100}$Sn [$(N-Z)/A$=0] would give a value of $W^{sur}_{max}$
close to zero. This seems unlikely and suggests that the true asymmetry 
dependence is nonlinear.  

We have therefore constructed non-local HF potentials for the protons in Sn isotopes, while keeping the dispersive part from the DOM fits fixed, apart from the well-defined non-locality enhancement~\cite{Mahaux91}.
The non-locality of the potential is of the standard Gaussian form~\cite{Perey62} used in Ref.~\cite{Dickhoff10} for ${}^{40}$Ca.
We also employ the dispersive part of the DOM fits to extrapolate these potentials to lighter and heavier Sn isotopes than considered in the DOM fit.
The non-local potentials were required to reproduce the position of the $0g_{9/2}$ proton level and where known, the mean square radius of the charge distribution~\cite{deVries1987}.
It is well known that HF calculations only succeed in reproducing the trend of the mean square radius of the charge distribution when an $A^{1/6}$ instead of a conventional $A^{1/3}$ radius dependence is employed~\cite{Waroquier1979}.
Employing this dependence for the non-local HF potential we are able to reproduce the mean square charge radius for ${}^{112}$Sn and ${}^{124}$Sn.
It was then only necessary to adjust the depth of the potential for each isotope in order to generate the required fit to the position of the $g_{9/2}$ levels and the charge radii.
The resulting depths exhibit an essentially linear $N-Z$ dependence.

\begin{figure}[tbp]
\begin{center}
\includegraphics[scale=.6]{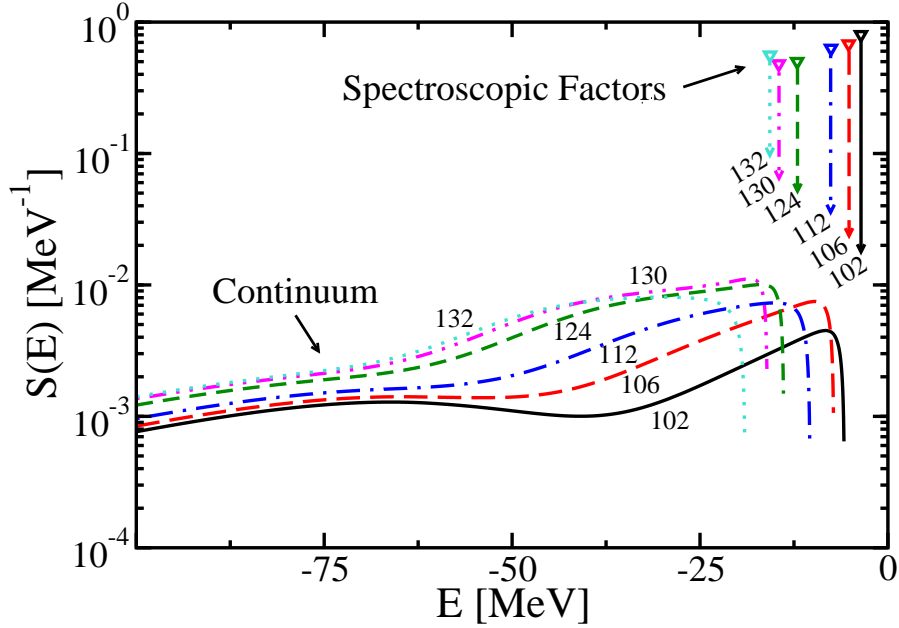}
\caption{Spectral function of the proton $g_{9/2}$ orbit in different Sn isotopes.
The curves represent the continuum contribution of the spectral function and are labeled by the appropriate mass number. Also indicated is the location of the $0g_{9/2}$ quasihole level in these different isotopes. The height of the corresponding vertical lines identifies the spectroscopic factor for each isotope.
} 
\label{fig:sfg92}
\end{center}
\end{figure}
In Fig.~\ref{fig:sfg92} the complete spectral function of the $g_{9/2}$ orbit is shown for a relevant selection of Sn isotopes. The results for different isotopes are labeled with corresponding mass number. 
The curves representing the strength in the continuum below the Fermi energy, clearly reflect the increase in the surface absorption derived from the elastic-proton-scattering data under the standard DOM assumption that surface imaginary potentials exhibit similar behavior above and below the Fermi energy~\cite{Mahaux91}. 
In the energy domain below the Fermi energy corresponding to the imaginary surface potential, a distinct increase in the strength can be observed with increasing neutron number.
Accompanying this increased strength, is a reduction of the corresponding $0g_{9/2}$ spectroscopic factor.
Quantitative results are reported in Table~\ref{Tbl:Snoccsf} for the spectroscopic factors ($S_{nl}$), strength in the continuum ($n^c_{nl}$), total occupation number ($n_{nl}$), where $nl$ refers to the non-local version of the DOM for this series of isotopes.
In addition, the occupation number ($n_l$) and spectroscopic factor ($S_l$) for the local DOM are listed that are obtained with approximate expression given in Eqs.~(\ref{eq:occHole}) and (\ref{eq:SF}) which originated in the work of Mahaux and Sartor~\cite{Mahaux91}. 
\begin{table}[bpt]
\caption{Spectroscopic factors and occupation numbers for the $0g_{9/2}$ proton orbit in Sn isotopes using the non-local ($nl$)  and local ($l$) versions of the DOM.}
\label{Tbl:Snoccsf}%
\begin{center}
\begin{tabular}{cccccc}
\hline \hline
Isotope & $S_{nl}$ & $n^c_{nl}$ & $n_{nl}$  & $n_l$ & $S_l$ \\
\hline
102 & 0.80 & 0.11 & 0.91 & 0.86 & 0.79 \\
106 & 0.68 & 0.17 & 0.85 & 0.81 & 0.68 \\
112& 0.63 & 0.20 & 0.83 & 0.74 & 0.63 \\
124 & 0.50 & 0.28 & 0.78 & 0.62 & 0.51 \\
130 & 0.48 & 0.30 & 0.78 & 0.60 & 0.49 \\
132 & 0.56 & 0.25 & 0.81 & 0.65 & 0.56\\
\hline \hline
\end{tabular}
\end{center}
\end{table}
The increase in the continuum contribution of the occupation number ends with ${}^{130}$Sn, on account of the larger gap between particle and hole states for the double-closed shell nucleus ${}^{132}$Sn.
We note that the reduction of the $g_{9/2}$ spectroscopic factor with increasing neutron number is accompanied by a weaker reduction of the occupation number.
This feature is consistent with the notion that increased surface absorption leads to removal of strength to both sides of the Fermi energy so that the reduction in the occupation should be less (and approximately half of the reduction of the spectroscopic factor for each isotope for a level very near the Fermi energy).
While the spectroscopic factors for the $0g_{9/2}$ level in the local DOM are consistent with the non-local results, there is a clear disagreement between the occupation numbers obtained by the two methods.
This confirms the conclusion of Ref.~\cite{Dickhoff10} that occupation numbers obtained from the approximate expressions of Eqs.~(\ref{eq:occHole}) and (\ref{eq:occPart}) may not always be accurate.
This is particularly true for a nominally empty level like the $g_{7/2}$, where the non-local version generated an occupation number of 0.15 in ${}^{130}$Sn, whereas the local result is 0.33.

The $N-Z$ behavior of the proton correlations obtained for Sn isotopes invites consideration of possible future experiments to confirm the trend predicted in Table~\ref{Tbl:Snoccsf}.
A consistent analysis of the ($d,{}^3$He) reaction employing a finite-range DWBA approach as in Ref.~\cite{Kramer01} might be able to shed light on the behavior of the proton $g_{9/2}$ spectroscopic factors by employing light and heavy radioactive Sn isotopes in inverse kinematics.
A serious difficulty will be the construction of appropriate optical potentials for these exotic reactions.
An alternative experimental approach might be to employ the $(d,n)$ reaction in inverse kinematics for these exotic isotopes and study the behavior of the $g_{7/2}$ spectroscopic factor for the addition of a proton.
The spectroscopic factor for this particle level tracks the one for the $g_{9/2}$ hole level reported in Table~\ref{Tbl:Snoccsf} within a few percent.

We note that the predicted behavior of the proton $g_{9/2}$ spectroscopic factor as a function of $N-Z$ is still mild compared to the deduced behavior of the removal strength using heavy-ion knockout reactions~\cite{Gade04,Gade08} in $sd$-shell nuclei.
The spectroscopic factors implied by these experiments are much smaller (or larger) far off stability than generated here for protons in Sn isotopes.
Indeed, since the implied physics is associated with surface phenomena~\cite{Fallon10}, one would expect that the remaining sp strength occurs in the domain where surface absorption takes place.
A spectroscopic factor of 0.2 for example in ${}^{132}$Sn would require a much larger surface absorption than obtained from the DOM extrapolation that generates the value of 0.56 in Table~\ref{Tbl:Snoccsf} and  correspondingly different elastic-scattering cross sections would be predicted.
The differences appear to be so large as to be amenable to experimental clarification.

\begin{figure}
\begin{center}
  \includegraphics[scale=0.4]{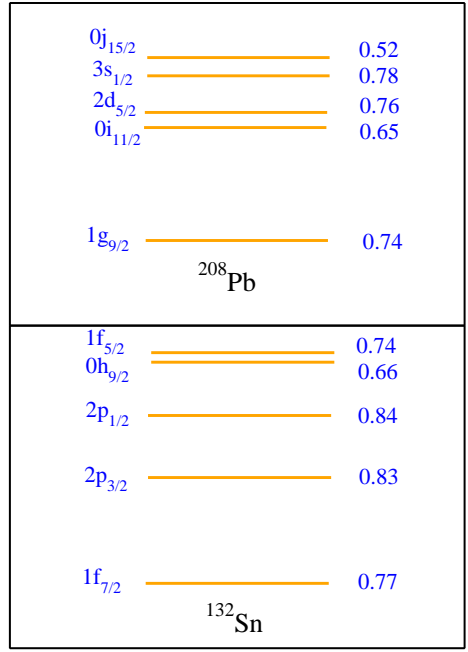}
\caption{ Properties of neutron states above the ${}^{208}$Pb core in the top panel compared with those above the 
${}^{132}$Sn core. In each panel, the levels are
labeled on the left and the corresponding spectroscopic factors are given on
the right. The levels and numbers are from the DOM implementation with a non-local HF potential~\protect{\cite{Dickhoff10}}.
} 
\label{fig:Levels132-208}
\end{center}
\end{figure}

Based on the analysis of elastic-scattering data the neutron properties appear to change considerably less than those of protons with increasing neutron number. 
Extrapolating the neutron potentials to ${}^{132}$Sn, using the non-local HF potential generates neutron particle states and related spectroscopic factors that are displayed in Fig.~\ref{fig:Levels132-208}.
A comparison with the corresponding neutron states above the ${}^{208}$Pb core demonstrates that the spectroscopic factors are of very similar nature confirming the double-closed-shell character of this exotic nucleus.

Neutron-transfer data on ${}^{132}$Sn in inverse kinematics generate unrealistic results when the traditional analysis of these experiments is performed, yielding unphysical spectroscopic factors that can be larger than 1~\cite{Jones2010,Jones2011} when normalizing these quantities as probabilities~\cite{Dickhoff08}. 
While the present analysis of these reactions yields useful information about the relative strength of these sp states and allows a comparison between ${}^{132}$Sn and ${}^{208}$Pb~\cite{Jones2010}, it is much more preferable to analyze these reactions in a way that generates physically reasonable outcomes.
A step in this direction was recently made in Ref.~\cite{Nguyen2011} and discussed in Sec.~\ref{sec:transfer}.

Further extrapolations towards the neutron drip line  in Sn isotopes yields an even more dramatic reduction of the proton spectroscopic factors.
Corresponding small reduction factors have also been extracted from heavy-ion knockout reactions for minority nucleons while the majority species appears to require almost no reduction~\cite{Gade04,Gade08}.
Transfer reactions appear to suggest a considerably smaller dependence of correlations on nucleon asymmetry~\cite{Lee10,Flavigny13} which is also suggested by FRPA \textit{ab initio} calculations~\cite{BaDi09}.
Such a DOM extrapolation was made for ${}^{154}$Sn~\cite{Moller95},  a possible candidate for the neutron drip line.
The proton $0g_{9/2}$ spectroscopic factor is a very sensitive function of the unknown separation energy of the last neutron~\cite{Charity14}.
It is therefore clear that the proximity of the neutron continuum has important consequences for the strength of correlations as measured by the valence proton spectroscopic factor.
The sensitivity of the reduction of the spectroscopic factor demonstrates the important role that the continuum could play in determining the size of the valence spectroscopic factor.
This feature was also pointed out in Ref.~\cite{Jensen11} on the basis of an \textit{ab initio} coupled-cluster calculation with a proper treatment of the continuum.
While such calculations point to a sizable reduction, they are not in agreement with the experimentally extracted reduction in stable closed-shell nuclei such as ${}^{16}$O~\cite{Leuschner94}.
The main effect of reducing the strength of the valence hole due to the presence of the continuum appears to be related to the stronger low-energy surface absorption which implies that the lost strength resides in the nearby continuum as illustrated in Fig.~\ref{fig:sfg92}. 
This also implies that the occupation number of such a valence orbit is much less sensitive to the proximity of the continuum~\cite{Mueller11}.

\subsection{Results for non-local absorptive potentials in ${}^{40}$Ca}
\label{sec:nlIM}

The discussion so far makes it very clear that in order to reproduce
ground-state properties, both a non-local imaginary self-energy and a
 non-local  real HF potential are required.
The first implementation of such an analysis for ${}^{40}$Ca was reported in Ref.~\cite{Mahzoon14}.
We now provide a more detailed description of the changes that are necessary in the conventional application of the DOM in order for the resulting potential to yield a realistic description of the sp properties below the Fermi energy.
In particular we refer to previous papers for a description of ingredients that have not changed from the purely local treatment of the DOM~\cite{Charity07,Mueller11} [see also Secs.~\ref{sec:param} and \ref{sec:localfits}].
The non-local treatment of the HF potential was discussed in Ref.~\cite{Dickhoff10} and reviewed in Sec.~\ref{sec:nlHF}.
The present form is taken as
\begin{eqnarray}
\Sigma_{HF}\left( \bm{r},\bm{r}' \right)   =   -V_{HF}^{vol}\left( \bm{r},\bm{r}'\right) 
+ V_{HF}^{wb}(\bm{r},\bm{r}') ,
\label{eq:HF}
\end{eqnarray}
where the volume term is given by
\begin{eqnarray}
\!\!\!\!\!\!\!\!\!\! V_{HF}^{vol}\left( \bm{r},\bm{r}' \right) = V_{HF}^0
\,f \left ( \tilde{r},r^{HF},a^{HF} \right ) 
\label{eq:HFvol} 
\left[ x_1 H \left( \bm{s}/\beta_{vol_1} \right) + (1-x_1) H \left( \bm{s}/\beta_{vol_2}\right) \right] ,
\end{eqnarray}
allowing for two different non-localities with different weights ($0 \le x_1 \le1$). 
We use the notation $\tilde{r} =(r+r')/2$ and $\bm{s}=\bm{r}-\bm{r}'$.
A Gaussian wine-bottle ($wb$) shape is introduced replacing the surface term of Ref.~\cite{Mueller11}
\begin{equation}
V_{HF}^{wb}(\bm{r},\bm{r}') = V_{wb}^0  \exp{\left(- \tilde{r}^2/\rho_{wb}^2\right)} H \left( \bm{s}/\beta_{wb} \right ).
\end{equation}
This Gaussian centered at the origin helps to represent overlap functions generated by simple potentials that reproduce corresponding Monte Carlo results~\cite{Brida11}. 
Non-locality is represented by a Gaussian form [see Eq.~(\ref{eq:gauss})]
first suggested in Ref.~\cite{Perey62}.
As usual, Woods-Saxon form factors are employed.
Equation~(\ref{eq:HF}) is supplemented by the Coulomb term determined from the experimental charge distribution and a local spin-orbit interaction as in Ref.~\cite{Mueller11}.

The introduction of non-locality in the imaginary part of the self-energy is well-founded theoretically both for LRC~\cite{Waldecker2011} as well as short-range ones~\cite{Dussan11} [see Secs.~\ref{sec:FRPA} and \ref{sec:SRC}]. 
Its implied $\ell$-dependence is essential in reproducing the correct particle number for protons and neutrons.
The non-local part of this imaginary component has the form
\begin{eqnarray}
\textrm{Im}\ \Sigma( \bm{r},\bm{r}',E) &=&   
-W^{vol}_0(E) f\left(\tilde{r};r^{vol};a^{vol}\right)H \left( \bm{s}/ \beta_{vol}\right)  
\nonumber
\\
&+& 4a^{sur}W^{sur}\left( E\right)H \left( \bm{s}/ \beta_{sur}\right) \frac{d}{d \tilde{r} }f(\tilde{r},r^{sur},a^{sur}) .
\hspace{0.5cm}
\label{eq:imnl}
\end{eqnarray}
We also include a local spin-orbit contribution as in Ref.~\cite{Mueller11}.
The energy dependence of the volume absorption has the form used in Ref.~\cite{Mueller11} whereas for surface absorption we employed the form of Ref.~\cite{Charity07}.
The solution of the Dyson equation below the Fermi energy was introduced in Ref.~\cite{Dickhoff10}.
The scattering wave functions are generated with the iterative procedure outlined in Ref.~\cite{Michel09} leading to a modest increase in computer time as compared to the use of purely local potentials.
Neutron and proton potentials are kept identical in the fit except for the Coulomb potential for protons.
See the supplemental material published with~\cite{Mahzoon14} for the numerical values of all parameters together with a list of all employed equations.

Included in the present fit are the same elastic-scattering data and level information considered in Ref.~\cite{Mueller11}.
In addition, we now include the charge density of ${}^{40}$Ca as given in Ref.~\cite{deVries1987} by a sum of Gaussians. 
Consideration is also given to data from the $(e,e'p)$ reaction at high missing energy and momentum obtained at Jefferson Lab.
We note that the spectral function of high-momentum protons per proton number is essentially identical for ${}^{27}$Al and ${}^{56}$Fe \cite{Rohe04} thereby providing a sensible benchmark for their presence in ${}^{40}$Ca. 
We merely aimed for a reasonable representation of these cross sections since their interpretation requires further consideration of the rescattering contributions~\cite{Barbieri04,Barbieri06}.
We did not include the results of the analysis of the $(e,e'p)$ reaction from NIKHEF~\cite{Kramer89} because the extracted  spectroscopic factors depend on the employed local optical potentials. We plan to reanalyze these data with our non-local potentials in a future study.

\begin{figure}[tbp]
\begin{center}
\includegraphics*[scale=0.6]{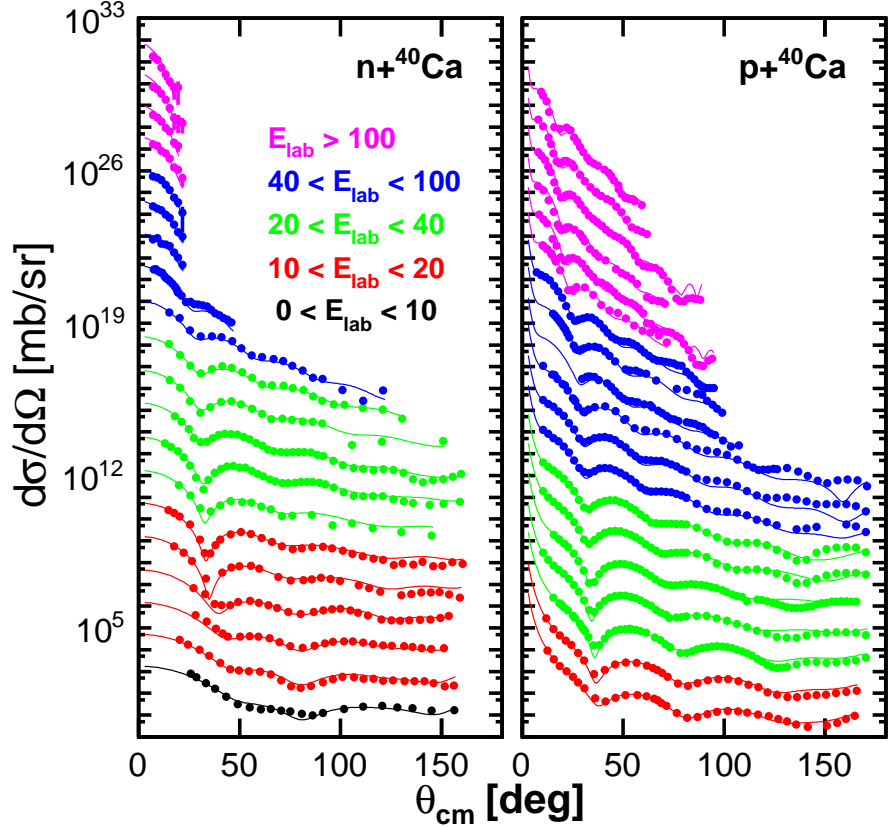}
\caption{Calculated and experimental elastic-scattering
angular distributions of the differential cross section $d\protect\sigma %
/d\Omega $. Panels shows results for \textit{n}+$%
^{40}$Ca and \textit{p}+ $^{40}$Ca. Data for each energy are offset for clarity with the lowest energy at the bottom and highest at the top of each frame. References to the data are given in Ref.~\protect\cite{Mueller11}.}
\label{fig:elast}
\end{center}
\end{figure}
Motivated by the work of Refs.~\cite{Waldecker2011,Dussan11}, we allow for different non-localities above and below the Fermi energy, otherwise the symmetry around this energy is essentially maintained by the fit.
The values of the non-locality parameters $\beta$ appear reasonable and range from 0.64 fm above to 0.81 fm below the Fermi energy for volume absorption.
These parameters are critical in ensuring that particle number is adequately described. 
We limit contributions to $\ell \le 5$ below $\varepsilon_F$~\cite{Dussan11} obtaining 19.88 protons and 19.79 neutrons. 
We note the extended energy domain for volume absorption below $\varepsilon_F$ to accommodate the Jefferson-Lab data.
Surface absorption requires non-localities of 0.94 fm above and 2.07 fm below $\varepsilon_F$.

\begin{figure}[bpt]
\begin{center}
\includegraphics*[scale=0.5]{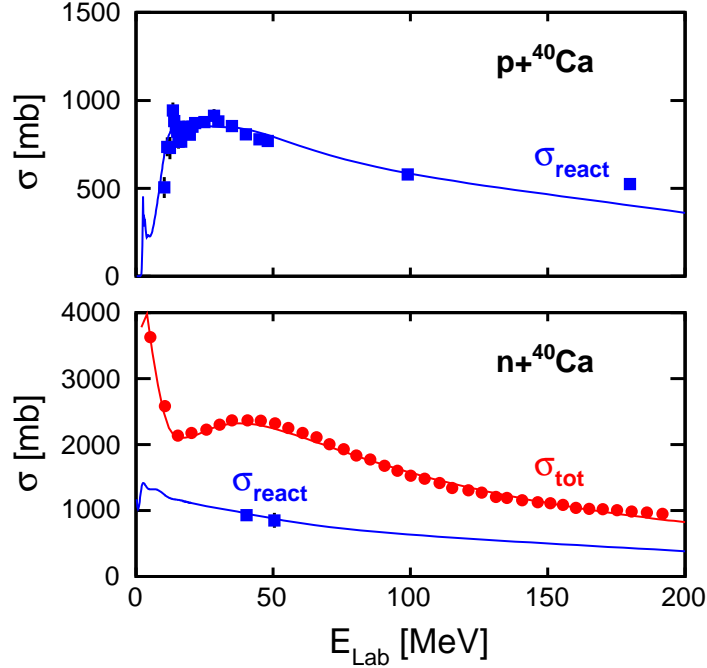}
\caption{Non-local DOM fits to the total reaction cross section are displayed as a
function of proton energy for $^{40}$Ca. Results for both total and reaction cross sections are shown for neutrons.}
\label{fig:totalc}
\end{center}
\end{figure}
The final fit to the experimental elastic-scattering data is shown in Fig.~\ref{fig:elast} 
while the fits to total and reaction cross sections are shown in Fig.~\ref{fig:totalc}.
In all cases, the quality of the fit is the same as in Refs.~\cite{Charity07} or \cite{Mueller11}.
This statement also holds for the analyzing powers.

Having established our description at positive energies is equivalent to earlier work, but now consistent with theoretical expectations associated with the non-local content of the nucleon self-energy, we turn our attention to the new results below the Fermi energy.
The spectral strength given in Eq.~(\ref{eq:specs}) as a function of energy is in good agreement with experimental information for the first few levels in the IPM~\cite{Mahzoon14}.
This includes the experimental location of the levels near the Fermi energy while
for deeply bound levels they correspond to the peaks obtained from $(p,2p)$~\cite{Jacob73} and $(e,e'p)$ reactions~\cite{FM84}.
The DOM strength distributions track the experimental results represented by their peak location and width.
\begin{table}[bp]
\begin{center}
\caption{Quasihole energies in MeV for neutron orbits in ${}^{40}$Ca near the Fermi energy compared with experiment.}
\label{Tbl:qpprop}%
\begin{tabular}{ccc}
\hline \hline
orbit  &DOM& experiment\\
\hline
$1p_{1/2}$ &  -3.47 & -4.20 \\
$1p_{3/2}$& -4.51 & -5.86 \\
$0f_{7/2}$ & -7.36 & -8.36 \\
$0d_{3/2}$ & -16.2 & -15.6 \\
$1s_{1/2}$ & -15.3 & -18.3 \\
\hline \hline 
\end{tabular}
\end{center}
\end{table}
Neutron sp energies are listed in Table~\ref{Tbl:qpprop} for levels near $\varepsilon_F$. The calculated levels exhibit a deviation of about 1 MeV from the experimental values similar to Ref.~\cite{Mueller11}, except for the 1$s_{1/2}$ orbital.

For the quasi-hole proton states we find spectroscopic factors of 0.78 for the $1s_{1/2}$ and 0.76 for the $0d_{3/2}$ level.
The location of the former level deviates from experiment as is true for neutrons and may require 
additional state dependence of the self-energy as expressed by poles nearby in energy~\cite{Dickhoff04}.
The analysis of the $(e,e'p)$ reaction in Ref.~\cite{Udias95} clarified that the treatment of non-locality in the relativistic approach leads to different distorted proton waves as compared to conventional non-relativistic optical potentials, yielding about 10-15\% larger spectroscopic factors.
Our current results are also larger by about 10-15\% than the numbers extracted in Ref.~\cite{Kramer89}.
Introducing local DOM potentials in the analysis of transfer reactions has salutary effects for the extraction of spectroscopic information of neutrons~\cite{Nguyen2011} and non-local potentials should further improve such analyses. 

\begin{figure}[bpt]
\begin{center}
\includegraphics*[scale=0.5]{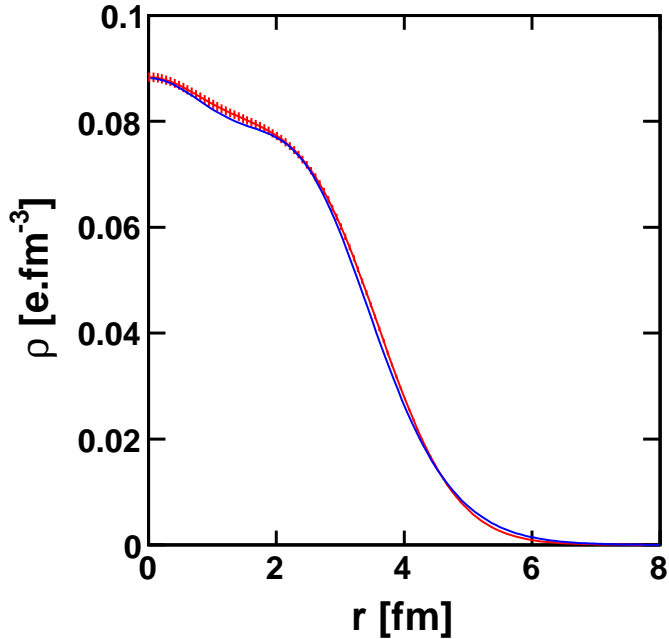}
\caption{Comparison of experimental charge density~\protect\cite{deVries1987} (hatched line) with the DOM fit (thin line).}
\label{fig:charged}
\end{center}
\end{figure}
In Fig.~\ref{fig:charged} we compare the experimental charge density of ${}^{40}$Ca (hatched line representing a 1\% error) with the DOM fit.
While some details could be further improved, it is clear that an excellent description of the charge density is possible in the DOM.
The correct particle number is essential for this result which in turn can only be achieved by including non-local absorptive potentials that are also constrained by the high-momentum spectral functions.
With only local absorption, one is  neither  able  to either generate a particle number close to 20 or describe the charge density accurately~\cite{Dickhoff10}.

\begin{figure}[btp]
\begin{center}
\includegraphics*[scale=0.6]{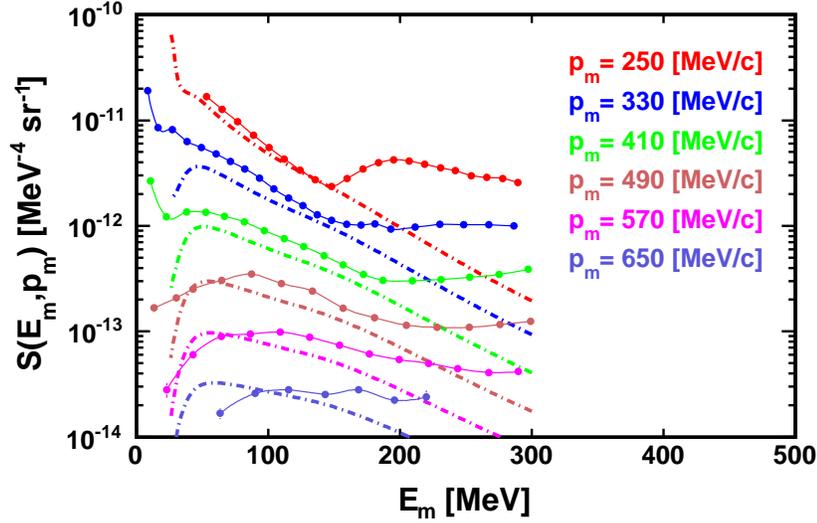}
\caption{Spectral strength as a function missing energy for different missing momenta as indicated in the figure. The data  are the average of the ${}^{27}$Al and ${}^{56}$Fe measurements from~\protect\cite{Rohe04A}.}
\label{fig:highk}
\end{center}
\end{figure}

We compare in Fig.~\ref{fig:highk} the spectral strength  for the high-momentum removal  with the Jefferson-Lab. data~\cite{Rohe04A}.
At the higher energies these data should not be reproduced in the DOM
calculations as intrinsic nucleon excitations must be considered.
To further improve the description, one would have to introduce an energy dependence of the radial form factors for the potentials. Nevertheless we conclude that an adequate description is generated which corresponds to 10.6\% of the protons having momenta above 1.4 $\textrm{fm}^{-1}$.

In conclusion we have demonstrated that the nucleon self-energy for ${}^{40}$Ca requires a non-local form and can then with reasonable assumptions represent all relevant sp properties of this nucleus.

\subsection{Spectral functions at positive and negative energies}
\label{sec:specf}

\begin{figure}[tbp]
\begin{center}
\includegraphics*[scale=0.5]{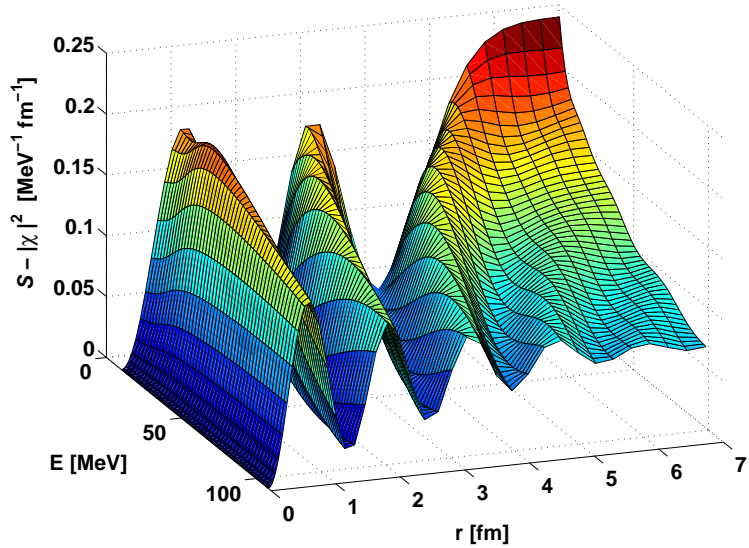}
\caption{Difference between the particle spectral function for $s_{1/2}$ and the contribution of the elastic-scattering wave function multiplied by $r^2$, as a function of both energy and position. Asymptotically with $r$, this difference is approximately constant and determined only by the inelasticity.}
\label{fig:Sandpsi}
\end{center}
\end{figure}

The calculations of spectral functions as outlined in Sec.~\ref{sec:equations} acquires additional significance with the implementation of non-local potentials.
In particular the positive energy part of the self-energy is thoroughly constrained by elastic-scattering data and provides detailed insight into the presence of some sp strength in the continuum for predominantly-bound levels.
We clarify the notion that scattered nucleons are precisely aware of the properties of the ground state in Fig.~\ref{fig:Sandpsi}.
The probability density $S_{\ell j}(r;E)$ of Eq.~(\ref{eq:specpr}) for adding a nucleon with energy $E$ at a distance $r$ from the origin for a given $\ell j$ combination is plotted after subtracting the absolute square of the 
elastic-scattering wave function given in Eq.~(\ref{eq:elwf}). 
This difference is plotted in Fig.~\ref{fig:Sandpsi} for different energies in the case of the $s_{1/2}$ partial wave.
Asymptotically at large distances, the influence of other open channels is represented by an almost constant shift whereas, inside the range of the potential, a pattern related to the absorptive properties of the potential and the orbits that are occupied emerges.
In particular, it is clear that the pattern shown in Fig.~\ref{fig:Sandpsi} demonstrates that the nucleon in this partial wave is aware of the almost full occupation of two bound $s_{1/2}$ states in the ${}^{40}$Ca 
ground state as two nodes are clearly visible at low energy thereby ensuring orthogonality of the scattered wave.
Similar plots for $\ell=2$ partial waves exhibit only one node and $\ell=4$ strength exhibits no nodes as would be expected for an essentially empty orbit.

We also present results for a microscopic calculation of the ${}^{40}$Ca self-energy obtained from the CDBonn interaction~\cite{cdbonn,cdbonna}.
Details have been provided in Ref.~\cite{Dussan11} and Sec.~\ref{sec:SRC}.
Because all ingredients of this calculation involve momentum-space quantities, the double folding in Eq.~(\ref{eq:specfunc}) is performed in momentum space utilizing overlap functions obtained in Ref.~\cite{Dussan11}. 
The experimentally-constrained non-local DOM potential of Ref.~\cite{Mahzoon14} was Fourier transformed to momentum space to allow the calculation of the off-shell reducible self-energy of Eq.~(\ref{eq:redSigma1}).
We have therefore employed the neutron self-energy for this calculation but the proton self-energy is identical apart from the Coulomb term.
Fourier transforming the spectral density according to Eq.~(\ref{eq:specpr}) allows further analysis and also to perform the folding with the bound DOM overlap functions obtained in coordinate space~\cite{Dickhoff10}.

\begin{figure}[tbp]
\begin{center}
\includegraphics*[scale=0.5]{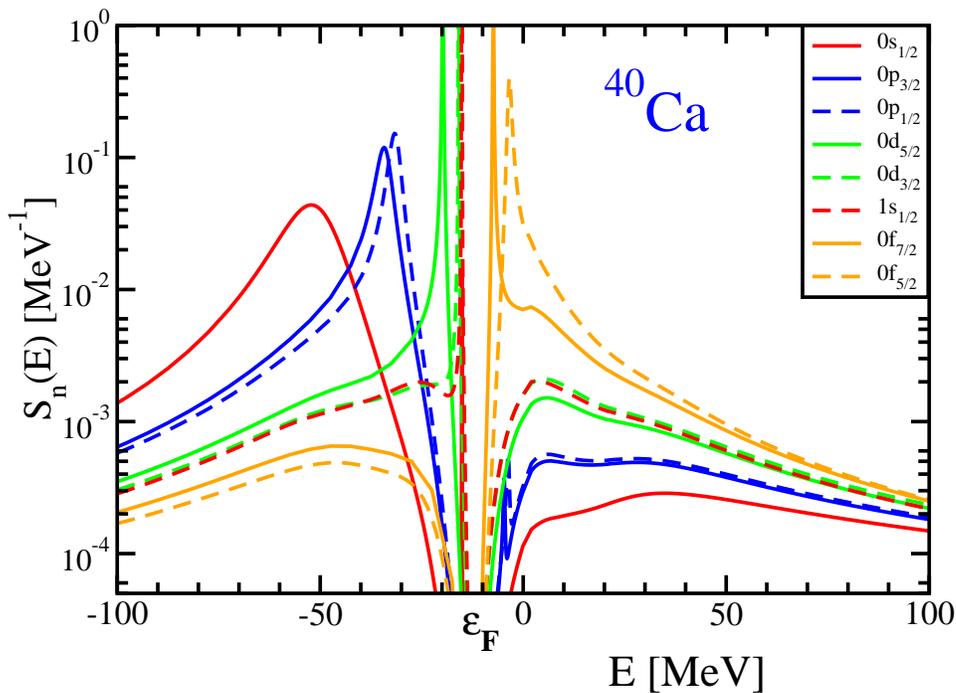}
\caption{Calculated spectral strength, both below and above the Fermi energy, for predominantly-bound orbits in ${}^{40}$Ca. The spectral strength is constrained by elastic-scattering data, level structure, charge density, particle number, and the presence of high-momenta below the Fermi energy~\protect\cite{Mahzoon14}.}
\label{fig:deplE}
\end{center}
\end{figure}

We display in Fig.~\ref{fig:deplE} the results of the DOM spectral function for the most relevant bound orbits in ${}^{40}$Ca including the hole spectral function from Ref.~\cite{Mahzoon14}.
Because the DOM analysis assumes that the imaginary part of the self-energy starts at $\varepsilon_F$, the  spectral strength is a continuous function of the energy.
The method of solving the Dyson equation for $E<0$ is very different than that for $E>0$.
The continuity of the curves at $E=0$ confirms the numerical aspects of both of these calculations. 
Below the Fermi energy, the spectral strength contains peaks associated with the $0s_{1/2},0p_{3/2},0p_{1/2},0d_{5/2},1s_{1/2},$ and $0d_{3/2}$ orbits with narrower peaks for orbits closer to the Fermi energy.
Their strength was calculated for the overlap functions associated with the location of the peaks by solving the Dyson equation without the imaginary part but with self-consistency for the energy of the real part~\cite{Dickhoff10}.
The strength of these orbits above the Fermi energy exhibits systematic features displaying more strength when the IPM energy is closer to the continuum threshold.
We make this observation quantitative by listing the integrated strength according to the terms of Eq.~(\ref{eq:sumr}) in Table~\ref{Tbl:depln}. 
For the depletion we integrate the strength from 0 to 200 MeV, 
the energy domain constrained by data in the DOM.
\begin{table}[bp]
\begin{center}
\caption{Occupation and depletion numbers for bound orbits in ${}^{40}$Ca. The $d_{nlj}[0,200]$ depletion numbers have only been integrated from 0 to 200 MeV. The fraction of the sum rule in Eq.~(\ref{eq:sumr}) that is exhausted, is illustrated by $n_{n \ell j} + d_{n \ell j}[\varepsilon_F,200]$. We also list the $d_{nlj}[0,200]$ depletion numbers for the CDBonn calculation in the last column.}
\label{Tbl:depln}%
\begin{tabular}{crrcl}
\hline \hline
orbit  & $n_{n \ell j}$ & $d_{n \ell j}[0,200]$ & $n_{n \ell j} + d_{n \ell j}[\varepsilon_F,200]$ & $d_{n_\ell j}[0,200]$ \\
& DOM & DOM & DOM & CDBonn \\
\hline
$0s_{1/2}$ & 0.926 & 0.032  & 0.958 & 0.035\\
$0p_{3/2}$& 0.914 & 0.047 & 0.961 & 0.036 \\
$1p_{1/2}$ &  0.906 & 0.051 &0.957 & 0.038 \\
$0d_{5/2}$ & 0.883 & 0.081 & 0.964 & 0.040 \\
$1s_{1/2}$ & 0.871 & 0.091 & 0.962 & 0.038 \\ 
$0d_{3/2}$ & 0.859 & 0.097 & 0.966 & 0.041 \\
$0f_{7/2}$ & 0.046 &  0.202 & 0.970 & 0.034 \\
$0f_{5/2}$ & 0.036  & 0.320 & 0.947 & 0.036 \\
\hline \hline
\end{tabular}
\end{center}
\end{table}
Results for the ``particle'' $0f_{7/2}$ and $0f_{5/2}$ spectral functions are also included in Fig.~\ref{fig:deplE} and  Table~\ref{Tbl:depln} noting that the strength in the continuum from 0 to 200 MeV now  rises to 0.202 and 0.320, respectively. 
From $\varepsilon_F$ to 0 the strength for these states is also included in the sum and decreases from 0.722 to 0.591, respectively.
This illustrates that there is a dramatic increase of strength into the continuum when the IPM energy approaches this threshold.
Such orbits correspond to valence states in exotic nuclei~\cite{Gade04,Jensen11,Charity14}.
The $1p_{3/2}$ and $1p_{1/2}$ spectral functions are not shown as they mimic the behavior of the $0f_{7/2}$ distribution but their presence causes the wiggles in the $0p_{3/2}$ and $0p_{1/2}$ spectral functions due to slight nonorthogonality.

\begin{figure}[tbp]
\begin{center}
\includegraphics*[scale=0.5]{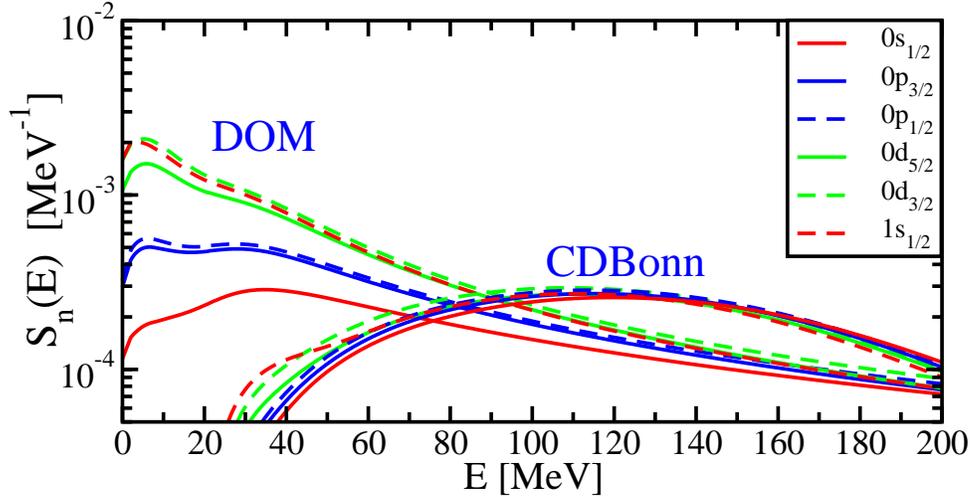}
\caption{Calculated spectral strength for mostly-occupied orbits in ${}^{40}$Ca from 0 to 200 MeV.
The CDBonn spectral functions exhibit mainly volume absorption.}
\label{fig:deplEX}
\end{center}
\end{figure}

This sensitivity to the separation from the continuum is associated with the pronounced surface absorption necessary to describe the elastic-scattering data in this energy range. 
At higher energies, volume absorption dominates and the strengths of the different orbits become similar as illustrated in Fig.~\ref{fig:deplEX}.
This figure also includes the CDBonn predictions which highlight the notion that SRC predominantly impact higher energies. 
While the CDBonn spectral function overestimates the DOM results above 100 MeV, it is quite likely that a somewhat harder interaction like Argonne V18~\cite{Wiringa1995} would move some of this excess strength to higher energy~\cite{Muther05}.

The fraction of the sum rule of Eq.~(\ref{eq:sumr}) for the DOM in Table~\ref{Tbl:depln} indicates that a few percent  of the strength occurs at energies higher than 200 MeV.
Theoretical work associates such strength with SRC~\cite{Vonderfecht91}.
No surface absorption is present in the microscopic CDBonn results and their depletions in Table~\ref{Tbl:depln} correspond to a uniform strength distribution for all orbits consistent with the SRC interpretation as illustrated in Fig.~\ref{fig:deplEX}.
As for the DOM, about 3-5\% of the strength occurs above 200 MeV when the occupation numbers of Ref.~\cite{Dussan11} are included. 

Our analysis clarifies that elastic-scattering data, combined with a complete treatment of sp properties below the Fermi energy, provide a quantitative demonstration for the presence of some continuum strength for predominantly-bound states.
This conclusion is possible because scattering and structure data are analyzed using the full non-local treatment of the DOM~\cite{Mahzoon14} which guarantees the proper treatment of sum rules like Eq.~(\ref{eq:sumr}).
The location of the strength closely tracks the absorptive properties of the self-energy and exhibits a pronounced dependence on the separation of the IPM energy to the continuum.
An \textit{ab initio} calculation of the nucleon self-energy, based on the CDBonn potential which only treats SRC,  generates a modest depletion without any pronounced dependence on the location of the orbit.
Both the CDBonn calculation and the DOM analysis allow for a few percent of strength beyond 200 MeV.
Our results therefore illustrate the importance of measuring elastic-nucleon-scattering data for exotic nuclei in inverse kinematics.
A non-local DOM analysis can then directly assess how correlations for nucleons change when the drip lines are approached.
Analysis of transfer reactions with ingredients of the non-local DOM treatment may shed further quantitative light on neutron properties in such systems~\cite{Nguyen2011,Timofeyuk13,Titus14}.

\subsection{Importance of 2N and 3N contribution to the energy of the ground state}
\label{sec:energy}
Some properties of light nuclei can nowadays be calculated in an exact manner 
with different techniques starting from a realistic NN interaction.
An example is the application of several methods to the 
calculation of the ground-state energy of ${}^4\mathrm{He}$, reported 
in~\cite{kama01}.
In addition, the low-lying states of nuclei up to $A=12$~\cite{pvw02} 
can be described with the Green's function Monte 
Carlo (GFMC) method~\cite{Pieper01}.
Such methods are able to explain many aspects of the low-energy spectra 
of light nuclei, starting from a realistic interaction.
Many details are further improved by including a three-body interaction 
between the nucleons.
In all cases studied so far, the calculated energy of the ground state is 
always above the experimental value when only two-body interactions are 
included: a clear indication for the need of an overall attractive 
three-body force which corresponds to about 1.5 MeV per particle attraction for the GFMC calculations~\cite{Pieper01}.
The two-body interaction used in the GFMC method is the AV18~\cite{Wiringa1995} which contains a substantial repulsive core capable of generating an appropriate amount of SRC~\cite{Wiringa14} demanded by exclusive electron scattering experiments~\cite{Subedi08}.
With the addition of  three-body interactions, an excellent description of $p-$shell nuclei is obtained including the ground state of ${}^{12}$C.
We note that the AV18 is a local interaction which facilitates its implementation in the GFMC framework.

The issue of high-momentum components has recently been addressed in the context of developments of the non-local implementation of the DOM~\cite{Mahzoon14} discussed in Sec.~\ref{sec:nlIM}.
In Fig.~\ref{fig:highk} the results for the  spectral strength for high-momentum removal obtained in Ref.~\cite{Mahzoon14} are compared with the Jefferson-Lab. data~\cite{Rohe04A}.
An adequate description is generated where 10.6\% of the protons have momenta above 1.4 $\textrm{fm}^{-1}$.
The energy sum rule for the ground state~\cite{Dickhoff08} can be expressed as in Eq.~(\ref{eq:erule}) which is valid only when  two-body interactions are present in the Hamiltonian.
In practice it is convenient to perform this calculation in momentum space employing the momentum distribution 
$n_{\ell j}(k)$ and  spectral function $S_{\ell j}(k;E)$  for a given $\ell j$ combination.
The results for ${}^{40}$Ca quoted below are corrected for center-or-mass effects in the form given in Ref.~\cite{Dieperink74}.
A binding energy of 7.91 MeV/$A$ is obtained that is much closer to the experimental value of 8.55 MeV/$A$ compared to the result for the local DOM results~\cite{Dickhoff10a}.
The constrained presence of the high-momentum nucleons is responsible for this change~\cite{Muther95}.
The 7.91 MeV/$A$ binding incorporates the contribution to the ground-state energy from two-body interactions including a kinetic energy of 22.64 MeV/$A$ and was not part of the fit.
This empirical approach therefore leaves about 0.64 MeV/$A$ of attraction for higher-body interactions but probably should be accompanied by an error of similar size at this stage. In interpretating this value, one should consider the  
modified energy sum rule in the presence of a three-body interaction  $\hat{W}$ \cite{Carbone13}
\begin{eqnarray}
\!\!\!\!\!\!\!\!\!\!\!\!\!\!\!\!\!\!\!\!\!\!\!\!\!\!\!\!\! E^N_0 
= \frac{1}{2\pi} \int_{-\infty}^{\varepsilon_F^-} \!\! dE\
\sum_{\alpha,\beta} \left\{ \bra{\alpha} T \ket{\beta}
+\ E\ \delta_{\alpha,\beta}
\right\} \textrm{ Im } G(\beta,\alpha;E) 
 - \frac{1}{2} \bra{\Psi^N_0} \hat W \ket{\Psi^N_0} ,
\label{eq:erule3}
\end{eqnarray}
where the last term denotes the explicit contribution from the three-body interaction. This equation clarifies that in addition an explicit  \textit{repulsive} contribution of the three-body expectation value of 1.28 MeV/$A$ is required on account of the extra minus sign for the $\hat W$ term.
This analysis can be compared to the approximately 1.5 MeV/$A$ \textit{attraction} contribution needed for light nuclei
in the GFMC calculations \cite{Pieper01}.
A 1.5 MeV/A contribution from three-body interactions, whether attractive or repulsive, is still small compared to the 30 MeV/$A$ attraction from the two-body terms. This suggests that higher-order contributions are minimal.
A discussion of the implications for the problem of nuclear saturation properties can be found in Ref.~\cite{Dickhoff16}.


\section{Conclusions and outlook}
\label{sec:outlook}
In this review the current status of the DOM has been presented. 
A substantial extension compared to its earlier implementation~\cite{Mahaux91} has been documented.
In particular, it is now possible to include elastic-scattering data up to 200 MeV in the analysis.
By simultaneously describing sequences of isotopes, a sensible extrapolation of DOM potentials can be generated towards the respective drip lines.
Successful applications have been made to all magic and semi-magic nuclei for Ca isotopes and heavier systems.
Predictions for future experiments are therefore possible which can be performed to calibrate such extrapolations.
In this sense we can speak of data-driven extrapolations.

To gain further insights into the functional forms for the DOM potentials that are compatible with underlying theory, two studies have been reviewed that generate a nucleon self-energy starting from a realistic nucleon-nucleon interaction.
In one approach the FRPA method has been applied to closed-shell Ca isotopes to clarify the role of low-energy or LRC on the nucleon self-energy.
Two conclusions are important for DOM work, namely the strength of absorptive potentials above and below the Fermi energy associated with LRC break the symmetry that was assumed in earlier DOM work~\cite{Mahaux91} and also these imaginary potentials are non-local.
The latter feature is also an important conclusion of the calculation for the ${}^{40}$Ca self-energy that treats SRC.
Non-local imaginary potentials are therefore on a solid theoretical footing and deserve implementation in all future work related to optical potentials.
In addition, we note that causality is another intrinsic property that has been neglected in most applications of optical potentials but provides a critical ingredient of the DOM allowing it to link nuclear structure and nuclear reactions.

A related motivation for the implementation of non-local potentials is the possibility to increase the quality of the description of ground-state properties, like particle number, charge density, the one-body density matrix, and related spectral densities and spectral functions.
The restoration of the non-locality in the HF potential accompanied by a rescaling of the imaginary potentials that remain local, is in itself not sufficient to allow a successful description of observables related to the ground state. We have documented that particle number and the charge density cannot be reproduced when local imaginary potentials are employed even when the HF potential is properly non-local.

The main transformation necessary to accomplish a successful description of ground-state properties is therefore the implementation of a fully non-local dispersive potential with separate absorption properties above and below the Fermi energy.
We have reviewed the first application of this new DOM potential to ${}^{40}$Ca which successfully describes elastic-scattering data and all available quantities that are experimentally accessible related to the ground state of this nucleus.
These properties include an accurate description of the charge density, and other spectroscopic information related to sp properties.

Accompanying the non-local fit has been the possibility to compute sensible particle spectral functions that are constrained by elastic-scattering data.
The presence of sp strength associated with orbits that are mostly occupied in the ground state of ${}^{40}$Ca is unambiguously demonstrated in the continuum domain of elastic-scattering processes.
This strength is unavoidable once it is realized that the description of elastic scattering requires the presence of absorptive potentials.
The DOM fit demonstrates that LRC associated with surface physics play an important role in the dramatic dependence of the integrated strength in the continuum upon the distance of the bound state to this continuum.
Such features may be relevant in the analysis of nuclei with loosely bound nucleons in the vicinity of the respective drip lines.

The constraints on the DOM provided by the presence of high-momentum protons demonstrated in Jefferson-Lab. experiments, suggests that about 10\% of the nucleons in ${}^{40}$Ca have momenta beyond the mean-field domain.
Their importance for the energy of the ground state was documented.
Furthermore, an analysis of the contribution of the two-body interaction to the energy of the ground state employing the energy sum rule requires a sensible reduction of the contribution of three-body forces comparable in magnitude to what has been found in light nuclei.

The initial application of DOM ingredients in the analysis of transfer reactions demonstrates beneficial consequences with regard to the interpretation of such data for extracting spectroscopic strength.
Applications of the analysis of the $(d,p)$  reaction on ${}^{48}$Ca at several energies clarifies that an energy-independent value of the extracted spectroscopic strength is possible with DOM potentials and overlap functions contrary to the traditional approach.
For the exotic ${}^{132}$Sn nucleus, the extrapolated DOM potentials lead to sensible results for the extraction of spectroscopic strength also improving on the traditional approach.
Furthermore, recent work on employing non-local potentials in the description of transfer reactions emphasizes the importance of this intrinsic property of optical potentials.
Non-local potentials generate distorted neutron or proton wave in the interior of the nucleus that are suppressed with respect to those that are generated by local potentials.
This intriguing feature may explain why our current values of spectroscopic factors for the removal of valence protons for ${}^{40}$Ca are about 0.15 larger than those extracted by the NIKHEF group employing local optical potentials.
We are therefore presently revisiting the analysis of $(e,e'p)$ reaction including non-local DOM potentials to shed further light on this issue.

In the immediate future we will conclude the DOM analysis of ${}^{48}$Ca incorporating an accurate fit of the charge density of this nucleus.
The intriguing possibility exists that data related to neutrons in this system may provide constraints that help determine the neutron distribution~\cite{Dickhoff16b}.
The resulting neutron skin will then provide supplementary insight into this quantity not available with other approaches.
An analysis of ${}^{208}$Pb is therefore another high priority.
Corresponding DOM fits to sequences of oxygen isotopes are also planned as elastic proton scattering on some unstable isotopes can be studied in inverse kinematics.

\section*{Acknowledgement}
The material presented in this review is based mostly upon work supported by the U.S. Department of Energy, Office of Science, Office of Nuclear Physics under Award number DE-FG02-87ER-40316 and by the U.S. National Science Foundation under grants PHY-0968941 and PHY-1304242.
The authors gratefully acknowledge the productive collaborations with Carlo Barbieri, Helber Dussan, Jon Mueller,  N. B. Nguyen, Filomena Nunes, Arturo Polls, Alaina Ross, Lee Sobotka, Luke Titus, Dimitri Van Neck, and Seth Waldecker who contributed to several aspects of the topics discussed in this review.

\section*{References}
\bibliographystyle{iopart-num}
\bibliography{DOM_review}

\end{document}